\documentclass[numberedappendix,appendixfloats,apj,twocolappendix]{emulateapj}
\pdfoutput=1
\usepackage{CJK}
\usepackage{multirow}
\usepackage{natbib}
\usepackage{amsmath}
\usepackage{hyperref}
\hypersetup{colorlinks = true, citecolor = blue}
\newcommand{\sersic}{S\'{e}rsic}
\newcommand{\galfit}{\texttt{GALFIT}}
\newcommand{\ellipse}{\texttt{ellipse}}
\makeatletter

\newcommand{\Rmnum}[1]{\expandafter\@slowromancap\romannumeral #1@}
\makeatother
\shorttitle{An Optimal Strategy for Bulge-to-disk Decomposition}
\shortauthors{Gao \& Ho}
\begin{document}
\begin{CJK*}{UTF8}{gkai}
\title{An Optimal Strategy for Accurate Bulge-to-disk Decomposition of Disk Galaxies}

\author{Hua Gao (高桦)\altaffilmark{1,2} and Luis C. Ho\altaffilmark{2,1}}

\altaffiltext{1}{Department of Astronomy, School of Physics, Peking University, Beijing 100871, China}
\altaffiltext{2}{Kavli Institute for Astronomy and Astrophysics, Peking University, Beijing 100871, China}

\begin{abstract}
  The development of two-dimensional (2D) bulge-to-disk decomposition techniques has shown their advantages over traditional one-dimensional (1D)
  techniques, especially for galaxies with non-axisymmetric features. However, the full potential of 2D techniques has yet to be fully
  exploited. Secondary morphological features in nearby disk galaxies, such as bars, lenses, rings, disk breaks, and spiral arms, are seldom accounted
  for in 2D image decompositions, even though some image-fitting codes, such as \galfit{}, are capable of handling them. We present detailed,
  2D multi-model and multi-component decomposition of high-quality \(R\)-band images of a representative sample of nearby disk galaxies selected from
  the Carnegie-Irvine Galaxy Survey, using the latest version of \galfit{}. The sample consists of five barred and five unbarred galaxies, spanning
  Hubble types from S0 to Sc. Traditional 1D decomposition is also presented for comparison. In detailed case studies of the 10 galaxies, we
  successfully model the secondary morphological features. Through a comparison of best-fit
  parameters obtained from different input surface brightness models, we identify morphological features that significantly impact bulge
  measurements. We show that nuclear and inner lenses/rings and disk breaks must be properly taken into account to obtain accurate bulge
  parameters, whereas outer lenses/rings and spiral arms have a negligible effect. We provide an optimal strategy to measure bulge parameters of
  typical disk galaxies, as well as prescriptions to estimate realistic uncertainties of them, which will benefit subsequent decomposition of a larger
  galaxy sample.
\end{abstract}

\section{Introduction}
\label{sec:introduction}

Bulges of disk galaxies, along with ellipticals as their counterparts on the other end of the Hubble sequence, play a central role in understanding
galaxy formation and evolution. Bulges were once recognized as small ellipticals living in the center of disks, since they bear similarities with
ellipticals in many aspects of their observational properties \citep{1962ApJ+ELS+Galaxy_Collapse,
  1974IAUS+de_Vaucouleurs+statistics_structure_dynamics,1977egsp.conf+Faber+chem_comp_old_stel_pop,1977ARA&A+Gott+gal_form_theory,
  1999fgb+Renzini+Origin_of_bulge}. They were both thought to form out of rapid, violent processes such as gravitational collapse
\citep{1962ApJ+ELS+Galaxy_Collapse, 2016ASSL+Bournaud+clump_high-z_disk_bulge_growth} and galaxy mergers
\citep{1977egsp+Tmoore+Mergers_and_Consequences}. However, as observations improved, bulges revealed a diversity of observational properties that
suggest distinct formation paths. Some bulges show younger stellar populations, more flattened stellar light distribution, and more rotation-dominated
kinematics (see \citealp{1997ARA&A+Wyse+bulge_heterogeneity,2004ARA&A+Kormendy+Pseudobulges} for classical reviews, and
\citealp{2016ASSL+Laurikainen+bulge_book} for a recent review). This diversity in bulge characteristics led to a re-evaluation of the simple picture of
bulge formation. In addition to violent processes, it has been increasingly appreciated that secular evolution, facilitated by non-axisymmetries in the
galaxy potential, is able to transport gas with low angular momentum to galaxy centers or to heat disk stars to rise above the disk plane,
consequently building up bulge-like components that resemble disks rather than merger-built ellipticals (e.g.,
\citealp{1981A&A+Combes+boxy_bulge_thick_bar,1981seng.proc+Kormendy+structure_barred_galaxy,1982SAAS+Kormendy+Obs_gal_stru_dyn,
  1993IAUS+Kormendy+disky_bulge_and_origin,1993RPPh+Sellwood+Bar_Dynamics,1996FCPh+Buta+Galactic_Rings,2004ARA&A+Kormendy+Pseudobulges,
  2005MNRAS+Athanassoula+Boxy_Bulges_Simulation,2014RvMP+Sellwood+secular_evo_disk_gal, 2016MNRAS+Tonini+fast_secular_evolution}). A new
terminology---the pseudobulge---was invented to distinguish bulges that are disk-like from classical bulges. The recognition of
pseudobulges, alongside the discovery of pure disk galaxies, both of which are vulnerable to the overwhelming effects of major mergers predicted by
\(\mathrm{\Lambda CDM}\), poses challenges to the canonical hierarchical clustering and merging scenario
\citep{2005RMxAC+Kormendy+growth_pseudobulges_problem_lambdaCDM,2008IAUS+Kormendy+growth_pseudobulges_problem_lambdaCDM,
  2008ASPC+Kormendy+growth_pseudobulges_problem_lambdaCDM,2010ApJ+Kormendy+bulgeless_lambdaCDM,2016ApJ+Sachdeva+survival_pure_disk}. Moreover, the
interplay between galaxy spheroids (bulges and ellipticals) and their central supermassive black holes has attracted much interest \citep[and
references therein] {2013ARA&A+Kormendy+coevolution_BHs_host}. Indeed, bulges record the evolutionary history of galaxy assembly and host physical
processes that govern galaxy evolution from small to large scales. The importance of bulges warrants robust quantitative measurements.

Parametric fitting of galaxy surface brightness has long proved to be a powerful tool to quantify galaxy spheroids
\citep{1948AnAp+de_Vaucouleurs+Profile,1959HDP+de_Vaucouleurs+de_Vauc_exp,1968adga+Sersic+Sersic_Function,1970ApJ+Freeman+Exp_Disk,
  1977ApJ+Kormendy+Kormendy_relation,1977ApJ+Kormendy+first_BD_Decom}. Many important scaling relations have been established as a byproduct of
parametric fitting, namely the Kormendy relation \citep{1977ApJ+Kormendy+Kormendy_relation}, the fundamental plane \citep{1987ApJ+Djorgovski+FP,
  1987nngp+Faber+FP1}, and empirical correlations between bulges and black holes \citep{2013ARA&A+Kormendy+coevolution_BHs_host}. Furthermore, scaling
relations help to differentiate spheroids formed through distinct pathways (e.g., \citealp{1999ApJ+Carollo+FP_PB_Lower_Sur,2008AJ+Fisher+Sersic_Index,
  2009MNRAS+Gadotti+SDSS_FP,2009ApJS+Kormendy+E_Sph+E_E_1D,2010ApJ+Fisher+Scaling_Relation_Bulges}). The success of parametric fitting is
indisputable. However, the assumption of analytic functions without a strong physical basis for such fitting is one vital but unavoidable
shortcoming. Non-parametric methods specifically designed for bulge-to-disk decomposition that rely solely on the distinct apparent ellipticities of
the bulge and disk are most effective for highly inclined galaxies (e.g., \citealp{1986AJ+Kent+BD_dec_non_para,1987AJ+Capaccioli+BD_dec_non_para,
  1990A&A+Scorza+BD_dec_non_para,1990A&A+Simien+BD_dec_non_para}). In relatively face-on cases, even non-parametric methods need to impose some
constraints on the profiles of the component in order to separate them. For instance,
\texttt{DiskFit}\footnote{\url{http://www.physics.rutgers.edu/~spekkens/diskfit/}} employs a hybrid scheme that assumes a parametric profile for the
bulge component but none for the bar and disk \citep{2003AJ+Barnes+DiskFit_ori,2007AJ+Reese+DiskFit_bar,2015arXiv+Sellwood+DiskFit}.  Other
non-parametric strategies, such as decomposing the image into a series of basis functions (Gaussians: \citealp{2002MNRAS+Cappellari+MGE}; wavelets:
\citealp{1998ipda+Starck+image_proc_data_analy}; shapelets: \citealp{2003MNRAS+Refregier+shapelets1,2003MNRAS+Refregier+shapelets2}), only suffice to
characterize the global surface brightness of galaxies; they are incapable of separating individual structural components.  Thus, in order to
decompose the individual structural components of disk galaxies, there are no practical alternatives to parametric fitting.

There are two categories of parametric techniques: one-dimensional (1D) fitting of surface
brightness profile of galaxies and two-dimensional (2D) fitting of galaxy images. 1D fitting
was the exclusive technique in early studies (e.g., \citealp{1977ApJ+Kormendy+Kormendy_relation,1977ApJ+Kormendy+first_BD_Decom,1979ApJ+Burstein+BD_dec_1D,
  1985ApJS+Kent+BD_decomp_1D}), and it is employed widely still (e.g., \citealp{2005ApJ+Erwin+Type_III,2006ApJS+Ferrarese+E_Sph_1D,
  2008AJ+Erwin+Disk_Classes,2008AJ+Fisher+Sersic_Index,2009ApJ+Fisher+PB_Growth,2009ApJS+Kormendy+E_Sph+E_E_1D,2010ApJ+Fisher+Scaling_Relation_Bulges,
  2015MNRAS+Erwin+Composite_Bulges,2016ApJS+Savorgnan+SBH_spheroids}), owing to its simplicity and perhaps computational speed. Despite its
advantages, 1D fitting does have some shortcomings. There is no consensus on how to extract the surface brightness profile---the basic input for 1D
fitting---from the galaxy image. One can extract azimuthally averaged profiles by fitting elliptical isophotes to images or, alternatively, one can
extract the radial profiles through a cut along the galaxy major or minor axis. Each method has its own pros and cons. Azimuthally averaged profiles make
full use of images but isophote twists may introduce ambiguities.  Major/minor axis cuts lose much information but they are useful when one wishes to
emphasize or de-emphasize certain components (e.g., bars).  Most crucially, 1D fitting cannot preserve spatial information such as variations in
ellipticities and orientations of structural components. Although radial ellipticity (\(\epsilon\)) and position angle (PA) profiles are measured in
the isophotal analysis, this information, which can help break the degeneracy between structural components (e.g., bulges usually appear rounder than
disks, bars are more flattened and often have different PAs compared to other components), cannot be used in the fitting.  Thus, 1D fitting is intrinsically
less capable of handling multi-component fits.  Moreover, 1D fitting cannot properly account for the smearing effects of the image point-spread
function (PSF) because 1D convolution does not conserve flux.

All these shortcomings can be overcome in 2D fitting, where ambiguities in extracting surface brightness profile naturally vanish, full spatial
information can be retained, and the effects of PSF smearing can be properly taken into account by convolution of 2D PSF images with model
images. Development of 2D image fitting tools began almost 30 years ago (e.g., \citealp{1989MNRAS+Shaw+first_BD_dec_2D,1995ApJ+Byun+BD_dec_2D_vs_1D,
  1996A&A+de_Jong+BD_dec_2D}) and has increasingly flourished in recent years (e.g.,
\texttt{GIM2D}\footnote{\url{http://www.astro.uvic.ca/~simard/GIM2D/}}: \citealp{1998ASPC+Simard+GIM2D1,2002ApJS+Simard+GIM2D2};
\texttt{BUDDA}\footnote{\url{http://www.sc.eso.org/~dgadotti/budda.html}}: \citealp{2004ApJS+de_Souza+BUDDA1,2008MNRAS+Gadotti+BUDDA2}; 
\galfit{}\footnote{\url{https://users.obs.carnegiescience.edu/peng/work/galfit/galfit.html}}: \citealp{2002AJ+Peng+GALFIT1,2010AJ+Peng+GALFIT2};
\texttt{BDBAR}: \citealp{2004MNRAS+Laurikainen+BDBAR,2005MNRAS+Laurikainen+multi_comp_dec_S0}; \texttt{GASP2D}: \citealp{2008A&A+Mendez-Abreu+GASP2D1,
  2010A&A+Mendez-Abreu+GASP2D2}; \texttt{IMFIT}\footnote{\url{http://www.mpe.mpg.de/~erwin/code/imfit/}}: \citealp{2015ApJ+Erwin+IMFIT}).  Many direct
comparisons have been made between 1D and 2D techniques.  Idealized galaxy image simulations have shown that 2D fitting recovers structural parameters
better than 1D fitting (e.g., \citealp{1995ApJ+Byun+BD_dec_2D_vs_1D, 1996A&A+de_Jong+BD_dec_2D}). The emergence of 2D fitting tools makes it practical
to fit non-axisymmetric galaxy features, such as lopsidedness, bars, and spiral arms (see \citealp{2010AJ+Peng+GALFIT2} for successful examples).
Despite the great potential of such tools, few studies attempt to explore beyond the two basic bulge and disk components (e.g.,
\citealp{2003ApJ+Gadotti+inner_disk_barred_galaxy, 2006MNRAS+Allen+BD_Decomposition_10095,2015MNRAS+Meert+BD_dec_2D_SDSS,
  2016MNRAS+Kennedy+BD_decomp_GAMA_multi_band,2016ApJS+Kim+SDSS_BD_dec}; but see \citealp{1996A&A+de_Jong+BD_dec_2D,2004MNRAS+Laurikainen+BDBAR,
  2005MNRAS+Laurikainen+multi_comp_dec_S0,2006AJ+Laurikainen+BD_dec_S0,2008MNRAS+Gadotti+BUDDA2,2009MNRAS+Gadotti+SDSS_FP,2014ApJ+Kim+Disk_Break,
  2015MNRAS+Head+multi_decom_coma,2015ApJS+Salo+S4G_multi_comp_dec} for aggressive examples). Bars, if not properly modeled, are known to introduce
major uncertainties in bulge parameters (e.g., \citealp{2004MNRAS+Laurikainen+BDBAR,2005MNRAS+Laurikainen+multi_comp_dec_S0,
  2008MNRAS+Gadotti+BUDDA2}). Still, lenses, rings, disk breaks, and spiral arms, common morphological features in nearby disk galaxies, are seldom
accounted for in 2D bulge-to-disk decomposition studies. These features are often considered gentle perturbations or minor accessories to the dominant
underlying galaxy surface brightness, having marginal impact on the principal structural components. However, some studies prove
otherwise. \citet{2005MNRAS+Laurikainen+multi_comp_dec_S0,2006AJ+Laurikainen+BD_dec_S0} show that lenses are essential components that need to be
modeled in S0s, and \citet{2014ApJ+Kim+Disk_Break} demonstrate that disk breaks are crucial to derive accurate bulge structural parameters.

We are interested in quantifying the fundamental parameters of the bulge component, namely its total magnitude (\(m\)), effective surface brightness
(\(\mu_{{e}}\)), effective radius (\(r_{{e}}\)), shape of its surface brightness profile as characterized by a \citet{1968adga+Sersic+Sersic_Function}
index (\(n\)), and apparent ellipticity (\(\epsilon\)). Based on these parameters, we can infer the luminosities, colors, stellar masses, and star
formation rates of the bulge. \sersic{} indices are commonly used to distinguish pseudobulges from classical bulges \citep{2008AJ+Fisher+Sersic_Index,
  2009ApJ+Fisher+PB_Growth,2010ApJ+Fisher+Scaling_Relation_Bulges}. In addition, fundamental plane correlations (e.g., \(\mu_{\mathrm{e}}\) vs.
\(r_{{e}}\)) can also be used to differentiate bulge types \citep{2009MNRAS+Gadotti+SDSS_FP}. Apparent ellipticities of spheroids are related to their
kinematics through the \(V/\sigma\text{--}\epsilon\) diagram (e.g., \citealp{1977ApJ+Illingworth+rotation_13ellipticals,
  1978MNRAS+Binney+rotation_elliptical,1982ApJ+Kormendy+rotation_bulges,1993IAUS+Kormendy+disky_bulge_and_origin,2004ARA&A+Kormendy+Pseudobulges}).
While these photometric parameters are commonly derived from bulge-to-disk decomposition, their error budget is often poorly quantified. Subjective
evaluation of galaxy surface brightness models is one of the major sources of systematic error, in cases where galaxies are well-resolved and have
sufficiently high signal-to-noise ratio (S/N). For example, as mentioned above, ignoring bars or disk breaks will cause noticeable biases in bulge parameters.

This study aims to clarify quantitatively whether secondary morphological features (lenses, rings, disk breaks, spiral arms) need to be included in 2D
image decomposition of disk galaxies.  We experiment with 10 representative disk galaxies selected from the Carnegie-Irvine Galaxy Survey (CGS;
\citealp{2011ApJS+Ho+CGS1}). We use \galfit{} to perform detailed 2D multi-model and multi-component decomposition of high-quality \(R\)-band
images. We start from the simplest surface brightness models that only account for major luminous components, and then gradually increase the
complexity of the models to include secondary morphological features. We pay special attention to variations of bulge parameters in response to
different input models and try to estimate their model-induced uncertainties, aiming to identify which morphological features are essential for
adequate 2D surface brightness models. In addition, 1D bulge-to-disk decomposition is also conducted in the traditional manner for comparison. We find
that 1D fitting is not adequate for most cases. The simplified assumption of exponential disks is generally not satisfactory for the purpose
  of accurately measuring bulges. Specifically, we show that, to achieve robust bulge parameters, nuclear lenses/rings, inner lenses/rings, and disk
breaks need to be properly treated using 2D fitting, while outer lenses/rings and spiral arms do not. This paper lays the groundwork for subsequent
decomposition of the entire CGS disk galaxy sample, with the aim of deriving more accurate demographics of bulges in the local universe. Note that in
this paper we do not attempt to distinguish bulge types (classical or pseudobulge) or the physics behind their appearance; we only focus on
measurements of the structural parameters of photometric bulges. Due to the limited resolution of the data, we also make no effort to separate
possible subcomponents within bulges (i.e., composite bulges: \citealp{2001A&A+Prugniel+two_pop_bulge, 2010ApJ+Kormendy+PB_embedded_in_boxy_bulge,
  2012ApJ+Barentine+two_PB_boxy_bulge,2015MNRAS+Erwin+Composite_Bulges}).

The paper is structured as follows. Details of the sample are described in Section~\ref{sec:training-sample}. Section~\ref{sec:methodology} gives an
overview of decomposition methodology, for both 1D and 2D fitting. Section~\ref{sec:bulge-disk-decomp} presents the decomposition results of each
galaxy. In Section~\ref{sec:discussion}, we discuss the relative importance of various morphological features and propose an optimal strategy for the
bulge-to-disk decomposition.  Finally, we summarize the findings of this study in Section~\ref{sec:summary}.

\section{Sample and Data}
\label{sec:training-sample}

\begin{figure*}
  \plotone{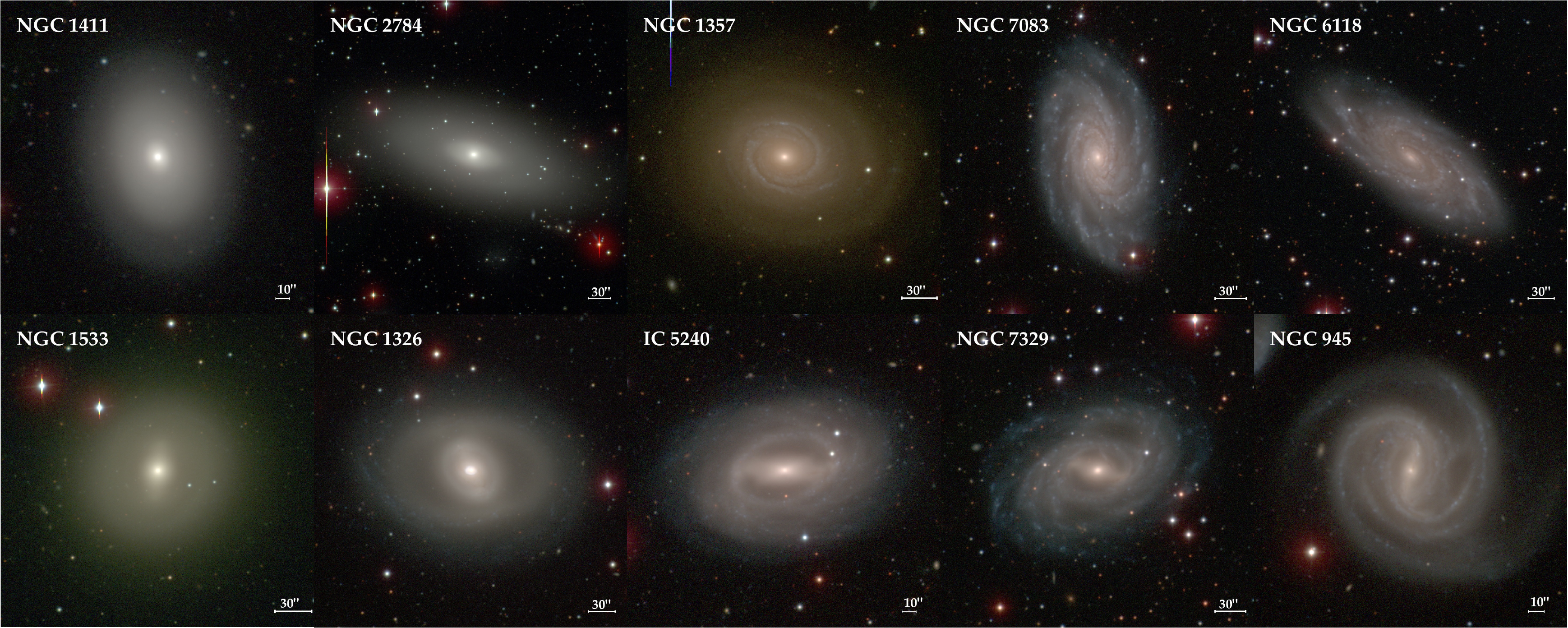}
  \caption{Image atlas of the training sample, including five unbarred galaxies in the top row and five barred ones below. Their
    morphological types range from S0 to Scd.\label{fig:atlas}}
\end{figure*}

\citet{2011ApJS+Ho+CGS1} initiated the CGS program to investigate the wealth of information stored in the structures of a statistically complete and
unbiased sample of 605 bright galaxies in the southern sky (Figure~\ref{fig:atlas}). The CGS sample is defined by $B_{T}\leq12.9\,\mathrm{mag}$ and
$\delta<0\arcdeg$, without any reference to morphology, size, or environment. The sample consists of 17\% ellipticals, 18\% S0 and S0/a, 64\% spirals,
and 1\% irregulars. The galaxies are nearby (median \(D_{L}=24.9\,\mathrm{Mpc}\)), luminous (median \(M_{B_{T}}=-20.2\,\mathrm{mag}\)), and angularly
large (median \(B\)-band isophotal diameter \(D_{25}=3\arcmin.3\)). Details of the observations and data reduction are given in
\citet{2011ApJS+Ho+CGS1}, so here we only present basic properties of the sample that concern image fitting. Images were taken in $B$, $V$, $R$, and
$I$ bands with a field of view of $8\arcmin.9\times8\arcmin.9$, using a CCD camera that has a decent pixel scale of
\(0.259\,\mathrm{arcsec~pixel^{-1}}\). The majority of the images are of high quality. The median seeing is $1\arcsec.17$, $1\arcsec.11$,
$1\arcsec.01$, and $0\arcsec.96$ for the $B$, $V$, $R$, and $I$ bands, respectively, and the corresponding median surface brightness depth is 27.5,
26.9, 26.4, and 25.3\,mag~arcsec$^{-2}$.  These characteristics combine to make an ideal sample for detailed structural decomposition. This study
focuses on the \(R\)-band images, which offer the best balance between image depth, spatial resolution, and less severe sensitivity to dust absorption
and young stars.  We avoid the \(I\)-band data, whose PSF suffers from the red halo effect \citep{2013ApJ+Huang+CGS3}.

For the purposes of this paper, which targets a representative ``training set'' of disk galaxies that can serve as a guide for the rest of CGS, we
need to select galaxies that show wide range of secondary morphological features, including lenses, rings, disk breaks, and spiral arms. For these
morphological features to be well recognized, our targets must not be highly inclined. In the meantime, we need to make sure that they span a
considerable range over the Hubble sequence. We select a sample of 10 galaxies, comprising five
unbarred and five barred galaxies, with Hubble types
ranging from S0 to Scd.  Figure~\ref{fig:atlas} shows three-color composite images of all galaxies in the sample.  The basic properties of the sample
are summarized in Table~\ref{tab:sample_property}.  Average \(B_{{T}}\), \(D_{{L}}\), and \(D_{25}\) of the sample are \(12.11\,\mathrm{mag}\),
\(26.3\,\mathrm{Mpc}\), and \(3\arcmin.6\), respectively, in fair agreement with the parent sample.

\begin{deluxetable}{lccccc}
  \tablecaption{Basic Properties of the Training Sample \label{tab:sample_property}}
  \tablecolumns{6}
  \tablehead{\multicolumn{1}{c}{\multirow{2}{*}{Name}} & \colhead{\(B_{{T}}\)} & \colhead{Leda} & \colhead{RC3} & \colhead{\(D_{25}\)} &
    \colhead{\(D_{{L}}\)}  \\
    \colhead{} & \colhead{(mag)} & \colhead{Type} & \colhead{Type} & \colhead{(\arcmin{})} & \colhead{(Mpc)} \\
    \colhead{(1)} & \colhead{(2)} & \colhead{(3)} & \colhead{(4)} & \colhead{(5)} & \colhead{(6)} }
  \startdata
  NGC 1411&12.19&E/S0&SA(r)0\(^{-}\)&2.71&15.5\\
  NGC 2784&11.19&S0&SA(s)0\(^{0}\)&5.39&\phn{}8.5\\
  NGC 1357&12.44&Sab&SA(s)ab&3.28&24.7\\
  NGC 7083&11.92&Sbc&SA(s)bc&3.70&33.9\\
  NGC 6118&12.30&Sc&SA(s)cd&4.43&23.1\\
  NGC 1533&11.82&E/S0&SB0\(^{-}\)&3.18&18.4\\
  NGC 1326&11.53&S0/a& (R)SB(r)0\(^{+}\)&4.34&16.9\\
  IC\quad\, 5240&12.69&SBa&SB(r)a&2.78&21.4\\
  NGC 7329&12.17&SBb&SB(r)b&3.77&42.6\\
  NGC \phn{}945&12.89&SBc&SB(rs)c&2.42& 58.2 
  \enddata
  \tablecomments{Column 1: galaxy name. Column 2: \(B\)-band total magnitude, from HyperLeda. Column 3: Hubble type, from HyperLeda. Column 4: Hubble
    type, from the Third Reference Catalog of Bright Galaxies (RC3; \citealp{1991Springer+de_Vaucouleurs+RC3}). Column 5: diameter at
    \(\mu_{B}=25.0\,\mathrm{mag~arcsec}^{-2}\). Column 6: luminosity distance. All quantities extracted from Tables~1 and 3 in
    \citet{2011ApJS+Ho+CGS1}.}
\end{deluxetable}

\section{Methodology}
\label{sec:methodology}

\subsection{1D Bulge-to-disk Decomposition}
\label{sec:1d-bulge-disk}

Nearby disk galaxies commonly show features such as bars, broken disks, lenses, and rings. However, these features are not taken into account in 1D
decomposition; our hands are tied by the inherent inability of 1D fitting to utilize all the spatial information contained in the full surface
brightness distribution.  Attempting to model these features along with the bulge and disk is dangerous without constraints from the PA and
\(\epsilon\). The components are degenerate with each other, making the fitting results highly unreliable, even if we can achieve good-looking fitting
residuals.

An often-adopted practical approach is to exclude from the fit part of the surface brightness profile that does not conform to the assumed
two-component (i.e.\ bulge+disk) model. This strategy was adopted by \citet{2008AJ+Fisher+Sersic_Index}. We follow a similar approach for our 1D
decomposition. We assume that the galaxies in the sample follow the surface brightness profile
\begin{equation}
  \label{eq:1D}
  \Sigma\left(r\right)=\Sigma_{{e}}\exp\left[-\kappa\left(\left(\frac{r}{r_{{e}}}\right)^{1/n}-1\right)\right]+
  \Sigma_{0}\exp\left(-\frac{r}{r_{{s}}}\right),
\end{equation}
regardless of the degree of observed complexity. The \sersic{} function describes the surface brightness profile of the bulge, where \(r_{{e}}\) is
the effective radius, \(\Sigma_{{e}}\) is the surface brightness at \(r_{{e}}\), and \(n\) is the S\'ersic index; \(\kappa\) is related to \(n\) by
the incomplete-gamma function \(\Gamma\left(2n\right)=2\gamma\left(2n,\kappa\right)\) \citep{2005PASA+Graham+ref_of_sersic}. The exponential function
describes the disk surface brightness profile \citep{1970ApJ+Freeman+Exp_Disk}, where \(\Sigma_{0}\) and \(r_{{s}}\) are the central surface
brightness and scale length, respectively. The observed surface brightness profiles are derived by fitting elliptical isophotes to sky-subtracted
images using the IRAF\footnote{IRAF is distributed by the National Optical Astronomy Observatory, which is operated by the Association of Universities
  for Research in Astronomy (AURA), Inc., under cooperative agreement with the National Science Foundation.} task \ellipse{}
\citep{1987MNRAS+Jedrzejewski+ellipse_fitting}. Along with the azimuthally averaged surface brightness profile, \ellipse{} outputs the radial
\(\epsilon\) and PA profiles of the isophotes. Special considerations about this process are discussed in Appendix~\ref{sec:technical-biases-1d},
while Appendix~\ref{sec:direct-appr-meas} explains how we measure the sky level and its uncertainties.

To determine the free parameters in Equation~(\ref{eq:1D}), we utilize the MPFIT\footnote{\url{http://purl.com/net/mpfit}} package
\citep{2009ASPC+Markwardt+MPFIT} in the IDL environment to perform nonlinear least-squares fitting of the model to the observed surface brightness
profile, truncated at \(1\sigma\) above the sky. MPFIT is based on the Levenberg-Marquardt algorithm. The data are weighted by their measurement
uncertainty.  \citet{2016ApJS+Savorgnan+SBH_spheroids} consider such an S/N-based weighting scheme biased, and they prefer to assign no weight to the
data; we address this issue in Appendix~\ref{sec:technical-biases-1d}. As the smearing effects of the PSF cannot be properly treated in 1D fitting,
then, following common practice, we simply exclude the data inside the seeing disk from the fitting process. Parameter errors reported by the program are
derived from the covariance matrix. The ellipticity and PA of the best-fit bulge and disk are computed as averaged values over their dominant part of
the radial \(\epsilon\) and PA profiles, after the fitting is completed. The flux of both components follows from
\begin{align}
  \label{eq:bul_disk_flux}
  F_{\mathrm{bulge}}&=2\pi r_{{e}}^{2}\Sigma_{{e}}e^{\kappa}n\kappa^{-2n}\Gamma\left(2n\right)\left(1-\epsilon\right), \\
  F_{\mathrm{disk}} &=2\pi r_{{s}}^{2}\Sigma_{0}\left(1-\epsilon\right).
\end{align}
We compute the total flux \(T\) of the galaxy by integration of the observed surface brightness profile truncated at \(1\sigma\) above the sky. Then the
bulge-to-total ratio (\(B/T\)) and disk-to-total ratio (\(D/T\)) are directly computed by dividing the flux of the respective components by the total
flux of the galaxy. Note that in our 1D fitting the sum of \(B/T\) and \(D/T\) is not necessarily close to 1. This arises from the fact that not all
data in the observed surface brightness profile participate in the fitting; the data inside the seeing disk are excluded, and other parts of the
profile could be further excluded due to the presence of bars, lenses, rings, etc. Hence, the flux of the model does not necessarily match the observed
flux. Even when no data are excluded from the fit, there is no guarantee that the sum of \(B/T\) and \(D/T\) should be $\sim$1. We have to remind
readers that fluxes of components are computed using their averaged \(\epsilon\), whereas the total fluxes of galaxies are computed using the overall
\(\epsilon\) profile. Therefore, the sum of component fluxes is not necessarily equivalent to the total flux of the galaxy even when the model perfectly
fits the data, especially for galaxies that exhibit strongly varying \(\epsilon\) profiles. This is caused by the nature of 1D fitting, which is not
directly fitting the sum of all components in images but, instead, the profiles extracted from images. We do not try to correct such irregularities
throughout this paper because we are not mainly concerned about \(D/T\).

Errors reported by the fitting program give fair estimates of the uncertainties introduced by sky subtraction, since the uncertainties of sky level
measurements were propagated into computation of measurement uncertainties of the observed surface brightness profiles. This is confirmed by measuring
variations of best-fit parameters by deliberately subtracting from the image the measured sky level \(\pm1\sigma\), and comparing them to errors
reported by the fitting programs. Another source of uncertainty originates from the range of excluded data. For most cases, we are not able to
unambiguously determine the start and end point of the excluded ranges; for example, bars smoothly blend with bulges so that the separation of these
two components can be unclear. We empirically estimate the uncertainty introduced by our subjective choice of excluded ranges by manually perturbing
them and examining their influence on the best-fit bulge parameters. Finally, we sum up the two kinds of uncertainties in quadrature.

\subsection{2D Bulge-to-disk Decomposition}
\label{sec:2d-bulge-disk}

We use \galfit{} 3.0.5 to perform 2D multi-component decomposition. \galfit{} is a highly flexible and fast image-fitting algorithm originally
designed to extract structural components from well-resolved \textit{Hubble Space Telescope} images of nearby galaxies \citep{2002AJ+Peng+GALFIT1}.
It has been widely used on many surveys, both ground-based (\citealp{2015MNRAS+Meert+BD_dec_2D_SDSS,2016MNRAS+Meert+BD_dec_2D_SDSS,2013ApJ+Huang+CGS3,
  2016ApJ+Huang+CGS4,2016ApJS+Kim+SDSS_BD_dec}) and space-based (\citealp{2015ApJS+Salo+S4G_multi_comp_dec,2016arXiv+Davari+BD_decomp_high_z}),
idealized image simulations (e.g., \citealp{2007ApJS+Haussler+GALFIT_GIM2D,2013MNRAS+Meert+fit_simul_SDSS_gal,
  2014ApJ+Davari+robustness_size_simul_highz_gal,2016ApJ+Davari+fit_simul_highz_gal}), and on many studies of individual objects (e.g.,
\citealp{2013ApJ+Gu+Shell_Es}). \galfit{} carries a large box of analytic functions (e.g., \sersic{}, exponential, modified Ferrer, Moffat,
King). Researchers can use these functions to construct models with an arbitrary number of components, with possibly different centers, whose parameters can
be totally free, constrained, or fixed. Model components can be modified by Fourier modes, bending modes, coordinate rotation, and truncation,
simultaneously or separately. All these features help to create realistic-looking galaxy models, not only for regular and isolated galaxies, but even
for irregular galaxies, merging pairs, and overlapping galaxies. In this paper we restrict our attention to regular galaxies.

\galfit{} requires users to provide a data image, a PSF image, an optional mask image, and an input model of surface brightness. Sigma images are
internally generated by the code. The input data image is not sky-subtracted. We make use of the full data image because we plan to fit the sky level
simultaneously with the galaxy (see Appendix~\ref{sec:indir-appr-meas} for details). PSF images and mask images were prepared in
\citet{2011ApJS+Ho+CGS1}. The convolution box diameter is set to 40--80 times the seeing disk, as suggested in \galfit{}
FAQ\footnote{\url{https://users.obs.carnegiescience.edu/peng/work/galfit/TFAQ.html}}.

The key input ingredient for \galfit{} is the surface brightness model of the galaxy. There is no universally accepted input model. As suggested in
\citet{2010AJ+Peng+GALFIT2}, we build up complexities gradually, starting from the simplest model assumption, namely bulge+disk for unbarred galaxies
and bulge+bar+disk for barred galaxies. These model assumptions are commonly adopted in the literature as final solutions, mainly for
the sake of simplicity and ease of interpretation.  We attempt to achieve stable solutions for these simplified models first, and then slowly increase
the complexity and number of parameters as required.

Although there are many analytic functions available in \galfit{}, we restrict ourselves to a few of them, as described below. We adopt the
\sersic{} function for the surface brightness profile of the bulge, as is commonly done, and therefore our results can be compared to those of
previous studies:
\begin{equation}
  \label{eq:sersic}
  \Sigma\left(r\right)=\Sigma_{{e}}\exp\left[-\kappa\left(\left(\frac{r}{r_{{e}}}\right)^{1/n}-1\right)\right].
\end{equation}
The special case of $n=1$ corresponds to the standard exponential profile of the disk:
\begin{equation}
  \label{eq:expdisk}
  \Sigma\left(r\right)=\Sigma_{0}\exp\left(-\frac{r}{r_{{s}}}\right).
\end{equation}
The notations in Equation~(\ref{eq:sersic}) and Equation~(\ref{eq:expdisk}) are consistent with those in Equation~(\ref{eq:1D}). If the disk profile
is more complicated than a simple exponential (e.g., \citealp{1977ApJ+Kormendy+first_BD_Decom,1979A&AS+van_der_Kruit+non_exp_disk}), it can be
substituted by another profile or a combination of profiles.  As for the bar, we choose the modified Ferrer profile,
\begin{equation}
  \label{eq:ferrer}
  \Sigma\left(r\right)=\Sigma_{0}\left(1-\left(r/r_{\mathrm{out}}\right)^{2-\beta}\right)^{\alpha}, 
\end{equation}
which is defined within \(r\leq r_{\mathrm{out}}\) and is otherwise 0; \(\Sigma_{0}\) is the central surface brightness, \(\alpha\) governs the
sharpness of the outer truncation, and \(\beta\) describes the central flatness of the profile. Some studies fix \(\alpha\) to 2 or 2.5 to ensure a
sharp truncation of the bar (e.g., \citealp{2005MNRAS+Laurikainen+multi_comp_dec_S0,2015ApJS+Salo+S4G_multi_comp_dec}); however, we find that
\(\alpha\) hardly affects the best-fit parameters of the other components, and hence we allow the range \(\alpha\), \(0\leq\alpha\leq5\). If
\(\alpha>5\), we fix it to 2, while \(\beta\) is always a free parameter. We also find that \(r_{\mathrm{out}}\) is correlated with \(\alpha\):
\(r_{\mathrm{out}}\) is larger when \(\alpha\) increases. Thus, \(r_{\mathrm{out}}\) may not be a fair characterization of bar length when the
best-fit bar component has an unrealistically extended outskirt (large \(\alpha\)). Since this paper only focuses on bulge parameters, we leave this
issue to be clarified in future studies. Besides galaxy components, we also include a component to fit the sky level, which is represented by a
first-order bivariate polynomial,
\begin{equation}
  \label{eq:sky}
  \Sigma_{\mathrm{sky}}\left(x,y\right)=\Sigma_{\mathrm{sky}}\left(x_{{c}},y_{{c}}\right)+\left(x-x_{{c}}\right)
  \frac{\mathrm{d}\Sigma_{\mathrm{sky}}}{\mathrm{d}x}+\left(y-y_{{c}}\right)\frac{\mathrm{d}\Sigma_{\mathrm{sky}}}{\mathrm{d}y},
\end{equation}
where \(\left(x_{{c}},y_{{c}}\right)\) is the geometric center of the image, and \(\mathrm{d}\Sigma_{\mathrm{sky}}/\mathrm{d}x\) and
\(\mathrm{d}\Sigma_{\mathrm{sky}}/\mathrm{d}y\) are the sky flux gradient along each dimension of the image.

\begin{figure}
  \epsscale{1.17}
  \plotone{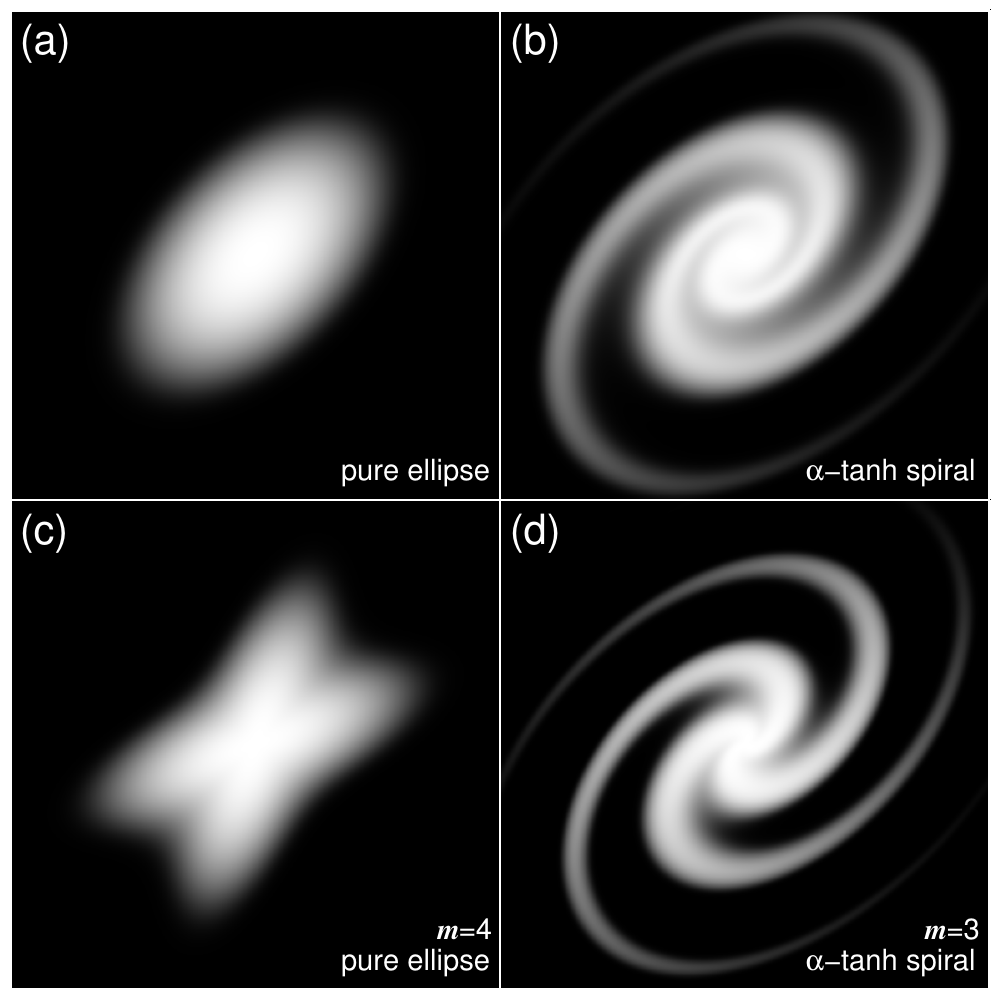}
  \caption{Examples of how various azimuthal functions shape \sersic{} profiles: (a) axisymmetric model; (b) power-law spiral model; (c) axisymmetric
    model modified by an \(m=4\) Fourier mode; and (d) power-law spiral model modified by an \(m=3\) Fourier mode. \label{fig:spir_illu}}
\end{figure}

The default azimuthal shape for each galaxy component is the traditional generalized ellipse,
\begin{equation}
  \label{eq:gener_ell}
  r\left(x,y\right)=\left(|x-x_{0}|^{C_{0}+2}+\left|\frac{y-y_{0}}{q}\right|^{C_{0}+2}\right)^{\frac{1}{C_{0}+2}},
\end{equation}
where \(\left(x_{0},y_{0}\right)\) is the centroid of the ellipse, the \(x\)-axis is aligned with the major axis of the ellipse, \(q\) is the axis ratio,
and \(C_{0}\) controls the diskyness or boxyness of the isophote. \(C_{0}\) is a hidden parameter unless it is invoked.  In this study,
\(C_{0}\equiv 0\), so the generalized ellipse simplifies to a pure ellipse,
\begin{equation}
  \label{eq:pure_ell}
    r\left(x,y\right)=\left(\left(x-x_{0}\right)^{2}+\left(\frac{y-y_{0}}{q}\right)^{2}\right)^{\frac{1}{2}}.
\end{equation}
The azimuthal function can be modified by coordinate rotation when fitting spiral disks. We adopt power-law--hyperbolic tangent coordinate rotation
(power-law spiral) instead of logarithmic--hyperbolic tangent rotation (logarithmic spiral); while both forms give equivalently good fits, the winding
scale radius parameter of the logarithmic spiral usually hits the parameter boundary (i.e., the parameter is infinitesimally small).  In brief, the
functional dependence of the power-law spiral is given by
\begin{equation}
  \label{eq:coor_rot}
  \theta\left(r\right)=\theta_{\mathrm{out}}\tanh \left(r_{\mathrm{in}},r_{\mathrm{out}},\theta_{\mathrm{incl}},\theta_{\mathrm{P.A.}}^{\mathrm{sky}};r\right)
  \times\left[\frac{1}{2}\left(\frac{r}{r_{\mathrm{out}}}+1\right)\right]^{\alpha}.
\end{equation}
The detailed analytic form of the hyperbolic function is lengthy and not of interest here; readers
can consult Appendix A in \citet{2010AJ+Peng+GALFIT2} for details. Rotation is largely controlled by
the \(\tanh\) function when \(r<r_{\mathrm{out}}\), and the asymptotic behavior beyond
\(r_{\mathrm{out}}\) is governed by the power-law term, which is characterized by the power-law
slope \(\alpha\).  The cumulative rotation angle roughly at \(r_{\mathrm{out}}\),
\(\theta_{\mathrm{out}}\), indicates how tightly the spiral arms wind. They define
\(r_{\mathrm{in}}\) to satisfy \(\theta\left(r_{\mathrm{in}}\right)\approx 20\arcdeg\); as
\(\theta\left(r\right)\) almost remains constant when \(r<r_{\mathrm{in}}\), a positive
\(r_{\mathrm{in}}\) produces a bar-like pattern that bridges the spiral arms at approximately
\(r_{\mathrm{in}}\). In cases of unbarred galaxies, \(r_{\mathrm{in}}\) is always set to 0.  The
inclination angle of the disk is \(\theta_{\mathrm{incl}}\), and
\(\theta_{\mathrm{P.A.}}^{\mathrm{sky}}\) is the sky PA. These two parameters together determine how
the spiral disk is projected onto the sky plane. Moreover, the pure ellipse or the coordinate
rotation can be modified by Fourier modes to create more complicated and realistic-looking
models. The Fourier modes perturb a pure ellipse in a way depicted by
\begin{equation}
  \label{eq:fourier_mode}
  r\left(x,y\right)=r_{0}\left(x,y\right)\left(1+\sum_{m=1}^{N}a_{m}\cos\left(m\left(\theta+\phi_{m}\right)\right)\right),
\end{equation}
where \(r_{0}\) is the unperturbed radius, \(a_{m}\) is the amplitude for mode \(m\),
\(\theta=\arctan\left((y-y_{0})/\left((x-x_{0})q\right)\right)\), and \(\phi_{m}\) is the phase angle relative to \(\theta\). For most cases, we use
coordinate rotation in conjunction with the Fourier modes to reproduce realistic spiral arms. Except for one case, the Fourier modes are seldom used
together with the pure ellipse; we apply an \(m=4\) Fourier mode to the \sersic{} bulge component of IC~5240 to model its X-shaped bulge.
Figure~\ref{fig:spir_illu} illustrates how we break away from axisymmetry by altering a pure ellipse using power-law coordinate rotation or/and
Fourier modes. Each panel of Figure~\ref{fig:spir_illu} is a prototype of realistic models that will show up from time to time in
Section~\ref{sec:bulge-disk-decomp} (e.g., panel (a) for all components of the starting model for
every galaxy; panel (b) for the disk of NGC~6118; panel (c) for the bulge of IC~5240; and panel (d)
for the disk of NGC~7083).

\begin{figure}
  \epsscale{1.18}
  \plotone{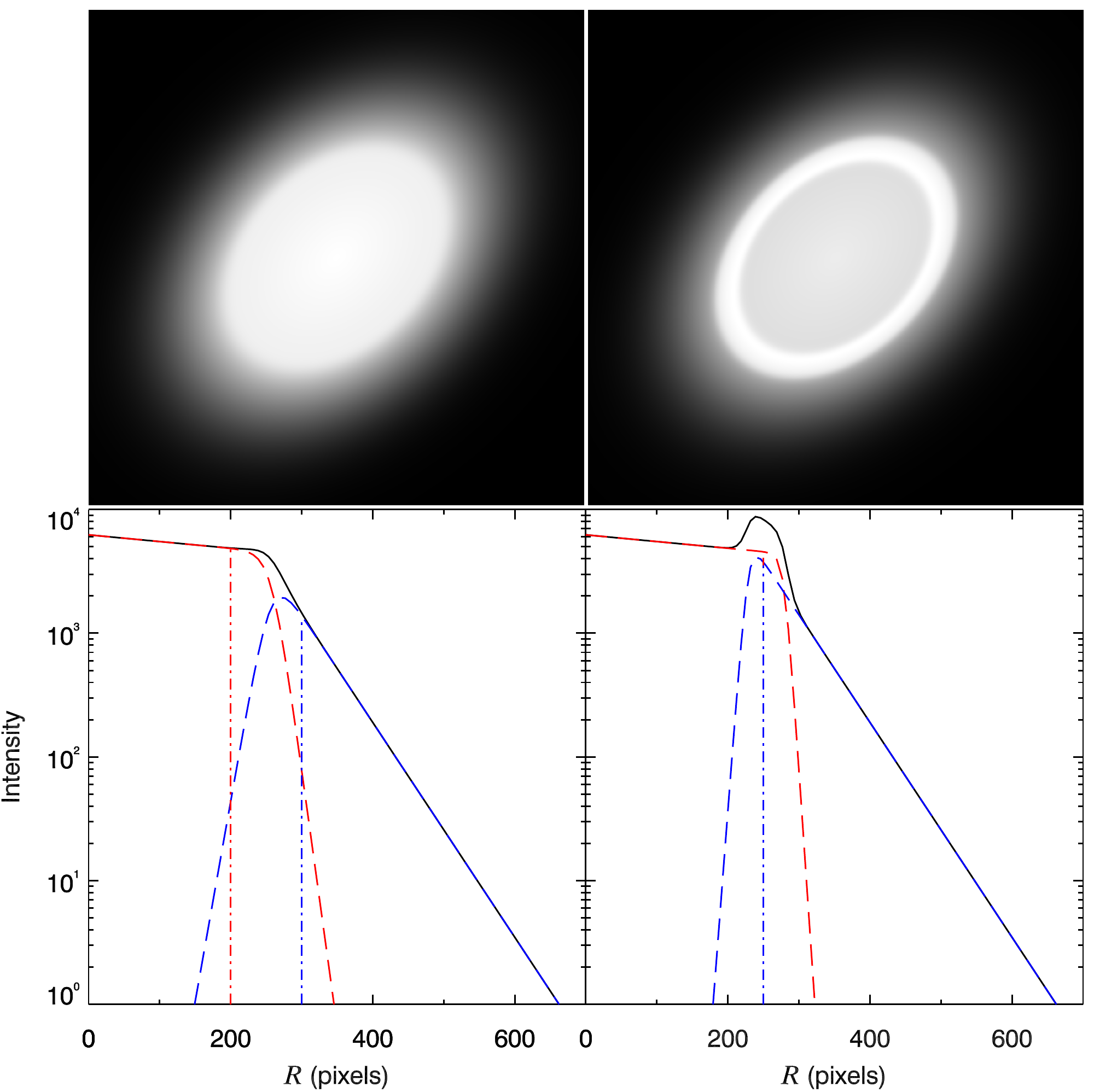}
  \caption{Examples of composite profiles. Left: loosely linked exponential profiles (lower) and their 2D manifestation (upper). Right: tightly linked
    exponential profiles (lower) and their 2D manifestation (upper). The size of both images is \(1000\times1000\,\mathrm{pixels}\). In both bottom panels,
    red dashed lines and blue dashed lines represent inner and outer components of the composite profiles, respectively. Black solid lines are overall
    profiles, which are the sum of the inner and outer components. Vertical dotted-dashed lines mark \(r_{\mathrm{break}}\) for the truncated components,
    where their profiles start to deviate from their original ones. Note that in the right-side example the inner and outer components
    share the same \(r_{\mathrm{break}}\). The values of \(\bigtriangleup r_{\mathrm{soft}}\) in the left-side and right-side examples are 100\,pixels and
    50\,pixels, respectively. \label{fig:trunc_illu}}
\end{figure}

In addition to the functions mentioned above, the truncation function is extensively used to model disk breaks and rings. The truncation function is
basically a hyperbolic tangent function, and its functional dependence on various parameters is given schematically by 
\begin{equation}
  \label{eq:truncation}
  P\left(x,y\right)=\tanh\left(x,y;x_{0},y_{0},r_{\mathrm{break}},\bigtriangleup r_{\mathrm{soft}},q,\theta_{\mathrm{P.A.}}\right),
\end{equation}
where \((x_{0},y_{0})\) is the center, \(q\) is the axis ratio, and \(\theta_{\mathrm{P.A.}}\) is the PA of the truncation function. These
three parameters are hidden by default; if not specified, their values are inherited from the component that is modified by the truncation
function. \(r_{\mathrm{break}}\) is the break radius where the truncated model flux drops to 99\% of its original flux at this radius. The softening
length \(\bigtriangleup r_{\mathrm{soft}}\) is defined as \(r_{\mathrm{soft}}-r_{\mathrm{break}}\) or \(r_{\mathrm{break}}-r_{\mathrm{soft}}\) for
outer truncation or inner truncation, respectively, where \(r_{\mathrm{soft}}\) is the radius where the truncated model flux drops to 1\% of its
original flux at this radius. Its detailed analytic form is lengthy and is not of immediate interest; readers can consult Appendix B in
\citet{2010AJ+Peng+GALFIT2} for details. Components are modified by the truncation function by multiplying \(P\) or \(1-P\) with the original flux
distribution for inner or outer truncation, respectively. In contrast with the azimuthal functions that allow for the possibility to break from
axisymmetry but still preserve the original meaning of the key parameters of the radial profiles, the truncation function can alter both the radial
profile and azimuthal shape of components, possibly altering the original meaning of key parameters. Hence we limit application of the truncation
functions only to the disk component, mainly to create composite radial profiles to account for disk breaks. In this case, inner and outer disk
components share the same truncation function but in opposite manner (outer truncation and inner truncation); the truncation function actually links
two truncated components. Such a composite profile has an inner part described by a certain analytic function and an outer part that behaves as
another, and how smoothly the two are bridged depends on the truncation function that links them. Moreover, the overlap region of the two parts can
naturally produce ring-like features. Figure~\ref{fig:trunc_illu} gives a schematic illustration of the two usages of the truncation function in our
study. One is to create Type \Rmnum{2} disk profiles (left panels; \citealp{1970ApJ+Freeman+Exp_Disk,2006A&A+Pohlen+Disk_Classes,
  2008AJ+Erwin+Disk_Classes}); the other models rings as well as truncated disks (right panels). Applications of these two prototypes appear in
Section~\ref{sec:bulge-disk-decomp} (e.g., NGC~7083 for the left-side example, and all barred galaxies with inner rings for the right-side example).
We show that producing rings is just a matter of how abruptly one part of the composite profile engages another, while smooth transitions result only in disk
breaks.\footnote{There is one exceptional case where we model the inner ring of NGC~1533 as an individual truncated component as well as byproduct of
  modeling disk breaks. However, this model only serves as a reference to show that bulge parameters are not sensitive to how we choose to model
  rings.} In addition, we will show that, in the case of NGC~1411, lenses can be modeled as exponential subsections, which is mathematically
  the same approach to model disk breaks. Although we show that disk breaks, lenses, and rings, along with the underlying disk, can be modeled
  mathematically interchangeably, we do not imply that these features are intrinsically the same morphological phenomena, or that they are necessarily
  coupled with each other. For example, the disk breaks in NGC~7083 and NGC~6118 are not accompanied by lenses or rings. In turn, when lenses or rings are
  present, whether the underlying disk is broken or not does not matter in any case (see the final two models of NGC~1411 and NGC~1533 in
  Section~\ref{sec:bulge-disk-decomp}). There are variants of the truncation function available in \galfit{}, such as radial truncation, length
truncation, height truncation, and inclined or non-inclined truncations, which will not be discussed here. We only use radial truncation for
axisymmetric components, and radial non-inclined truncation for spiral components throughout this paper.

Initial guesses of free parameters are roughly estimated through detailed inspection of images and isophotal analysis. Initial guesses of the sky levels
are obtained by the direct approach (see Section~\ref{sec:direct-appr-meas} for details). We pay close attention to the best-fit sky level measured by
\galfit{}.  We find that the best-fit sky levels are quite close to their initial guesses, and the sky flux gradient is generally small. Subsequent
refinements of the input model differ from galaxy to galaxy, which will be discussed in detail in Section~\ref{sec:bulge-disk-decomp}.

The flux of each component is directly computed from its model image after fitting is completed. In contrast to 1D fitting, the flux ratio for each
component is computed by dividing by the total model flux instead of the total data flux. Because the data image can be contaminated by foreground
sources, measurement of total flux from the data image is not straightforward. One way to measure this quantity is demonstrated in
Section~\ref{sec:1d-bulge-disk}, which is to integrate the observed surface brightness profiles. We consider the total flux of 2D best-fit models to
be a good approximation of the total flux of the data image. Many studies show that even single-component models suffice to recover global properties
(e.g., half-light radius and total magnitude) of galaxies with multi-component configuration (e.g., \citealp{2010AJ+Peng+GALFIT2,
  2013MNRAS+Meert+fit_simul_SDSS_gal,2014ApJ+Davari+robustness_size_simul_highz_gal,2016ApJ+Davari+fit_simul_highz_gal}).

Sky subtraction presents a major source of uncertainty \citep{2013ApJ+Huang+CGS3}, which is not properly captured in the formal errors of the best-fit
model parameters. We follow the empirical approach of \citet{2013ApJ+Huang+CGS3} to estimate the uncertainties of the bulge parameters by measuring
variations of the model parameters by perturbing the sky levels around \(\pm1\sigma\) of the best-fit sky level. This works for most, but not all,
cases. In some instances, we can only obtain lower limits to the true uncertainties of the bulge by manually adjusting other components while still
generating plausible-looking models. The range that allowed possible input models bracket serves as a measure of the model-induced uncertainties (see
discussion in Section~\ref{sec:discussion}).

\section{Decompositions of Individual Objects}
\label{sec:bulge-disk-decomp}

\subsection{NGC 1411}
\label{sec:ngc-1411}

NGC~1411 is an S0 galaxy of particular interest due to its complicated structures. It has a comprehensive manifestation of various types of lenses in
disk galaxies. \citet{1994cag+Sandage} identified a three-zone luminosity distribution with a ring that signifies the edge of an inner
lens. \citet{2013seg+Buta+gal_morph} recognized a nuclear lens, an inner lens, and an outer lens on a \(K_{s}\)-band image of the galaxy, based on
marginal change of the \(B-V\) color profile at the edge of each lens. \citet{2015ApJS+Buta+Morphology_S4G} reached a similar conclusion using
mid-infrared images from the \textit{Spitzer} Survey of Stellar Structure in Galaxies (S\(^{4}\)G; \citealp{2010PASP+Sheth+S4G}). However, visual
classification of a \(K_{s}\)-band image by \citet{2011MNRAS+Laurikainen+NIRS0S} missed the outer lens. On our CGS \(R\)-band image of the galaxy, the
inner lens and the nuclear lens clearly stand out, but the outer lens is hard to discern. The nuclear lens, unlike nuclear rings and nuclear bars that
unambiguously point to presence of pseudobulges, is not considered as part of the photometric bulge due to its unclear physical nature. Therefore, in
this case the secondary morphological features are the nuclear and inner lenses.

\begin{figure*}
  \epsscale{1.17}
  \plotone{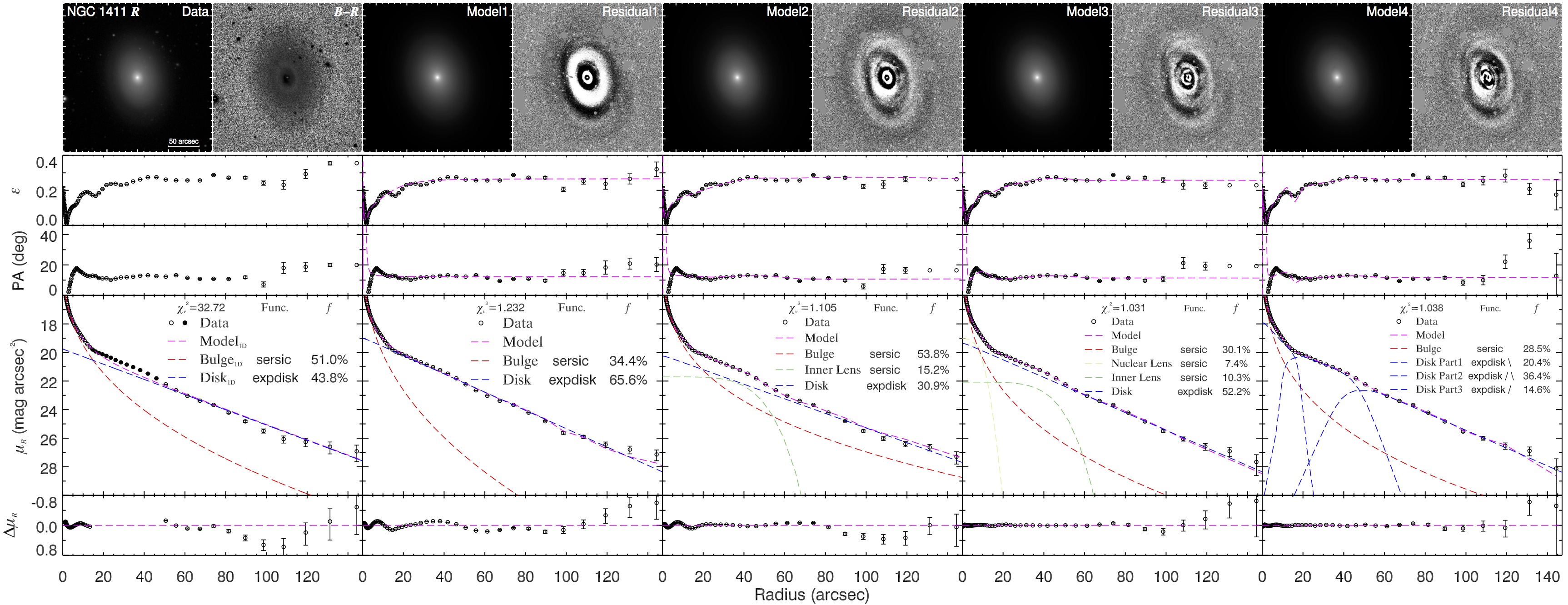}
  \caption{Best-fit 1D/2D models and isophotal analysis of NGC~1411. The top row shows, from left to
    right, the grayscale \(R\)-band image, the \(B-R\) color index map (darker means redder), and
    the 2D model image and the residual image in pairs. All the images are cropped to have the same
    size of 1.5\(D_{25}\) and are centered on the galaxy. The bottom row displays the fitting
    results of the 1D surface brightness profile (first column) or the 1D illustration of the 2D
    image fitting. From top to bottom, the panels show the radial profiles of ellipticity
    (\(\epsilon\)), position angle (PA), \(R\)-band surface brightness (\(\mu_{R}\)), and fitting
    residuals (\(\bigtriangleup \mu_{R}\)). Each column represents a best-fit model; from left to
    right, they are ordered by increasing dimension (1D--2D) and increasing complexity in the 2D
    models. The profiles of the data images, the model images, and the individual components are
    encoded consistently with different symbols, line styles, and colors, which are explained in the
    legends. The filled circles denote data that are excluded in the 1D fitting of the surface
    brightness profile. Note that the surface brightness profiles of the overall 2D models are
    generated by fixing the geometric parameters to those of the data surface brightness profiles,
    and the surface brightness profiles of the 2D components are generated along their major axes;
    hence, the model surface brightness profiles are not a simple summation of those of their
    individual components. The text on the right side of the legends gives brief information of each
    component; from left to right, each column describes the functions of the radial profiles
    (\sersic{}, expdisk, and Ferrer), whether they are complete or truncated (blank for complete,
    ``\textbackslash'' for outer truncation, and ``/'' for inner truncation), their azimuthal shape
    functions (blank for pure ellipse and "pow" for power-law spiral), and their light
    fractions. Fourier modes are specifically hidden to save space, and their application, if any,
    will be mentioned in the main text. Usually they only serve to improve the
    residuals.\label{fig:NGC1411}}
\end{figure*}

Although in principle the inner lens and nuclear lens should be excluded from 1D fitting of the surface brightness profile (Figure~\ref{fig:NGC1411}),
we opt not to do so. Excluding such a large portion of the profile (5\arcsec{}--50\arcsec{}) produces highly uncertain fits that are very sensitive
to the exact choice of excluded radii. We only exclude the part of the profile that is dominated by the inner lens (15\arcsec{}--50\arcsec{}), and
we estimate the uncertainties of the best-fit parameters (Table~\ref{tab:NGC1411}) by expanding and contracting the excluded range through shifting
the start point by 5\arcsec{} and the end point by 10\arcsec{} on a logarithmically spaced surface brightness profile. Despite our conservative choice
of the excluded range, we find that the error bars of the 1D best-fit parameters are still quite large, and we expect the real uncertainties to be
even larger.

For the 2D models, we follow the general strategy of building up the complexity step by step. We first fit a two-component model (Model1).  Clearly
the inner lens and the nuclear lens stand out in the residual image. Then we add a \sersic{} function to represent the inner lens on top of the
best-fit two-component model and refit the galaxy (Model2), but the nuclear lens is still not included.  Model3 adds another \sersic{} function to
model the nuclear lens. So far, we have successfully modeled all the identified components. Lastly (Model4), we successfully reproduce the appearance of
the nuclear and inner lenses as exponential subsections of the disk surface brightness that are linked by truncation functions.

The 1D best-fit bulge parameters show considerable deviation compared with those derived from 2D analysis, although there is significant overlap
within their uncertainties.  Given that there are complicated technical issues in 1D fitting (see Section~\ref{sec:technical-biases-1d} for detailed
discussion), it is difficult to track down the exact source of the discrepancy.

Model2 gives an exceptionally strong bulge, with \(B/T\) higher by a factor of \(\sim2/3\) compared with the other three 2D models. This unrealistic
intermediate product is caused by the inclusion of the inner lens, which suppresses the disk component, and by the absence of the nuclear lens, which
allows the bulge to expand. This model is simply an intermediate step, one to highlight the necessity of simultaneously fitting the two lens
components. Model3 produces a similar bulge compared with that of Model1, except for their different \sersic{} indices. We observe that the disk
component of Model1 actually accounts for the inner lens, and perhaps the bulk of the nuclear lens. Apart from $n$, it is not surprising
that both models yield very similar bulge parameters (i.e., \(m_{R}\), \(B/T\), and \(\mu_{{e}}\)); however, whether it is due to the fact that
  the disk component in Model1 accidentally takes out most of the nuclear lens light or that the nuclear lens only carries a small fraction
  of the total galaxy light is unclear, since we lack an adequate intermediate model. By comparison of Model3 to Model1, we learn that when the
  nuclear lens is not properly modeled, it at least affects measurements of bulge \sersic{} index. The key lesson here is this: if one is interested
in accurately measuring bulge \sersic{} indices, one \textit{must} pay special attention to even minor luminous components that are intimately
localized with the bulge. Model3 and Model4 describe the same degree of complexity, albeit in different ways, and the best-fit parameters of the two
models agree quite well. They demonstrate that how we treat the two lenses---as superimposed components on the underlying disk or as subsections of
the disk---does not greatly affect the bulge measurement, so long as the lenses are not neglected.

The large number of free parameters in our final 2D models may seem daunting and excessive. However, the number of free parameters is not a fair
measure of the degree of degeneracy. There is an excellent point made by \citet{2010AJ+Peng+GALFIT2}: fitting many well-isolated stars simultaneously
is as robust as fitting a single star. In the case of the structural components of our 2D models, they are, of course, not isolated but well
resolved and well defined by their different orientations and different light profiles. This is why they are readily identified by visual
examination. Nevertheless, we notice that the light fraction of the inner lens varies significantly from Model2 to Model3 (0.15--0.10). We argue that
the lens component in our model may not be robust against changes in other components, but it is quite robust when it is regarded as an accessory to
the underlying disk component. As seen in both Model2 and Model3, the inner lens component always fits well the surface brightness from
\(\sim20\arcsec\) to \(\sim50\arcsec\), with the help from the underlying disk component. This radial range is exactly dominated by the inner lens
morphology, as observed in the original image.

In summary: both the nuclear lens and inner lens should be properly modeled if we wish to properly measure all structural parameters of the
bulge. Specifically, the inner lens, as a large-scale and high-surface-brightness component, should be included in the model; otherwise, the bulge
flux may be biased significantly. The nuclear lens, when it is not properly modeled, can at least alter the bulge \sersic{} index
  dramatically.

\begin{deluxetable*}{lccccccc}
  \tablecaption{Best-fit Parameters for the Bulge of NGC~1411 \label{tab:NGC1411}}
  \tablecolumns{8}
  \tablehead{\multicolumn{1}{c}{\multirow{2}{*}{Model}} & \colhead{\(m_{R}\)} & \colhead{\(B/T\)} & \colhead{\(\mu_{{e},R}\)} &
    \colhead{\(n\)} & \colhead{\(r_{{e}}\)} & \colhead{\(\epsilon\)} & \colhead{PA} \\ \colhead{} & \colhead{(mag)} & \colhead{} &
    \colhead{(mag~arcsec\(^{-2}\))} & \colhead{} & \colhead{(\arcsec{})} & \colhead{} & \colhead{(\arcdeg{})}\\ \colhead{(1)} & \colhead{(2)} &
    \colhead{(3)} & \colhead{(4)} & \colhead{(5)} & \colhead{(6)} & \colhead{(7)} & \colhead{(8)}}
  \startdata
  Model0\tablenotemark{a} & \(11.38\pm0.37\) & \(0.510\pm0.156\) & \(18.32\pm0.66\) & \(2.56\pm1.27\) & \(6.00\pm2.09\) & \(0.089\pm0.017\) &
  \(23.22\pm8.59\) \\[0.05cm]
  \hline
  \rule{0pt}{3ex}Model1\tablenotemark{b} & \(11.83\substack{+0.01\\-0.01}\) & \(0.344\substack{+0.001\\-0.001}\) & \(17.54\substack{+0.01\\-0.01}\) &
  \(2.43\substack{+0.02\\-0.02}\) & \(3.54\substack{+0.03\\-0.02}\) & \(0.139\substack{+0.000\\-0.000}\) & \(13.30\substack{+0.00\\-0.00}\) \\[0.1cm]
  Model2\tablenotemark{c} & \(11.34\substack{+0.03\\-0.04}\) & \(0.538\substack{+0.020\\-0.011}\) & \(18.59\substack{+0.10\\-0.07}\) &
  \(3.60\substack{+0.12\\-0.08}\) & \(6.66\substack{+0.42\\-0.25}\) & \(0.164\substack{+0.002\\-0.001}\) & \(13.32\substack{+0.04\\-0.06}\) \\[0.1cm]
  Model3\tablenotemark{d} & \(11.98\substack{+0.09\\-0.11}\) & \(0.301\substack{+0.029\\-0.023}\) & \(17.73\substack{+0.23\\-0.19}\) &
  \(3.33\substack{+0.26\\-0.20}\) & \(3.30\substack{+0.49\\-0.35}\) & \(0.121\substack{+0.002\\-0.001}\) & \(7.12\substack{+0.47\\-0.44}\) \\[0.1cm]
  Model4\tablenotemark{e} & \(12.04\substack{+0.02\\-0.03}\) & \(0.285\substack{+0.008\\-0.005}\) & \(17.86\substack{+0.08\\-0.06}\) &
  \(3.61\substack{+0.10\\-0.07}\) & \(3.34\substack{+0.15\\-0.11}\) & \(0.109\substack{+0.001\\-0.000}\) & \(6.48\substack{+0.12\\-0.10}\)
  \enddata
  \tablecomments{Column 1: model identifier; Model0 is 1D model, while Model1--Model4 are 2D models. Column 2: \(R\)-band total magnitude. Column 3:
    bulge-to-total ratio. Column 4: surface brightness at effective radius. Column 5: \sersic{} index. Column 6: effective radius. Column 7:
    ellipticity. Column 8: position angle.}
  \tablenotetext{a}{Model configuration: Bulge+Disk.}
  \tablenotetext{b}{Model configuration: Bulge+Disk.}
  \tablenotetext{c}{Model configuration: Bulge+Inner Lens+Disk.}
  \tablenotetext{d}{Model configuration: Bulge+Nuclear Lens+Inner Lens+Disk.}
  \tablenotetext{e}{Model configuration: Bulge+Disk(Part1+Part2+Part3).}
\end{deluxetable*}

\subsection{NGC 2784}
\label{sec:ngc-2784}

\begin{figure*}
  \epsscale{1.17}
  \plotone{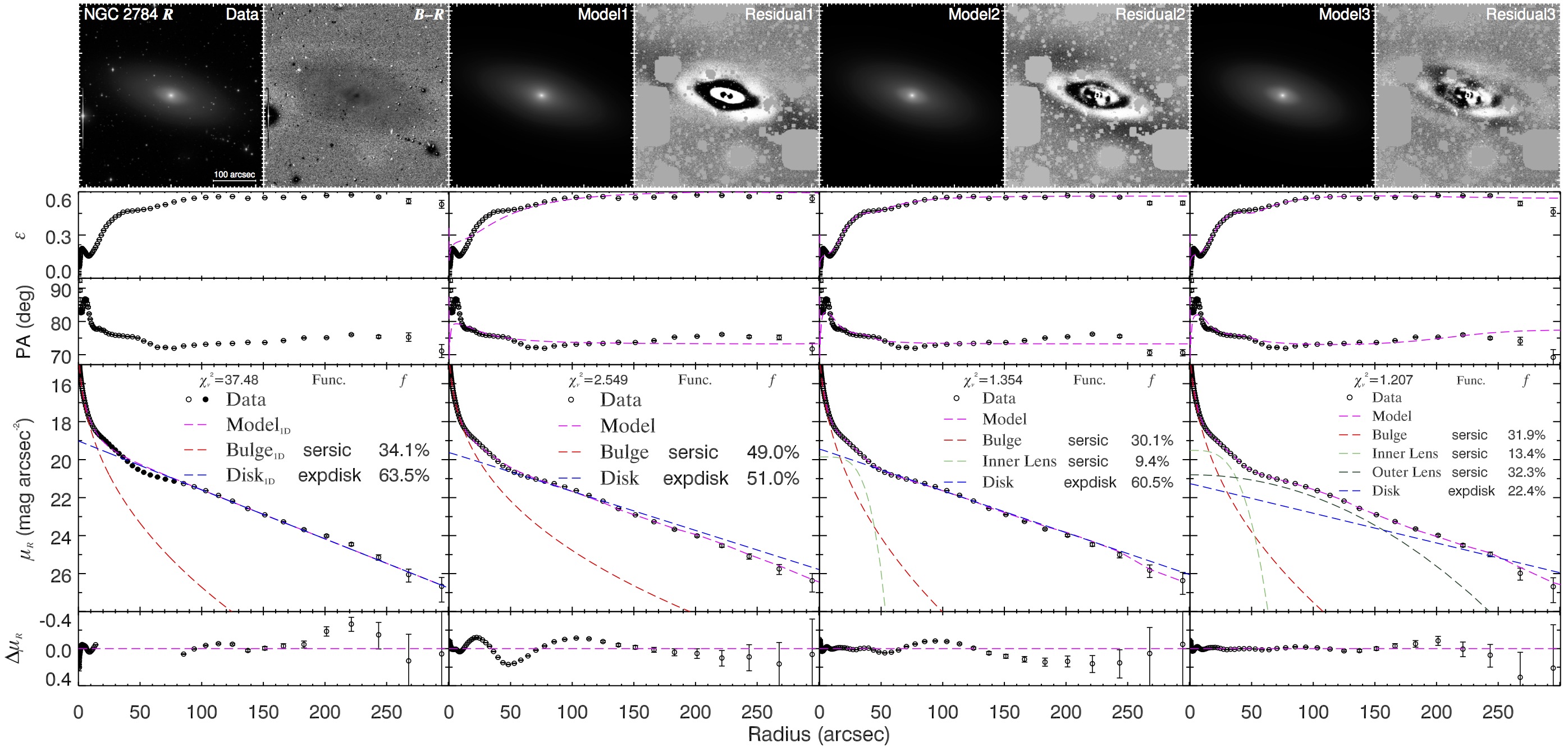}
  \caption{Best-fit 1D/2D models and isophotal analysis of NGC~2784. Same conventions as in Figure~\ref{fig:NGC1411}. \label{fig:NGC2784}}
\end{figure*}

NGC~2784 has not received as much attention as NGC~1411 because of its relatively simple structure. A three-zone S0 structure was found by
\citet{1994cag+Sandage}. \citet{2013seg+Buta+gal_morph}, as did \citet{1961hag+Sandage}, identified a nucleus, a lens, and an envelope.
The inner lens of the galaxy is evident in the CGS \(R\)-band image, while the outer lens (envelope) is rather vague but is readily seen in the
residual image (Figure~\ref{fig:NGC2784}).

For the 1D fit (Figure~\ref{fig:NGC2784}), we exclude the part of the surface brightness profile that is dominated by the inner and outer
lenses. Again, we find that the fitting results are sensitive to the choice of exclusion radii. The final best fit excludes the region between
15\arcsec{} and 80\arcsec{}, and the uncertainties (Table~\ref{tab:NGC2784}) are estimated by perturbing the these these values.  The large
uncertainties of the best-fit parameters are not caused by the aggressive range of excluded data. We experimented with excluding only the inner lens
(15\arcsec{}--45\arcsec{}), but the results were equally poor.

Our procedure for 2D modeling is similar to that adopted for NGC~1411 (Section~\ref{sec:ngc-1411}), except that NGC~2784 does not contain a nuclear
lens and instead has an outer lens. Model1 only includes the bulge and the disk component, and its residual image clearly shows the imprints of the
inner and outer lenses.  Model2 adds a \sersic{} function to represent the inner lens, and Model3 includes an outer lens as well.  We leave the
outer lens, which dominates only in the faint outer disk, to the last step because we expect it to impact the bulge less than the inner lens.

As seen in Table~\ref{tab:NGC2784}, the 1D best-fit parameters of the bulge overlap with those of the 2D models. Nevertheless, such uncertain results
are solutions of last resort, as we have better means to handle the structural complexities in 2D. The technical difficulties encountered in 1D
fitting of NGC~1411 and NGC~2784 are intrinsically the same---either we have to exclude a considerable part of the surface brightness profile to
mitigate perturbations from some minor luminous components (in these two cases, the lenses), which may result in uncertain best-fit parameters, or we
have to assume that the perturbations are negligible and leave the uncertainties introduced by them unquantified. Both solutions are unsatisfactory,
as shown in these two cases, and will be confirmed infeasible in more complicated galaxies in the rest of the sample. As for the discrepant results
presented by the first two 2D models (Model1 and Model2), they again confirm the importance of modeling lenses that are intimately overlapped with the
bulge.  In this case, the inner lens is actually incorporated into the bulge component in Model1, which leads to overestimates of \(B/T\). Comparison
of Model2 and Model3 confirms our expectation that modeling the outer lens has a minor impact on bulge parameters. Although the imprint of the outer
lens is prominent on the residual image of Model2, there is no systematic positive or negative residual pattern inside \(\sim50\arcsec\), which
indicates a reasonably good fit well beyond the bulge.

As in the case of NGC~1411, we observe that flux ratio of the inner lens changes significantly from Model2 to Model3 (0.094--0.134); the arguments we
presented in the previous case still hold here. The combination of the inner lens, outer lens, and the underlying disk component in Model3 describes
the disk surface brightness as well as does the combination of the inner lens and the underlying disk component in Model2. The variations of the inner
lens arise from the inclusion of the outer lens in Model3.

\begin{deluxetable*}{lccccccc}
  \tablecaption{Best-fit Parameters for the Bulge of NGC~2784 \label{tab:NGC2784}}
  \tablecolumns{8}
  \tablehead{\multicolumn{1}{c}{\multirow{2}{*}{Model}} & \colhead{\(m_{R}\)} & \colhead{\(B/T\)} & \colhead{\(\mu_{{e},R}\)} &
    \colhead{\(n\)} & \colhead{\(r_{{e}}\)} & \colhead{\(\epsilon\)} & \colhead{PA} \\ \colhead{} & \colhead{(mag)} & \colhead{} &
    \colhead{(mag~arcsec\(^{-2}\))} & \colhead{} & \colhead{(\arcsec{})} & \colhead{} & \colhead{(\arcdeg{})}\\ \colhead{(1)} & \colhead{(2)} &
    \colhead{(3)} & \colhead{(4)} & \colhead{(5)} & \colhead{(6)} & \colhead{(7)} & \colhead{(8)}}
  \startdata
  Model0\tablenotemark{a} & \(10.47\pm0.47\) & \(0.341\pm0.197\) & \(18.08\pm1.09\) & \(2.41\pm1.21\) & \(8.58\pm8.24\) & \(0.150\pm0.008\) &
  \(92.53\pm2.41\) \\[0.05cm]
  \hline
  \rule{0pt}{3ex}Model1\tablenotemark{b} & \(10.16\substack{+0.01\\-0.01}\) & \(0.490\substack{+0.002\\-0.002}\) & \(18.58\substack{+0.02\\-0.02}\) &
  \(2.80\substack{+0.02\\-0.02}\) & \(12.78\substack{+0.18\\-0.17}\) & \(0.250\substack{+0.002\\-0.002}\) & \(79.75\substack{+0.09\\-0.09}\) \\[0.1cm]
  Model2\tablenotemark{c} & \(10.67\substack{+0.01\\-0.01}\) & \(0.301\substack{+0.000\\-0.001}\) & \(17.81\substack{+0.01\\-0.01}\) &
  \(2.22\substack{+0.01\\-0.01}\) & \(7.10\substack{+0.04\\-0.04}\) & \(0.164\substack{+0.001\\-0.001}\) & \(83.68\substack{+0.07\\-0.07}\) \\[0.1cm]
  Model3\tablenotemark{d} & \(10.61\substack{+0.00\\-0.00}\) & \(0.319\substack{+0.003\\-0.003}\) & \(17.89\substack{+0.00\\-0.00}\) &
  \(2.30\substack{+0.00\\-0.00}\) & \(7.55\substack{+0.03\\-0.01}\) & \(0.173\substack{+0.001\\-0.000}\) & \(82.45\substack{+0.00\\-0.01}\) 
  \enddata
  \tablecomments{See Table~\ref{tab:NGC1411} for details.}
  \tablenotetext{a}{Model configuration: Bulge+Disk.}
  \tablenotetext{b}{Model configuration: Bulge+Disk.}
  \tablenotetext{c}{Model configuration: Bulge+Inner Lens+Disk.}
  \tablenotetext{d}{Model configuration: Bulge+Inner Lens+Outer Lens+Disk.}
\end{deluxetable*}

\subsection{NGC 1357}
\label{sec:ngc-1357}

NGC~1357 has been recognized as a spiral galaxy with two major tightly wound arms \citep{1994cag+Sandage}. It is classified as SA(s)ab in RC3, which
suggests considerable bulge prominence. The bulge has a smooth appearance and dominates at least inside \(\sim10\arcsec\). At around 20\arcsec{}, the
spiral disk starts to take over in morphology. Interestingly, the disk of NGC~1357 has a two-zone structure: the inner bright part (inside
\(\sim40\arcsec\)) is relatively blue and shows strong spiral arms, while the outer part of the disk is red and has no well-defined arms. This
morphology separation in the galaxy disk was also reported by \citet{1994cag+Sandage}. The abrupt change of disk color at the edge of inner bright
spiral arms is readily recognized in color profiles of the galaxy \citep[see][Figure~19.194]{2011ApJS+Li+CGS2}.  To summarize: the basic layout of NGC
1357 is the bright bulge, the inner blue disk with two well-defined spiral arms and the outer redder disk with weak spiral features.

Although the relatively face-on orientation of the galaxy blurs geometric differences of the bulge and the disk, we are helped by the varying
strengths of spiral features from inside to outside---from zero (bulge), strong (inner disk) to weak (outer disk). In addition, the surface brightness
of the inner bright disk is reminiscent of a lens, if one neglects the disturbance from the spiral arms. The inner disk appears as a shelf in the
surface brightness profile, well described by a low-\(n\) (\(n<1\)) \sersic{} function, which differs from a high-\(n\) \sersic{} function (bulge) and
an exponential function (the underlying disk). We take all these morphological features, which help to break degeneracies and ensures robust
decomposition, into account in our 2D fit.  However, this information is averaged out in 1D fitting.

\begin{figure*}
  \epsscale{1.17}
  \plotone{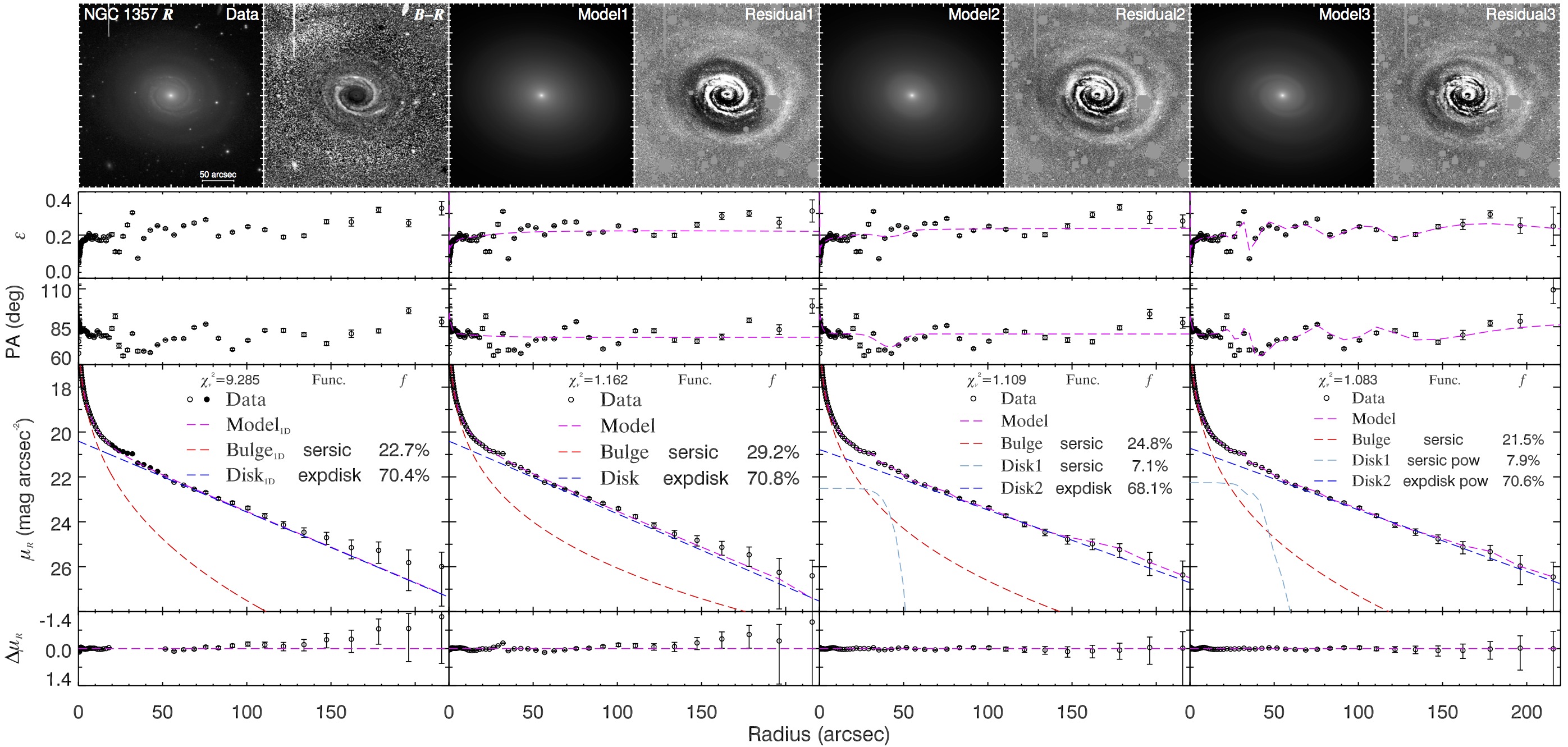}
  \caption{Best-fit 1D/2D models and isophotal analysis of NGC~1357. Same conventions as in Figure~\ref{fig:NGC1411}. \label{fig:NGC1357}}
\end{figure*}

For the 1D fit (Figure~\ref{fig:NGC1357}), we exclude data from 20\arcsec{} to 50\arcsec{} and estimate the uncertainties (Table~\ref{tab:NGC1357}) by
expanding and contracting the excluded range through shifting the start point by 5\arcsec{} and the end point by 10\arcsec{}.

For the 2D fit, we first neglect the separation of disk morphology and the spiral features and fit the galaxy with an axisymmetric, two-component
model (Model1). Then we include a \sersic{} function to represent the inner bright disk (Model2). Finally, we apply coordinate rotation to both disk
components to model their spiral arms (Model3). In addition, \(m=1\) and \(m=2\) Fourier modes are applied to both disk components, to achieve
slightly better residuals. Thanks to the extra component, Model2 and Model3 show better residuals than Model1. Nevertheless, the slightly better
residuals are not sufficient justification for the extra component. Instead, we invoke an extra component because (1) the inner bright disk and the
outer faint disk show different physical properties (i.e., they show different stellar populations and different surface brightness profiles), and (2)
if the extra component for the inner bright disk is absent, the bulge component will try to take the inner bright disk as part of itself (see Model1
in Figure~\ref{fig:NGC1357}). This is the reason why the best-fit bulge of Model1 is systematically overestimated compared with those of Model2 and
Model3.

The only difference between Model2 and Model3 is whether or not we apply coordinate rotation to the two disk components.  We notice that the two disk
components with coordinate rotation get stretched to some extent (i.e., larger \sersic{} index and
larger effective radius for a \sersic{} function, and
larger scale length for an exponential function). These effects can be understood by consideration of the fact that a spiral disk shows rises and falls in
surface brightness and ends up with a final fall in the outskirt. If one fits an axisymmetric model to a spiral disk, the model will compromise
between the rises and falls and ``see'' a sharp final fall; however, if the model is modified by coordinate rotation, the spiral model can deal with
rises and falls more naturally and will ``see'' a shallower final fall, because isophotes are able to rotate in this situation. Therefore, an
axisymmetric model for a spiral disk tends to have smaller \sersic{} index and smaller effective radius for a \sersic{} function, or smaller scale
length for an exponential function, compared with their true value. These trends also hold for the other five spiral galaxies in the sample.
Variations in the disk components result in variations in the bulge structural parameters. However, the bulges of Model2 and Model3 show minor
differences, especially when considering parameter error bars.

This is the first galaxy in our sample for which we have to deal with spiral features. However, these turn out to affect the bulge
parameters only slightly; modeling spiral features is time-consuming and unnecessary. The extra disk component acts similarly to the inner lens of
NGC~2784, namely that if it is not properly modeled, it will be incorporated, incorrectly, as part of the bulge.

\begin{deluxetable*}{lccccccc}
  \tablecaption{Best-fit Parameters for the Bulge of NGC~1357 \label{tab:NGC1357}}
  \tablecolumns{8}
  \tablehead{\multicolumn{1}{c}{\multirow{2}{*}{Model}} & \colhead{\(m_{R}\)} & \colhead{\(B/T\)} & \colhead{\(\mu_{{e},R}\)} &
    \colhead{\(n\)} & \colhead{\(r_{{e}}\)} & \colhead{\(\epsilon\)} & \colhead{PA} \\ \colhead{} & \colhead{(mag)} & \colhead{} &
    \colhead{(mag~arcsec\(^{-2}\))} & \colhead{} & \colhead{(\arcsec{})} & \colhead{} & \colhead{(\arcdeg{})}\\ \colhead{(1)} & \colhead{(2)} &
    \colhead{(3)} & \colhead{(4)} & \colhead{(5)} & \colhead{(6)} & \colhead{(7)} & \colhead{(8)}}
  \startdata
  Model0\tablenotemark{a} & \(12.26\pm0.15\) & \(0.227\pm0.042\) & \(20.18\pm0.27\) & \(3.09\pm0.26\) & \(9.33\pm1.59\) & \(0.160\pm0.005\) &
  \(82.59\pm0.43\) \\[0.05cm]
  \hline
  \rule{0pt}{3ex}Model1\tablenotemark{b} & \(12.02\substack{+0.30\\-0.08}\) & \(0.292\substack{+0.014\\-0.061}\) & \(21.10\substack{+0.29\\-0.73}\) &
  \(4.43\substack{+0.38\\-0.75}\) & \(14.93\substack{+2.44\\-5.26}\) & \(0.184\substack{+0.001\\-0.000}\) & \(79.96\substack{+0.18\\-0.03}\) \\[0.1cm]
  Model2\tablenotemark{c} & \(12.14\substack{+0.14\\-0.15}\) & \(0.248\substack{+0.021\\-0.019}\) & \(20.59\substack{+0.32\\-0.26}\) &
  \(3.79\substack{+0.32\\-0.26}\) & \(11.56\substack{+2.57\\-1.88}\) & \(0.188\substack{+0.000\\-0.000}\) & \(80.35\substack{+0.12\\-0.16}\) \\[0.1cm]
  Model3\tablenotemark{d} & \(12.30\substack{+0.09\\-0.11}\) & \(0.215\substack{+0.011\\-0.007}\) & \(20.23\substack{+0.22\\-0.16}\) &
  \(3.43\substack{+0.22\\-0.15}\) & \(9.32\substack{+1.38\\-0.91}\) & \(0.187\substack{+0.003\\-0.002}\) & \(79.16\substack{+0.28\\-0.02}\)
  \enddata
  \tablecomments{See Table~\ref{tab:NGC1411} for details.}
  \tablenotetext{a}{Model configuration: Bulge+Disk.}
  \tablenotetext{b}{Model configuration: Bulge+Disk. For this model, upper errors of \(B/T\), \(\mu_{e,R}\), \(n\) and \(r_{e}\),
    and lower errors of \(m_{R}\), \(\epsilon\) and PA are only lower limits.}
  \tablenotetext{c}{Model configuration: Bulge+Disk1+Disk2.}
  \tablenotetext{d}{Model configuration: Bulge+Spiral Disk1+Spiral Disk2.}
\end{deluxetable*}

\subsection{NGC 7083}
\label{sec:ngc-7083}

\begin{figure*}
  \epsscale{1.17}
  \plotone{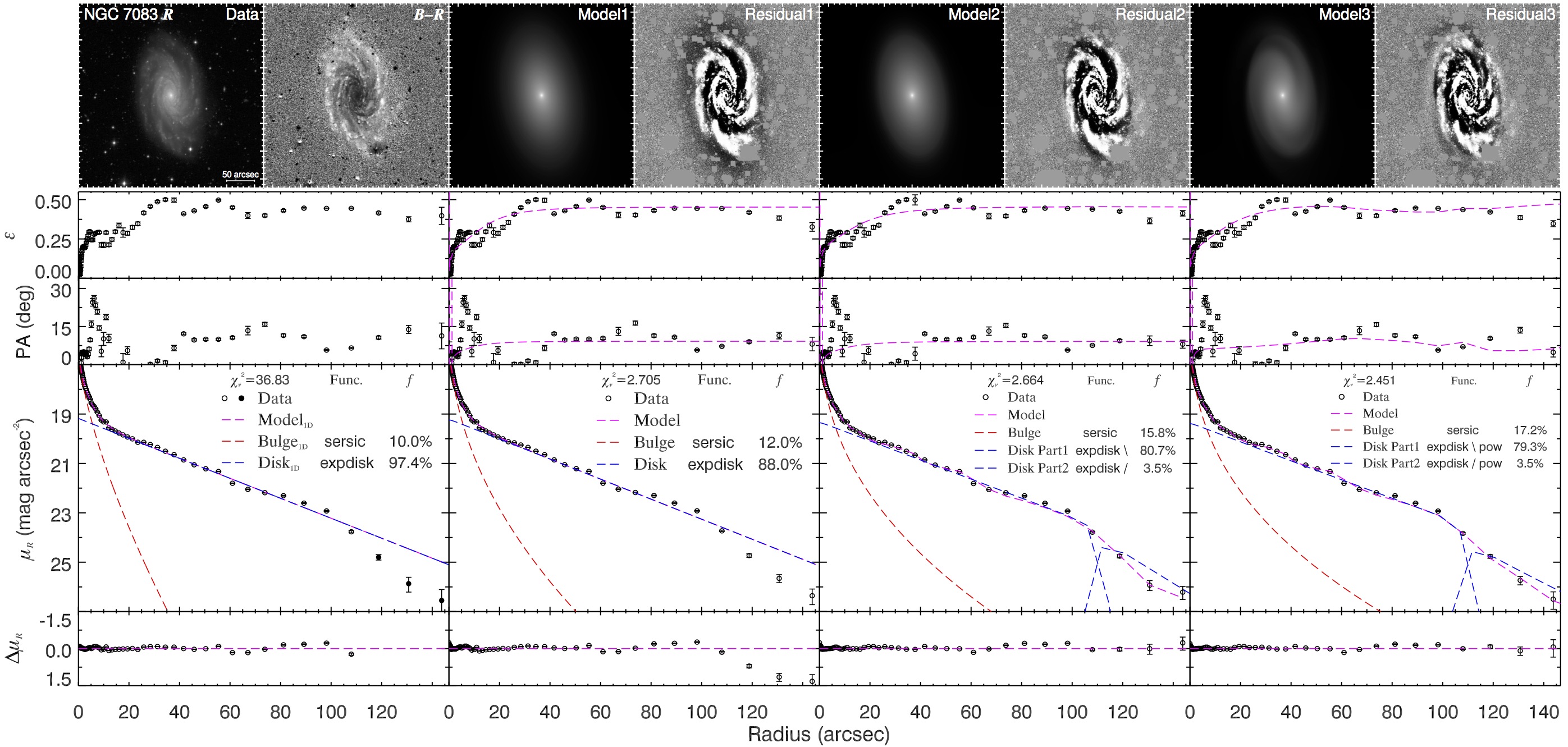}
  \caption{Best-fit 1D/2D models and isophotal analysis of NGC~7083. Same conventions as in Figure~\ref{fig:NGC1411}. \label{fig:NGC7083}}
\end{figure*}

The spiral arms of NGC~7083 have a filamentary appearance that causes ambiguity in identifying the
number of arms. A mixture of grand-design and fragmentary features in the spiral arms was reported
by \citet{1994cag+Sandage}. In near-infrared bands, NGC~7083 appears more likely to be a
grand-design spiral \citep{1998A&A+Grosbol+decoupling_spiral_disk,2002ApJS+Eskridge+OSUBGS}, while
it is a recognized as a multi-arm spiral in the \(B\) band
\citep{1998A&A+Grosbol+decoupling_spiral_disk}. On a CGS \(R\)-band image, intricate dust lanes that
trace the spiral pattern and star-forming knots are present. Regularity and smoothness of the spiral
arms are largely disturbed by these complexities. However, after subtracting a smooth model from the
original image, we are able to identify a three-arm structure on the residual image (see the
Residual1 panel of Figure~\ref{fig:NGC7083}).  Moreover, the \(R\)-band surface brightness profile
of the galaxy shows itself as a Type \Rmnum{2} disk profile.

We fit the 1D surface brightness profile (Figure~\ref{fig:NGC7083}) by excluding the outer truncated
part of the disk. As the exclusion of data occurs at the faint end of the profile, the fitting
results are barely changed even when we do not discard that part of the profile. In addition, the
break of the profile is sharp enough to be unambiguously identified. In this case, we conclude that
uncertainties arising from choices of the excluded range are marginal, and thus are not included in the
error budget (Table~\ref{tab:NGC7083}).

As usual, we start fitting the image from the simplest assumption, a \sersic{} bulge and an
exponential disk (Model1), regardless of the presence of the disk break and the spiral arms. In
Model2, the disk break is taken into account by introducing an extra exponential function and a
truncation function to link the two exponential functions. These are
constrained to have the same centroid, PA, and ellipticity, but they are free to have different
scale lengths. In our philosophy of model construction, they are meaningless entities as individual
components, and they only make sense when considered together. We refer to these subcomponents as
``Disk Part\textasteriskcentered{},'' to distinguish them from the case of NGC~1357, whose two disk
components are referred to as ``Disk\textasteriskcentered{}.'' In Model3, we apply coordinate
rotation to the disk components (including Part1 and Part2) to model the truncated three-arm spiral
disk. As regular coordinate rotation is only able to produce grand-design spirals, we invoke the
\(m=3\) Fourier mode to split the grand-design arms to mimic the three-arm appearance.

The bulge derived from the 1D fitting is systematically somewhat weaker, shallower, and smaller (in
terms of \(B/T\), \(n\), and \(r_{{e}}\), respectively) than those obtained from the 2D fits. The
reason for this is unclear.  We also note that the bulge strength gets enhanced from Model1 to
Model2. In Model1, the single exponential disk component has to compromise between the inner part
and the outer part of the broken disk. Thus, the down-bending outer part will cause the exponential
function to have shorter scale length and brighter central surface brightness compared with the true
value of the inner part, although the change of parameters should be small because the outer part of
the disk carries much less weight in the fitting compared with the inner part. In Model2, after
introducing a truncated exponential disk with smaller scale length to model the outer part, the
inner part is free to increase its scale length and to reduce its central surface brightness.
Therefore, the best-fit bulge of Model2 stretches outward to have larger \(B/T\), \(n\), and
\(r_{{e}}\) compared with that of Model1. As for the reason why the bulge structural parameters
varies when we apply coordinate rotation to the disk, this can be explained by the same argument
that was discussed in length in Section~\ref{sec:ngc-1357}. We find that the disk scale lengths
become larger for both parts of the disk component, which is consistent with the variations of
scale-length parameters for disk components that were observed in the previous case. Nonetheless,
the variations of bulge parameters due to the inclusion of spiral arms are marginal.

In this case, we confirm that spiral arms have a minor impact on the bulge parameters once the basic structure of the disk is determined. Disk
break, however, needs to be properly modeled.

\begin{deluxetable*}{lccccccc}
  \tablecaption{Best-fit Parameters for the Bulge of NGC~7083 \label{tab:NGC7083}} \tablecolumns{8}
  \tablehead{\multicolumn{1}{c}{\multirow{2}{*}{Model}} & \colhead{\(m_{R}\)} & \colhead{\(B/T\)} & \colhead{\(\mu_{{e},R}\)} &
    \colhead{\(n\)} & \colhead{\(r_{{e}}\)} & \colhead{\(\epsilon\)} & \colhead{PA} \\ \colhead{} & \colhead{(mag)} & \colhead{} &
    \colhead{(mag~arcsec\(^{-2}\))} & \colhead{} & \colhead{(\arcsec{})} & \colhead{} & \colhead{(\arcdeg{})}\\ \colhead{(1)} & \colhead{(2)} &
    \colhead{(3)} & \colhead{(4)} & \colhead{(5)} & \colhead{(6)} & \colhead{(7)} & \colhead{(8)}}
  \startdata
  Model0\tablenotemark{a} & \(13.05\pm0.04\) & \(0.100\pm0.002\) & \(19.23\pm0.02\) & \(1.47\pm0.02\) & \(5.09\pm0.09\) & \(0.192\pm0.013\) &
  \(3.39\pm0.64\) \\[0.05cm]
  \hline
  \rule{0pt}{3ex}Model1\tablenotemark{b} & \(12.88\substack{+0.03\\-0.03}\) & \(0.120\substack{+0.003\\-0.003}\) & \(19.58\substack{+0.04\\-0.05}\) &
  \(2.06\substack{+0.04\\-0.05}\) & \(6.02\substack{+0.19\\-0.21}\) & \(0.203\substack{+0.003\\-0.003}\) & \(4.83\substack{+0.15\\-0.17}\) \\[0.1cm]
  Model2\tablenotemark{c} & \(12.63\substack{+0.02\\-0.02}\) & \(0.158\substack{+0.001\\-0.002}\) & \(19.97\substack{+0.03\\-0.03}\) &
  \(2.41\substack{+0.02\\-0.03}\) & \(7.91\substack{+0.16\\-0.18}\) & \(0.224\substack{+0.002\\-0.002}\) & \(4.44\substack{+0.04\\-0.02}\) \\[0.1cm]
  Model3\tablenotemark{d} & \(12.52\substack{+0.01\\-0.01}\) & \(0.172\substack{+0.001\\-0.001}\) & \(20.14\substack{+0.02\\-0.02}\) &
  \(2.57\substack{+0.01\\-0.02}\) & \(8.73\substack{+0.13\\-0.13}\) & \(0.205\substack{+0.002\\-0.002}\) & \(6.48\substack{+0.02\\-0.03}\)
  \enddata
  \tablecomments{See Table~\ref{tab:NGC1411} for details.}
  \tablenotetext{a}{Model configuration: Bulge+Disk.}
  \tablenotetext{b}{Model configuration: Bulge+Disk.}
  \tablenotetext{c}{Model configuration: Bulge+Broken Disk(Part1+Part2).}
  \tablenotetext{d}{Model configuration: Bulge+Spiral Broken Disk(Part1+Part2).}
\end{deluxetable*}

\subsection{NGC 6118}
\label{sec:ngc-6118}

NGC~6118 is a late-type spiral galaxy with an apparently weak bulge. \citet{1964rcbg+de_Vaucouleurs} recognized three main spiral arms in its low
surface brightness disk. \citet{1994cag+Sandage} noticed the dominant grand-design spiral pattern that becomes fragmentary approaching the outer
part. \citet{2015ApJS+Buta+Morphology_S4G} classified it as a multi-arm spiral based on mid-infrared images from S\(^{4}\)G. Our identification of its
spiral pattern is consistent with that from \citet{1994cag+Sandage}. However, we refrain from modeling the branches at the tail of the grand-design
spiral arms and only focus on the principal spiral pattern. The galaxy also has a Type \Rmnum{2} disk profile, except that the break of
the profile is smoother compared with that of NGC~7083.

\begin{figure*}
  \epsscale{1.17}
  \plotone{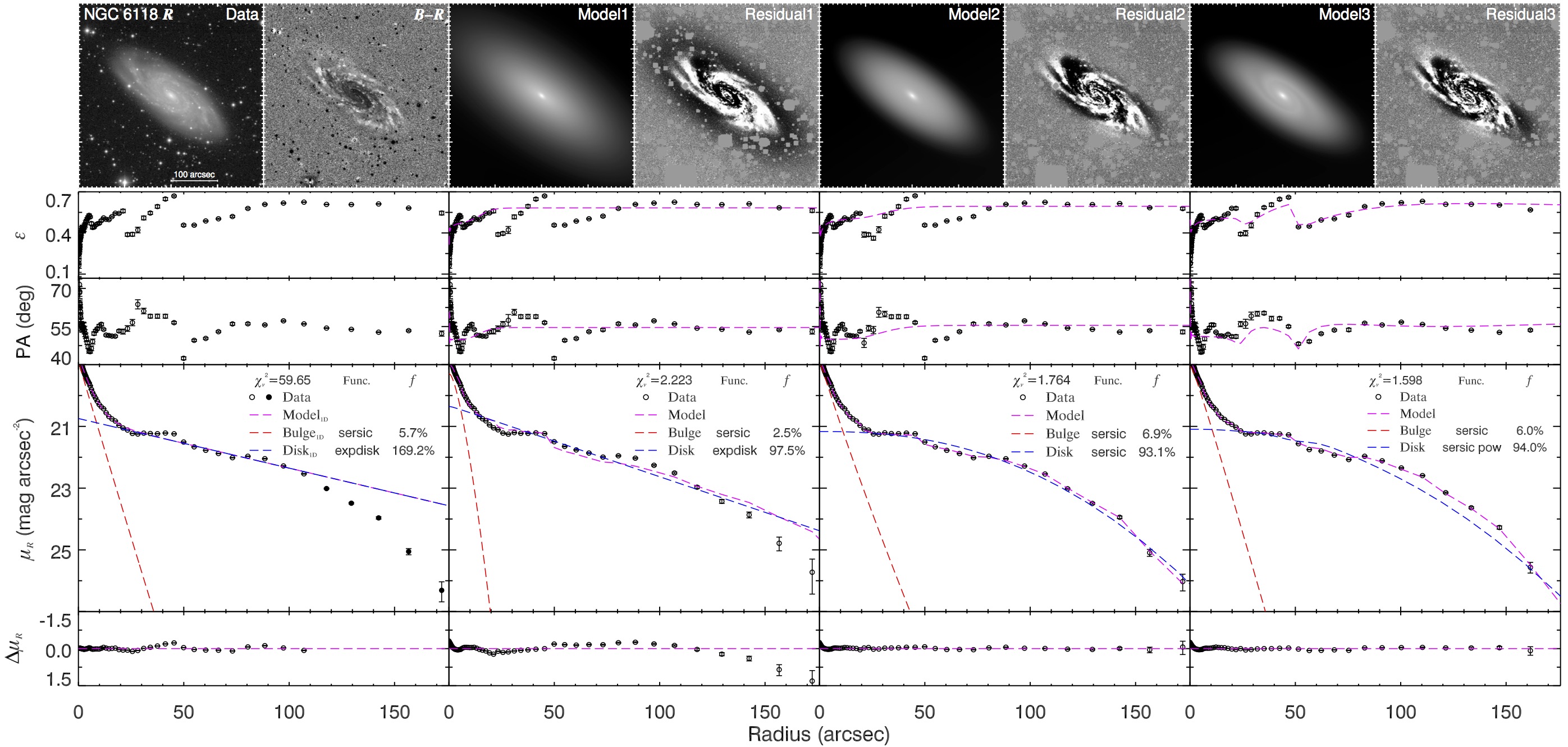}
  \caption{The best-fit 1D/2D models and isophotal analysis of NGC~6118. Same convention as in Figure~\ref{fig:NGC1411}. \label{fig:NGC6118}}
\end{figure*}

As NGC~6118 is qualitatively similar to NGC~7083, we decompose it in a similar fashion. We fit the
1D surface brightness profile (Figure~\ref{fig:NGC6118}; Table~\ref{tab:NGC6118}) by excluding data
beyond 110\arcsec{}, and we do not estimate the uncertainties caused by different excluded ranges.

The initial 2D model (Model1), consisting only of a single exponential for the disk, gives a bad fit to the outer disk, as expected.  In view of the
fact that the disk profile exhibits a characteristic \sersic{} shape and the smooth break precludes an unambiguous identification of the break radius,
we choose not to follow the case of NGC~7083 to construct a composite profile to model disk break. Instead, we replace the exponential disk of Model1
with a \(n<1\) \sersic{} disk to represent a broken disk in Model2.  In Model3, we apply coordinate rotation to account for spiral arms.

Except for \(\epsilon\) and PA, the best-fit parameters from the 1D and 2D decompositions are in good agreement.  This is at odds with the case of NGC
7083. Whether the good agreement is fortuitous or not is unclear; as with previous case studies, the output from 1D fitting is difficult to predict.
The \(B/T\) increases when we account for the disk break (from Model1 to Model2), consistent with
the behavior seen in NGC~7083.  Structural parameters of
the bulge change as expected when we apply coordinate rotation to the disk component, for the same reason that was discussed in
Section~\ref{sec:ngc-1357}.

We again find that disk break will cause the bulge to be underestimated if it is not properly modeled. And we confidently conclude that spiral
arms can be safely neglected.

\begin{deluxetable*}{lccccccc}
  \tablecaption{Best-fit Parameters for the Bulge of NGC~6118 \label{tab:NGC6118}}
  \tablecolumns{8}
  \tablehead{\multicolumn{1}{c}{\multirow{2}{*}{Model}} & \colhead{\(m_{R}\)} & \colhead{\(B/T\)} & \colhead{\(\mu_{{e},R}\)} &
    \colhead{\(n\)} & \colhead{\(r_{{e}}\)} & \colhead{\(\epsilon\)} & \colhead{PA} \\ \colhead{} & \colhead{(mag)} & \colhead{} &
    \colhead{(mag~arcsec\(^{-2}\))} & \colhead{} & \colhead{(\arcsec{})} & \colhead{} & \colhead{(\arcdeg{})}\\ \colhead{(1)} & \colhead{(2)} &
    \colhead{(3)} & \colhead{(4)} & \colhead{(5)} & \colhead{(6)} & \colhead{(7)} & \colhead{(8)}}
  \startdata
  Model0\tablenotemark{a} & \(14.10\pm0.16\) & \(0.057\pm0.009\) & \(20.74\pm0.13\) & \(1.02\pm0.07\) & \(7.97\pm0.58\) & \(0.408\pm0.045\) &
  \(56.82\pm3.55\) \\[0.05cm]
  \hline
  \rule{0pt}{3ex}Model1\tablenotemark{b} & \(14.91\substack{+0.03\\-0.03}\) & \(0.025\substack{+0.000\\-0.000}\) & \(20.30\substack{+0.01\\-0.01}\) &
  \(0.69\substack{+0.01\\-0.01}\) & \(5.21\substack{+0.09\\-0.08}\) & \(0.486\substack{+0.000\\-0.000}\) & \(49.22\substack{+0.08\\-0.08}\) \\[0.1cm]
  Model2\tablenotemark{c} & \(13.97\substack{+0.03\\-0.03}\) & \(0.069\substack{+0.003\\-0.002}\) & \(20.74\substack{+0.04\\-0.03}\) &
  \(1.12\substack{+0.03\\-0.03}\) & \(9.10\substack{+0.22\\-0.20}\) & \(0.505\substack{+0.000\\-0.001}\) & \(49.92\substack{+0.05\\-0.06}\) \\[0.1cm]
  Model3\tablenotemark{d} & \(14.11\substack{+0.01\\-0.01}\) & \(0.060\substack{+0.002\\-0.001}\) & \(20.56\substack{+0.01\\-0.01}\) &
  \(0.96\substack{+0.01\\-0.01}\) & \(8.10\substack{+0.08\\-0.08}\) & \(0.506\substack{+0.000\\-0.000}\) & \(51.00\substack{+0.03\\-0.02}\) 
  \enddata
  \tablecomments{See Table~\ref{tab:NGC1411} for details.}
  \tablenotetext{a}{Model configuration: Bulge+Disk.}
  \tablenotetext{b}{Model configuration: Bulge+Disk.}
  \tablenotetext{c}{Model configuration: Bulge+Broken Disk.}
  \tablenotetext{d}{Model configuration: Bulge+Spiral Broken Disk.}
\end{deluxetable*}

\subsection{NGC 1533}
\label{sec:ngc-1533}

\begin{figure*}
  \epsscale{1.17}
  \plotone{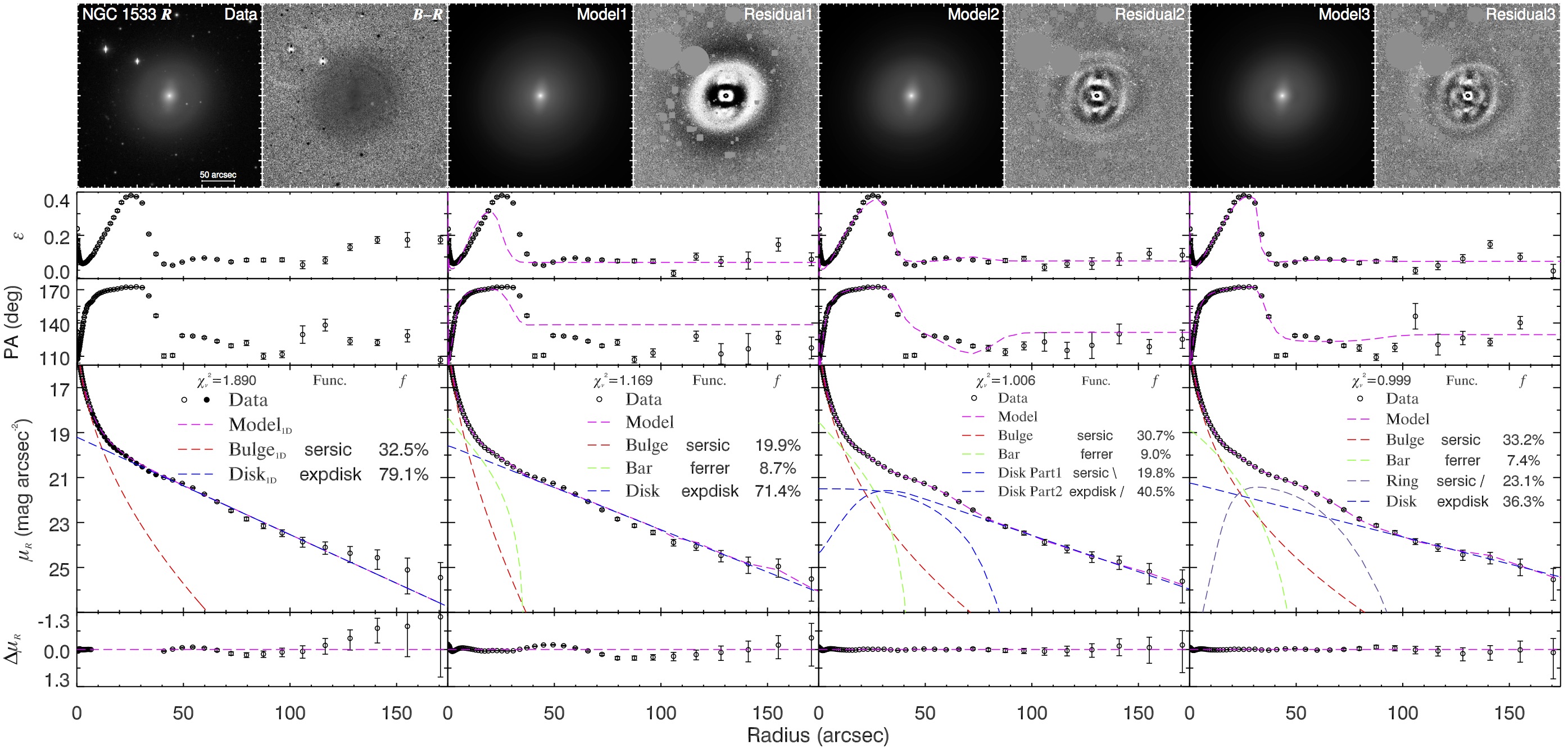}
  \caption{Best-fit 1D/2D models and isophotal analysis of NGC~1533. Same conventions as in Figure~\ref{fig:NGC1411}. \label{fig:NGC1533}}
\end{figure*}

NGC~1533 is the first barred galaxy in the sample. It is viewed in a nearly perfect face-on orientation and is characterized by the typical smooth
appearance of S0s. It has a prominent bulge, a round disk, and a short bar, which is the simplest configuration of barred galaxies.
\citet{1994cag+Sandage} reported possible spiral features at the edge of the disk, leading to a mixed classification of SB0/Sa. However, we observe no
sign of spiral features on the CGS \(R\)-band image of the galaxy, either on the original image or on the residual images (see
Figure~\ref{fig:NGC1533}). We notice a sharp edge in the light distribution beyond 70\arcsec{}, indicating the possible presence of a lens or a
large-scale ring. This is consistent with the identification of structural components by \citet{2006AJ+Laurikainen+BD_dec_S0} using a \(K_{s}\)-band
image.

As discussed in Section~\ref{sec:methodology}, we exclude the bar-dominated part of the surface
brightness profile when fitting it (Figure~\ref{fig:NGC1533}; Table~\ref{tab:NGC1533}). We omit data
from 7\arcsec{} to 40\arcsec{} and estimate the uncertainties by expanding and contracting the
excluded range by shifting the start point by \(2\arcsec\) or \(3\arcsec\) and the end point by
10\arcsec{}. We refrain from excluding the ring-dominating region to ensure stable fitting results,
for the same reason we did not exclude the nuclear lens in the case of NGC~1411.

The initial 2D model for barred galaxies differs from that for unbarred galaxies by introducing the
modified Ferrer function to represent the bar component. On the residual image of Model1, we observe
a positive ring pattern that signals the change of disk light profile from one side to
another. Inside the residual ring pattern, there is a conspicuous ``dark hole'' around the bulge,
indicating that the bulge may have been severely underestimated in this model. Interestingly, this
``hole'' has a comparable size with the bar, and the residual image shows negative residuals roughly
perpendicular to the bar major axis and positive residuals along the bar minor axis. All these
morphological features are natural consequences of bar-induced evolution. During development of the
bar, stellar orbits are rearranged to align with the major axis of the bar, and the bar potential
gets strengthened further. Consequently, disk stars within the bar influence are captured and gas
gets shocked or collides, inevitably inflowing to the galaxy center (e.g.,
\citealp{1982SAAS+Kormendy+Obs_gal_stru_dyn,1993RPPh+Sellwood+Bar_Dynamics}). The inner disk is
robbed of stars and gas, and of course shows a deficit in surface brightness.  The \(N\)-body
simulations of \citet{2002MNRAS+Athanassoula+Nbody_bar_halo_concentration} reproduced such an empty
region around the bar (see their Figures~2 and 3). These considerations lead us to conclude that a
single exponential function cannot describe the overall profile of the disk component of a barred
galaxy, from the faint outskirt all the way to the center. Therefore, in Model2 we break the disk
into two parts, as in the case of NGC~7083, and introduce a \sersic{} function to account for the
inner part of the disk with shallower light profile. We tried to use an exponential function for the
inner part of the disk, but it turns out to be so flat that its scale length is excessively
large. So we use a \sersic{} function with \(n<1\) instead. On the residual image of Model2, we
notice that the systematic negative residual pattern around the bulge disappears. In Model3, the
disk surface brightness model is constructed in a totally different way. The inner ring is
represented by an inner-truncated \sersic{} function. In conjunction with an underlying exponential
disk, these two components can also produce a seemingly broken disk. Model2 and Model3 are
equivalent in the sense that they describe complexities of the same level.

The 1D results agree with those of Model2 and Model3, except for the \sersic{} indices. As the disk is described more precisely from Model1 to Model2,
\(B/T\) grows as expected.  As mentioned above, since the disk components overestimate their light contribution in the central region in Model1, the
bulge component is consequently suppressed. The bulge parameters of Model2 and Model3 are consistent with each other, which reassures us that they are
not sensitive to the choice of model construction for the inner ring. The bar component in the three 2D models seems to have escaped the trap. On
account of the distinct PA and ellipticity of the bar, its parameters are quite stable in contrast to those of the bulge.  We also notice that the PA
of the bulge component varies by more than 16\arcdeg{} among the three 2D models. This is caused by the nearly face-on orientation of the galaxy. The
low ellipticities of the bulge and disk blurs their orientations. In any event, the PAs of bulges are not key parameters of interest.

Rings and lenses are physically different morphological components; inner rings arise as dynamical consequence of bar potential
\citep{1996FCPh+Buta+Galactic_Rings}, while inner lenses could be defunct bars \citep{1979ApJ+Kormendy+bar_lens_ring,
  2009ApJ+Laurikainen+bar_lens_oval}. Also, rings are considered to be localized while lenses have radial extent to the very center. However,
sometimes they are indistinguishable; for example, in this case the inner ring can also be considered an inner lens. They appear as shelves or end of
shelves on surface brightness profiles and thus deliver the same message---the slope of the surface brightness profile varies from one side to another
of a ring or edge of a lens. Moreover, we show that the overall disk surface brightness that includes lenses or rings can be modeled mathematically
interchangeably. As shown in this case, the inner ring can be modeled as accessory of the broken disk (Model3), or as a superimposed component on top
of the underlying disk (Model4), which is reminiscent of how the inner lens is accounted for in Model4 or Model3 of NGC~1411, respectively. Therefore,
we will make no effort to distinguish rings from lenses in barred galaxies, as they will be captured by the same modeling approach.

A constant ring-like positive residual pattern located 10\arcsec{} away from the centroid of the
galaxy shows up in the residual images of the three 2D models. It is reminiscent of the ``barlens''
proposed by \citet{2011MNRAS+Laurikainen+NIRS0S}. Indeed, the galaxy is classified as
(RL)SB(bl)0\arcdeg{} by \citet{2015ApJS+Buta+Morphology_S4G}, where bl is short for
barlens. \citet{2013MNRAS+Laurikainen+stat_structural_comp} studied the statistics of structural
components in early-type disk galaxies and concluded that barlenses might have evolve to become
inner lenses. A recent theoretical study on barlenses suggests that they are merely face-on versions
of boxy/peanut bulges \citep{2015MNRAS+Athanassoula+barlens}.
\citet{2014MNRAS+Laurikainen+barlens_xbulges} lent observational support to this scenario. If we
assume that the barlens is a boxy/peanut bulge viewed face-on, then we do not need to account for it
separately because it is part of the photometric bulge and in this study we do not separate
subcomponents of composite bulges.

In this case, we learn that the inner ring ought to be properly modeled, for it affects bulge measurements in the same way as
disk break does. This type of inner disk modification differs from the disk break often discussed.  Light deficit of the inner disk is
easily missed in azimuthally averaged 1D profiles, but disk break that usually occurs at faint outer regions is not.

\begin{deluxetable*}{lccccccc}
  \tablecaption{Best-fit Parameters for the Bulge of NGC~1533 \label{tab:NGC1533}}
  \tablecolumns{8}
  \tablehead{\multicolumn{1}{c}{\multirow{2}{*}{Model}} & \colhead{\(m_{R}\)} & \colhead{\(B/T\)} & \colhead{\(\mu_{e,R}\)} &
    \colhead{\(n\)} & \colhead{\(r_{e}\)} & \colhead{\(\epsilon\)} & \colhead{PA} \\ \colhead{} & \colhead{(mag)} & \colhead{} &
    \colhead{(mag~arcsec\(^{-2}\))} & \colhead{} & \colhead{(\arcsec{})} & \colhead{} & \colhead{(\arcdeg{})}\\ \colhead{(1)} & \colhead{(2)} &
    \colhead{(3)} & \colhead{(4)} & \colhead{(5)} & \colhead{(6)} & \colhead{(7)} & \colhead{(8)}}
  \startdata
  Model0\tablenotemark{a} & \(11.27\pm0.14\) & \(0.325\pm0.047\) & \(18.05\pm0.15\) & \(1.98\pm0.12\) & \(5.90\pm0.68\) & \(0.088\pm0.005\) &
  \(132.72\pm3.80\) \\[0.05cm]
  \hline
  \rule{0pt}{3ex}Model1\tablenotemark{b} & \(11.74\substack{+0.02\\-0.03}\) & \(0.199\substack{+0.000\\-0.001}\) & \(17.51\substack{+0.02\\-0.02}\) &
  \(1.66\substack{+0.02\\-0.01}\) & \(3.81\substack{+0.07\\-0.06}\) & \(0.061\substack{+0.002\\-0.002}\) & \(116.56\substack{+1.92\\-1.69}\) \\[0.1cm]
  Model2\tablenotemark{c} & \(11.29\substack{+0.03\\-0.04}\) & \(0.307\substack{+0.003\\-0.002}\) & \(18.24\substack{+0.02\\-0.01}\) &
  \(2.39\substack{+0.02\\-0.00}\) & \(6.01\substack{+0.15\\-0.08}\) & \(0.056\substack{+0.003\\-0.002}\) & \(123.14\substack{+6.75\\-4.32}\) \\[0.1cm]
  Model3\tablenotemark{d} & \(11.17\substack{+0.09\\-0.10}\) & \(0.332\substack{+0.003\\-0.012}\) & \(18.41\substack{+0.17\\-0.15}\) &
  \(2.57\substack{+0.15\\-0.13}\) & \(6.76\substack{+0.81\\-0.64}\) & \(0.053\substack{+0.000\\-0.000}\) & \(131.50\substack{+0.00\\-1.68}\) 
  \enddata
  \tablecomments{See Table~\ref{tab:NGC1411} for details.}
  \tablenotetext{a}{Model configuration: Bulge+Disk.}
  \tablenotetext{b}{Model configuration: Bulge+Bar+Disk.}
  \tablenotetext{c}{Model configuration: Bulge+Bar+Broken Disk(Part1+Part2).}
  \tablenotetext{d}{Model configuration: Bulge+Bar+Ring+Disk.}
\end{deluxetable*}

\subsection{NGC 1326}
\label{sec:ngc-1326}
\begin{figure*}
  \epsscale{1.17}
  \plotone{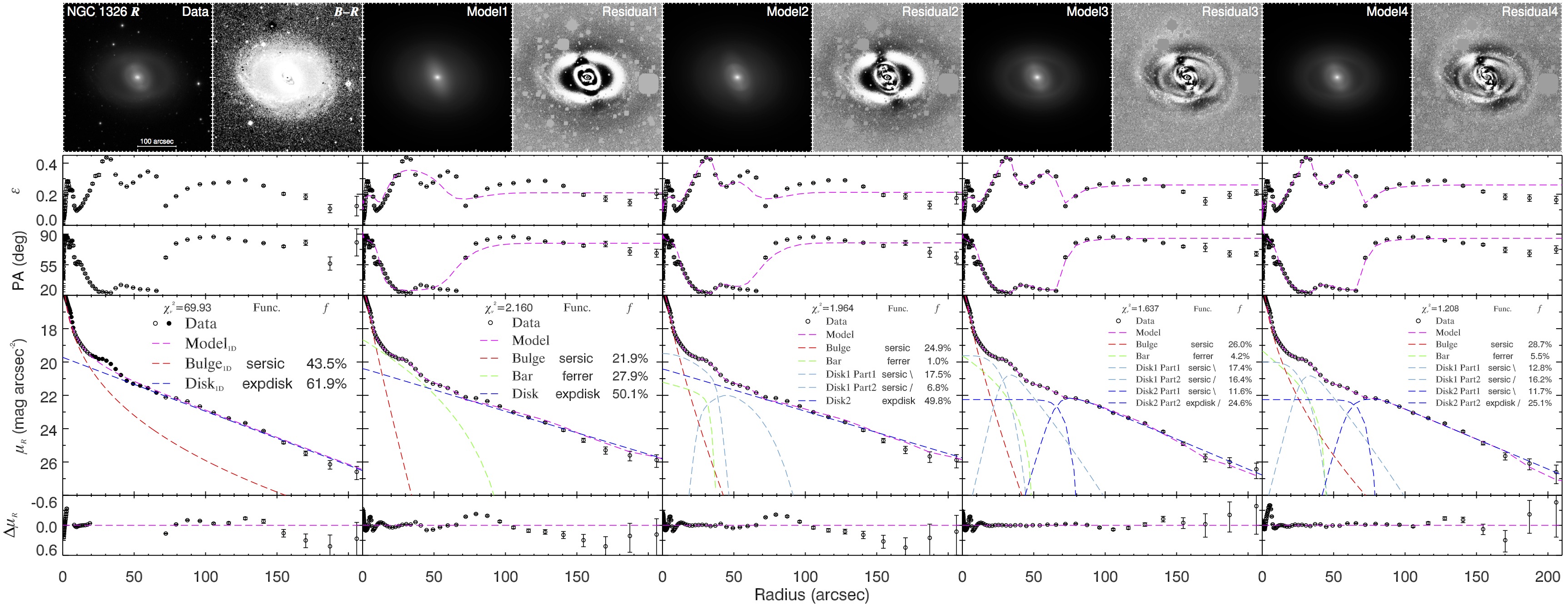}
  \caption{Best-fit 1D/2D models and isophotal analysis of NGC~1326. Same conventions as in Figure~\ref{fig:NGC1411}. \label{fig:NGC1326}}
\end{figure*}

NGC~1326, analogous to NGC~1411, has a comprehensive manifestation of various types of rings. The
presence of rings with different sizes suggests the presence of a bar, although it is not evident on
the \(R\)-band image due to dust lanes on the edges of the bar and the bright inner ring surrounding it
(see the data image and the color map in Figure~\ref{fig:NGC1326}). However, the morphological features that
obfuscate the visual appearance of the bar actually betray its presence: the inner ring is
believed to be associated with a bar-induced resonance, and gas inflow along the dust lanes, also a
dynamical consequence of the bar potential
\citep{1980ApJ+Sanders+gas_flow_barred_galaxy,2015ApJ+Li+simulation_nr}, which is fueling the
nuclear ring, should appear on the leading edges of the bar. The peak at \(30\arcsec\) in the
ellipticity profile further strengthens our belief that there is indeed a bar. Therefore, we
conclude that the presence of bar is unambiguous. Whether it can be robustly modeled with 2D fitting
remains to be seen. \citet{2015ApJS+Buta+Morphology_S4G} classify the galaxy as
(\(\mathrm{R_{1}SAB_{a}(r,bl,nr)0^{+}}\)), although we observe no sign of a barlens on their image
or ours. We further question whether a barlens can be recognized when the nuclear ring is
present.  Inspection of the intensity of the \(\mathrm{R_{1}}\) outer ring reveals that its concave
part (i.e., the part closest to the major axis of the bar) is brighter than the rest of
it. In 2D image fitting, the brighter part of the outer ring will induce the bar component to
stretch outward more than it should for flux compensation. Nuclear rings are considered to be
observational signatures of pseudobulges; they are a clear manifestation of the build-up of
  inner disks \citep{2004ARA&A+Kormendy+Pseudobulges}. Thus, it is legitimate to consider the
nuclear ring as part of the photometric bulge, although we will explore the impact of the nuclear
ring on bulge parameters.

We decompose the 1D surface brightness profile of the galaxy (Figure~\ref{fig:NGC1326}; Table~\ref{tab:NGC1326}) by excluding data from 3\arcsec{} to
8\arcsec{} (nuclear ring) and from 20\arcsec{} to 70\arcsec{} (bar and inner ring). We refrain from excluding the outer ring, or else few data will
be left. Uncertainties are measured by expanding and contracting the excluded range for the nuclear ring, by shifting the start point by 0.5\arcsec{}
and the end point by 1\arcsec{}. As it is impractical to further explore different excluded ranges for the bar and the inner ring, the uncertainties
of the 1D parameters should be regarded as lower limits to the true uncertainties.

The 2D models (Table~\ref{tab:NGC1326}) are constructed in a similar way as in the case of
NGC~1533. The bar component in Model1 is fitting both the bar and the inner ring because of their
similar sizes and orientations. Moreover, systematic negative residuals on both sides of the rings
are consistent with our recognition of the ring(lens)-like feature of NGC~1533. Based on what we
learned about treating the lenses of NGC~2784 and NGC~1411, we speculate that the outer ring is not
as crucial as the nuclear and inner rings for bulge parameter measurements. The nuclear ring is a
subtle feature that is hard to model; hence, we first fit the inner ring in Model2.  We model the
inner ring based on another underlying disk component because the inner ring has distinct PA and
\(\epsilon\) compared to the outer disk, on which the outer ring will be based. In Model2, the bar
and the inner ring are separated, and the systematic negative residual pattern inside the inner ring
in Model1 vanishes. The outer ring is accounted for in Model3 in the same way as the inner ring
was. We find that the difference between the bulge parameters of Model2 and Model3 is marginal,
which again justifies the deliberately assigned lower priority of the outer ring. In Model4, we mask
the nuclear ring and refit Model3 to the image. The \sersic{} index appears to be the most affected
parameter of the bulge. We mask the nuclear ring instead of modeling it because of the inability of
our method to separate the ring from the underlying component (i.e., the disk for inner/outer rings
or the bulge for nuclear rings).  However, we achieve our main objective, which is to isolate the
bulge from every other structural component.

The 1D parameters, even considering their errors, are fairly discrepant with the 2D ones. The trend for \(B/T\) to increase from Model1 to the more
sophisticated 2D models is obvious. This is for the same reason that was explained in the previous case---the disk light contribution in the central
region is overestimated by Model1. We note that the dramatically varying parameters of the bar component from Model2 to Model3 are worrisome. In
contrast with the strong bar in NGC~1533, the ill-defined bar of NGC~1326 is not capable of resisting the variation of the input model.  Nevertheless,
the robust bulge parameters are quite reassuring. In Model3 and Model4, the bar components project their ends outside the inner ring due to the pull
from the brighter part of the outer ring. The bar sizes are evidently overestimated, but the bulge parameters should be stable against such degeneracy
in the outskirts. This is intrinsically the same argument we made for assigning low priorities to outer lenses and rings. We find that, except for the
\sersic{} index, the nuclear ring has a negligible effect on the rest of the bulge parameters. The role of the nuclear ring in perturbing the surface
brightness of the bulge is similar to that of the nuclear lens of NGC~1411. The nuclear ring does not carry enough light to dramatically
alter \(B/T\); however, if it is not excluded from the fit, the bulge component will try to incorporate the nuclear ring, which will result in a
smaller \sersic{} index \(n\) because of the shallower light distribution of the ring. This may be responsible, at least in part, for the low
\sersic{} indices reported for pseudobulges that intrinsically have high \sersic{} indices, as nuclear rings are common
morphological indicators of pseudobulges \citep{2004ARA&A+Kormendy+Pseudobulges,2008AJ+Fisher+Sersic_Index, 2009ApJ+Fisher+PB_Growth,
  2010ApJ+Fisher+Scaling_Relation_Bulges}.

NGC~1326 also has an inner ring, apart from a nuclear and an outer ring. By exploring models with and without the inner ring, we reinforce our
previous finding that it is important to take into account the light deficit of the inner disk
induced by the bar. The outer ring has a negligible
effect on the bulge parameters. The nuclear ring has little impact on the bulge luminosity but may alter the \sersic{} index significantly.

\begin{deluxetable*}{lccccccc}
  \tablecaption{Best-fit Parameters for the Bulge of NGC~1326 \label{tab:NGC1326}} \tablecolumns{8}
  \tablehead{\multicolumn{1}{c}{\multirow{2}{*}{Model}} & \colhead{\(m_{R}\)} & \colhead{\(B/T\)} & \colhead{\(\mu_{{e},R}\)} &
    \colhead{\(n\)} & \colhead{\(r_{{e}}\)} & \colhead{\(\epsilon\)} & \colhead{PA} \\ \colhead{} & \colhead{(mag)} & \colhead{} &
    \colhead{(mag~arcsec\(^{-2}\))} & \colhead{} & \colhead{(\arcsec{})} & \colhead{} & \colhead{(\arcdeg{})}\\ \colhead{(1)} & \colhead{(2)} &
    \colhead{(3)} & \colhead{(4)} & \colhead{(5)} & \colhead{(6)} & \colhead{(7)} & \colhead{(8)}}
  \startdata
  Model0\tablenotemark{a} & \(10.73\pm0.30\) & \(0.435\pm0.109\) & \(18.86\pm0.75\) & \(3.06\pm0.77\) & \(10.01\pm3.77\) & \(0.108\pm0.023\) &
  \(73.99\pm4.78\) \\[0.05cm]
  \hline
  \rule{0pt}{3ex}Model1\tablenotemark{b} & \(11.42\substack{+0.00\\-0.00}\) & \(0.219\substack{+0.004\\-0.004}\) & \(17.07\substack{+0.00\\-0.00}\) &
  \(1.14\substack{+0.00\\-0.00}\) & \(4.31\substack{+0.00\\-0.00}\) & \(0.228\substack{+0.000\\-0.000}\) & \(83.73\substack{+0.00\\-0.01}\) \\[0.1cm]
  Model2\tablenotemark{c} & \(11.28\substack{+0.00\\-0.00}\) & \(0.249\substack{+0.004\\-0.003}\) & \(17.22\substack{+0.00\\-0.00}\) &
  \(1.32\substack{+0.00\\-0.00}\) & \(4.73\substack{+0.00\\-0.01}\) & \(0.209\substack{+0.000\\-0.001}\) & \(82.48\substack{+0.00\\-0.11}\) \\[0.1cm]
  Model3\tablenotemark{d} & \(11.29\substack{+0.00\\-0.00}\) & \(0.260\substack{+0.000\\-0.000}\) & \(17.20\substack{+0.00\\-0.00}\) &
  \(1.29\substack{+0.01\\-0.00}\) & \(4.72\substack{+0.00\\-0.00}\) & \(0.215\substack{+0.000\\-0.000}\) & \(83.37\substack{+0.00\\-0.01}\) \\[0.1cm]
  Model4\tablenotemark{e} & \(11.20\substack{+0.00\\-0.00}\) & \(0.287\substack{+0.001\\-0.000}\) & \(17.75\substack{+0.01\\-0.01}\) &
  \(2.01\substack{+0.01\\-0.01}\) & \(5.64\substack{+0.01\\-0.01}\) & \(0.198\substack{+0.000\\-0.000}\) & \(85.86\substack{+0.00\\-0.01}\)
  \enddata
  \tablecomments{See Table~\ref{tab:NGC1411} for details.}
  \tablenotetext{a}{Model configuration: Bulge+Disk.}
  \tablenotetext{b}{Model configuration: Bulge+Bar+Disk.}
  \tablenotetext{c}{Model configuration: Bulge+Bar+Broken Disk1(Part1+Part2)+Disk2.}
  \tablenotetext{d}{Model configuration: Bulge+Bar+Broken Disk1(Part1+Part2)+Broken Disk2(Part1+Part2). Errors for this model are only lower limits.}
  \tablenotetext{e}{This model has the same configuration as Model3, only that the nuclear ring is masked. Errors for this model are only lower
    limits.}
\end{deluxetable*}

\subsection{IC 5240}
\label{sec:ic-5240}

\begin{figure*}
  \epsscale{1.17}
  \plotone{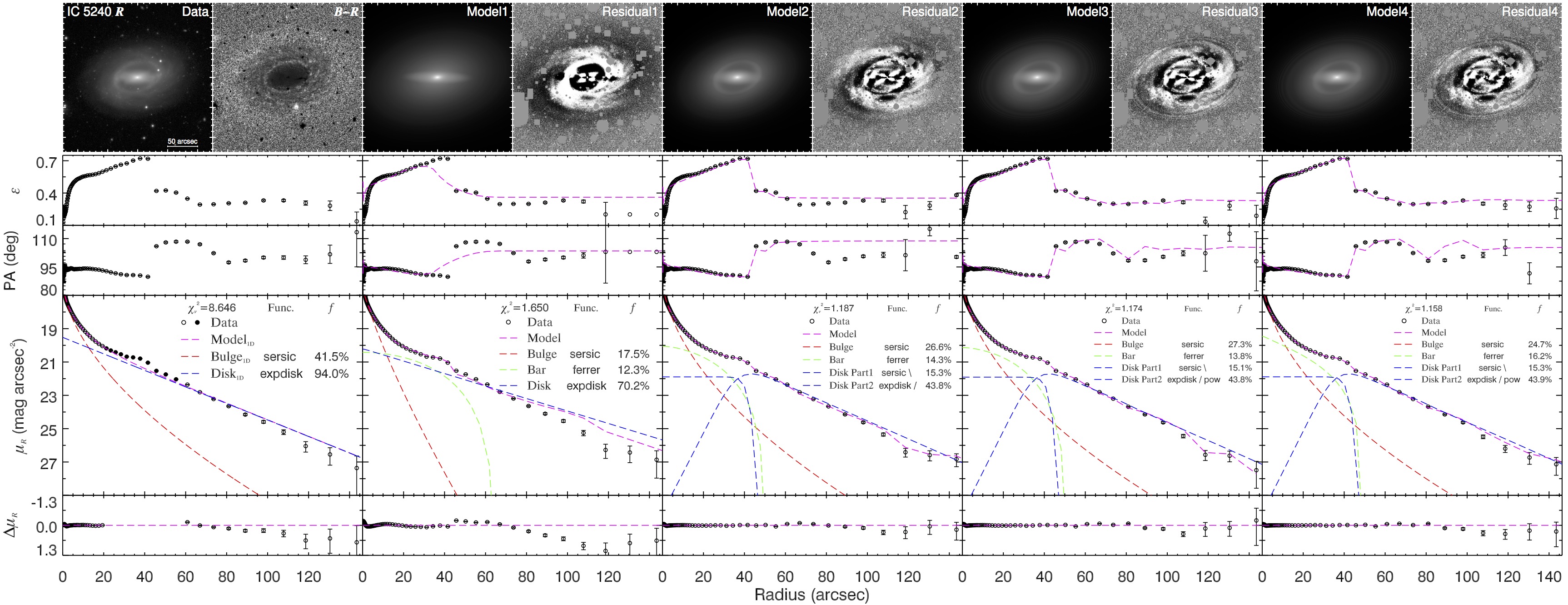}
  \caption{Best-fit 1D/2D models and isophotal analysis of IC~5240. Same conventions as in Figure~\ref{fig:NGC1411}. \label{fig:IC5240}}
\end{figure*}

Unlike NGC~1326, the prominent bar with the associated inner ring of IC~5240 is readily identified from the \(R\)-band image, while the principal
spiral pattern outside the inner ring is blurred. The spiral pattern recognized by \citet{1994cag+Sandage} is evident only on the residual images (see
the Residual1 and Residual2 panels of Figure~\ref{fig:IC5240}). The strong X-shape feature of the boxy/peanut bulge is visible even on the original image
and clearly stands out in the residual images. This boxy/peanut bulge seen at moderately inclined viewing angles has been studied in detail
\citep{2013MNRAS+Erwin+non_edge-on_Boxy_Bulges,2014MNRAS+Laurikainen+barlens_xbulges}. Obviously, the inner disk (inside the inner ring) exhibits a
surface brightness as faint as that of the outskirt of the disk. Moreover, the redder color (darker in the color map) of the inner disk indicates
that gas inside the inner ring has been depleted by the bar and therefore the inner smooth disk is not as vibrant as the outer spiral disk.

We fit the 1D surface brightness profile of the galaxy (Figure~\ref{fig:IC5240}) by excluding data from 20\arcsec{} to 55\arcsec{}, which is dominated
by the bar and the inner ring. The uncertainties of the best-fit parameters (Table~\ref{tab:IC5240}) are obtained by expanding and contracting the
excluded range by shifting the start point by 5\arcsec{} and the end point by 10\arcsec{}.

Similar to the previous two barred galaxies, 2D Model1 again yields a dark hole. There is also a conspicuous X-shape pattern associated with the
bulge. As before, we improve Model1 by breaking the disk component into two parts. Then the systematic negative residuals vanish in Model2. In Model3,
we apply coordinate rotation to the outer part of the disk to model the outer spiral pattern. Improvement of the residuals is marginal, and there is
little change in the bulge structural parameters, because the spiral feature is very weak. In Model4, we add an \(m=4\) Fourier mode to the \sersic{}
bulge component to model its X-shape feature. The amplitude of the \(m=4\) Fourier mode potentially serves as another metric to quantify the boxyness
of the bulge \citep{2016MNRAS+Ciambur+quanti_boxyness}.

Most of the 1D parameters show significant deviation from the 2D ones, especially for \(B/T\). Their
huge error bars are also worrisome. The dramatic variation of \(B/T\) from Model1 to Model2 further
strengthens the importance of replacing the default exponential law for the disk profile. By
comparison, the influence of spiral arms or X-shape features is relatively minor.  The spiral arms
stop at the inner ring and thus barely affect the surface brightness distribution inside.

\begin{deluxetable*}{lccccccc}
  \tablecaption{Best-fit Parameters for the Bulge of IC~5240 \label{tab:IC5240}} \tablecolumns{8}
  \tablehead{\multicolumn{1}{c}{\multirow{2}{*}{Model}} & \colhead{\(m_{R}\)} & \colhead{\(B/T\)} & \colhead{\(\mu_{{e},R}\)} &
    \colhead{\(n\)} & \colhead{\(r_{{e}}\)} & \colhead{\(\epsilon\)} & \colhead{PA} \\ \colhead{} & \colhead{(mag)} & \colhead{} &
    \colhead{(mag~arcsec\(^{-2}\))} & \colhead{} & \colhead{(\arcsec{})} & \colhead{} & \colhead{(\arcdeg{})}\\ \colhead{(1)} & \colhead{(2)} &
    \colhead{(3)} & \colhead{(4)} & \colhead{(5)} & \colhead{(6)} & \colhead{(7)} & \colhead{(8)}}
  \startdata
  Model0\tablenotemark{a} & \(12.12\pm0.22\) & \(0.415\pm0.079\) & \(19.54\pm0.25\) & \(1.74\pm0.19\) & \(9.56\pm1.81\) & \(0.333\pm0.028\) &
  \(93.60\pm0.21\) \\[0.05cm]
  \hline
  \rule{0pt}{3ex}Model1\tablenotemark{b} & \(12.88\substack{+0.00\\-0.01}\) & \(0.175\substack{+0.002\\-0.001}\) & \(18.64\substack{+0.01\\-0.00}\) &
  \(1.26\substack{+0.00\\-0.00}\) & \(5.57\substack{+0.03\\-0.00}\) & \(0.509\substack{+0.000\\-0.000}\) & \(94.50\substack{+0.02\\-0.02}\) \\[0.1cm]
  Model2\tablenotemark{c} & \(12.49\substack{+0.00\\-0.00}\) & \(0.266\substack{+0.001\\-0.001}\) & \(19.26\substack{+0.00\\-0.00}\) &
  \(1.86\substack{+0.00\\-0.00}\) & \(8.05\substack{+0.02\\-0.02}\) & \(0.499\substack{+0.000\\-0.000}\) & \(95.60\substack{+0.02\\-0.03}\) \\[0.1cm]
  Model3\tablenotemark{d} & \(12.46\substack{+0.00\\-0.00}\) & \(0.273\substack{+0.002\\-0.001}\) & \(19.29\substack{+0.00\\-0.00}\) &
  \(1.88\substack{+0.00\\-0.00}\) & \(8.27\substack{+0.02\\-0.01}\) & \(0.501\substack{+0.000\\-0.000}\) & \(95.43\substack{+0.02\\-0.02}\) \\[0.1cm]
  Model4\tablenotemark{e} & \(12.57\substack{+0.00\\-0.00}\) & \(0.247\substack{+0.001\\-0.002}\) & \(19.34\substack{+0.00\\-0.08}\) &
  \(1.93\substack{+0.00\\-0.00}\) & \(7.76\substack{+0.01\\-0.01}\) & \(0.486\substack{+0.000\\-0.000}\) & \(95.51\substack{+0.02\\-0.01}\)
  \enddata
  \tablecomments{See Table~\ref{tab:NGC1411} for details.}
  \tablenotetext{a}{Model configuration: Bulge+Disk.}
  \tablenotetext{b}{Model configuration: Bulge+Bar+Disk.}
  \tablenotetext{c}{Model configuration: Bulge+Bar+Broken Disk(Part1+Part2).}
  \tablenotetext{d}{Model configuration: Bulge+Bar+Broken Disk(Part1+Spiral Part2).}
  \tablenotetext{e}{Model configuration: Boxy Bulge+Bar+Broken Disk(Part1+Spiral Part2).}  
\end{deluxetable*}

\subsection{NGC 7329}
\label{sec:ngc-7329}

\begin{figure*}
  \epsscale{1.17}
  \plotone{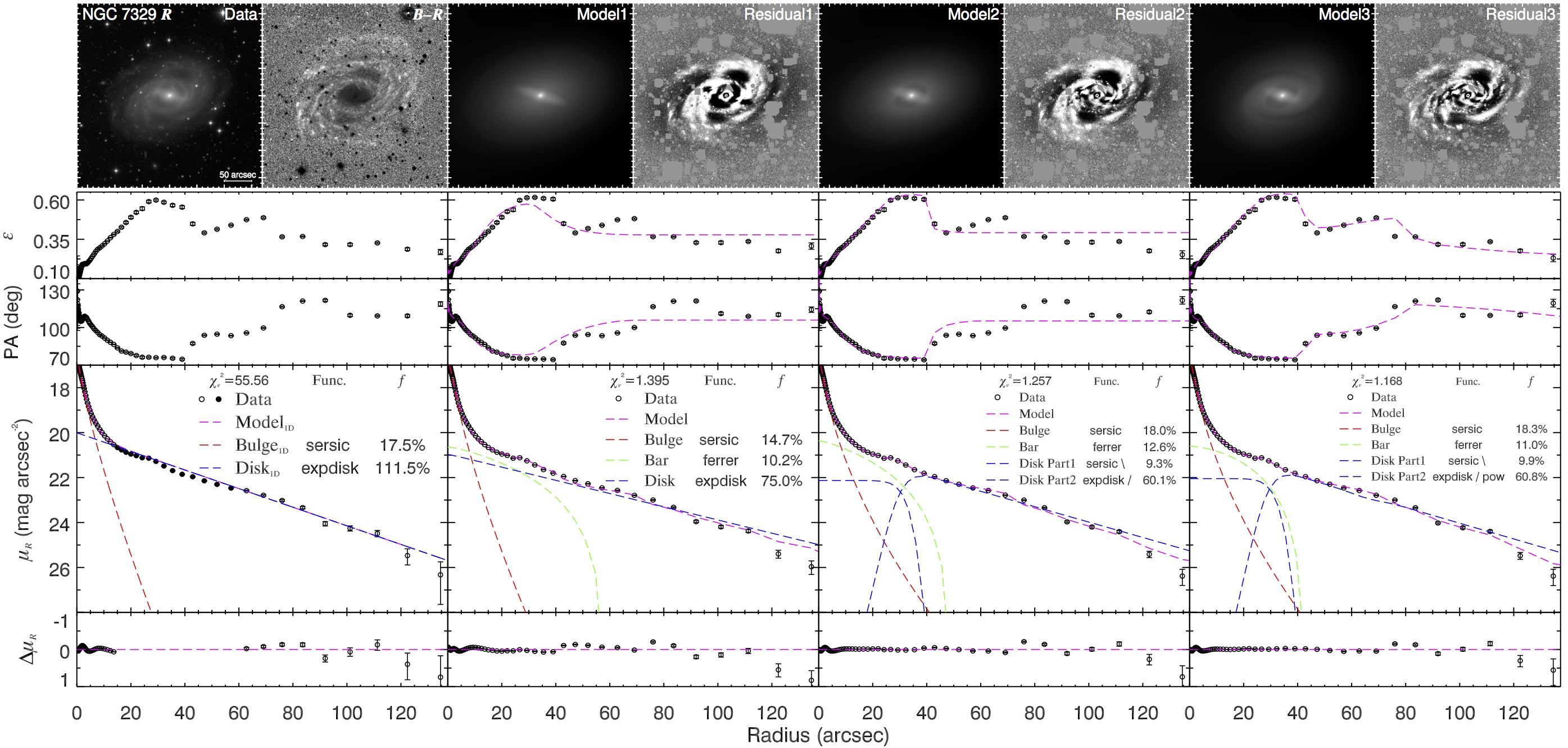}
  \caption{The best-fit 1D/2D models and isophotal analysis of NGC~7329. Same convention as in Figure~\ref{fig:NGC1411}. \label{fig:NGC7329}}
\end{figure*}

NGC~7329 is structurally qualitatively similar to IC~5240. The inner ring also demarcates the inner
red disk to the outer, vigorously star-forming disk (see the color map in Figure~\ref{fig:NGC7329}),
except that NGC~7329 does not contain a boxy/peanut bulge and its spiral pattern is more
evident. However, the galaxy may contain a barlens component as seen on the residual image (see the
Residual1 panel of Figure~\ref{fig:NGC7329}) which, as the central thickened part of bar, has been
argued to be intrinsically the same as a boxy/peanut bulge. A principal two-arm spiral feature is
readily recognized on the \(R\)-band image, and it fragments into multiple arms when approaching
outward. Analogous to the situation in NGC~6118, only the main two-arm spiral feature is modeled in
our decomposition.

The 1D and 2D decompositions (Figure~\ref{fig:NGC7329}; Table~\ref{tab:NGC7329}) are conducted as for IC~5240. We decompose the surface brightness
profile of the galaxy by excluding data from 15\arcsec{} to 60\arcsec{}, which is dominated by the bar and the inner ring. The uncertainties are
obtained by expanding and contracting the excluded range by shifting the start point by 5\arcsec{} and the end point by 10\arcsec{}.  The ubiquitous
empty region around the bar shows up again in this case, as seen on the residual image of Model1.  Model2 corrects this bias but leaves the spiral
features to the next model. Model3 applies coordinate rotation to the outer part of the disk to reproduce the grand-design spiral arms. Note that the
errors of the 1D parameters are quite stable compared with those of the other three barred galaxies in the sample. Even without excluding data from
the radial profile, a \sersic{}+exponential model fits the surface brightness very well, and the fitting results are consistent with those presented
here. The 1D parameters of the bulge are consistent with the 2D ones, except for the \sersic{} indices and ellipticities. The tendency for \(B/T\) to
increase from Model1 to Model2 also appears, although to a much smaller degree. We again find that the spiral arms have negligible effects on the
bulge parameters.

We note that the bar components of Model2 and Model3 extend into the outer disk. This is caused by
the spiral arms in the outer disk and the ansae at the ends of the bar. The light of the outer disk
is not uniformly distributed in all azimuthal directions but rather is concentrated in the
arms. When the outer disk is modeled by an axisymmetric model without coordinate rotation, it will
compromise between the rises and falls in the disk light.  Consequently the bar component will
stretch outward to compensate for the ``rises'' that are not perfectly fitted by the axisymmetric
outer disk component. However, the extension of the bar into the outer disk is limited because it
cannot bend to trace the spiral arms (see Model2). The size of the bar component in Model3 is
shortened because the spiral outer disk is properly modeled, even though the ansae at the ends of
the bar still induce the bar component to stretch into the outer disk.

This example reinforces the need to properly treat the empty region surrounding the bar. We further verify that, as already shown for unbarred
galaxies, spiral arms can be neglected insofar as their effects on the bulge parameters are concerned.

\begin{deluxetable*}{lccccccc}
  \tablecaption{Best-fit Parameters for the Bulge of NGC~7329 \label{tab:NGC7329}}
  \tablecolumns{8}
  \tablehead{\multicolumn{1}{c}{\multirow{2}{*}{Model}} & \colhead{\(m_{R}\)} & \colhead{\(B/T\)} & \colhead{\(\mu_{{e},R}\)} &
    \colhead{\(n\)} & \colhead{\(r_{{e}}\)} & \colhead{\(\epsilon\)} & \colhead{PA} \\ \colhead{} & \colhead{(mag)} & \colhead{} &
    \colhead{(mag~arcsec\(^{-2}\))} & \colhead{} & \colhead{(\arcsec{})} & \colhead{} & \colhead{(\arcdeg{})}\\ \colhead{(1)} & \colhead{(2)} &
    \colhead{(3)} & \colhead{(4)} & \colhead{(5)} & \colhead{(6)} & \colhead{(7)} & \colhead{(8)}}
  \startdata
  Model0\tablenotemark{a} & \(13.36\pm0.18\) & \(0.175\pm0.027\) & \(18.83\pm0.24\) & \(1.28\pm0.20\) & \(3.72\pm0.53\) & \(0.165\pm0.009\) &
  \(107.05\pm0.60\) \\[0.05cm]
  \hline
  \rule{0pt}{3ex}Model1\tablenotemark{b} & \(13.40\substack{+0.00\\-0.00}\) & \(0.147\substack{+0.003\\-0.003}\) & \(18.62\substack{+0.01\\-0.01}\) &
  \(1.45\substack{+0.01\\-0.01}\) & \(3.40\substack{+0.01\\-0.01}\) & \(0.250\substack{+0.000\\-0.000}\) & \(107.83\substack{+0.09\\-0.08}\) \\[0.1cm]
  Model2\tablenotemark{c} & \(13.24\substack{+0.00\\-0.00}\) & \(0.180\substack{+0.003\\-0.004}\) & \(18.97\substack{+0.00\\-0.00}\) &
  \(1.86\substack{+0.00\\-0.00}\) & \(4.10\substack{+0.00\\-0.00}\) & \(0.257\substack{+0.000\\-0.000}\) & \(108.60\substack{+0.22\\-0.19}\) \\[0.1cm]
  Model3\tablenotemark{d} & \(13.22\substack{+0.00\\-0.00}\) & \(0.183\substack{+0.003\\-0.003}\) & \(18.95\substack{+0.00\\-0.00}\) &
  \(1.84\substack{+0.01\\-0.00}\) & \(4.10\substack{+0.01\\-0.01}\) & \(0.256\substack{+0.000\\-0.000}\) & \(106.62\substack{+0.04\\-0.03}\)
  \enddata
  \tablecomments{See Table~\ref{tab:NGC1411} for details.}
  \tablenotetext{a}{Model configuration: Bulge+Disk.}
  \tablenotetext{b}{Model configuration: Bulge+Bar+Disk.}
  \tablenotetext{c}{Model configuration: Bulge+Bar+Broken Disk(Part1+Part2).}
  \tablenotetext{d}{Model configuration: Bulge+Bar+Broken Disk(Part1+Spiral Part2).}
\end{deluxetable*}

\subsection{NGC 945}
\label{sec:ngc-945}

\begin{figure*}
  \epsscale{1.17}
  \plotone{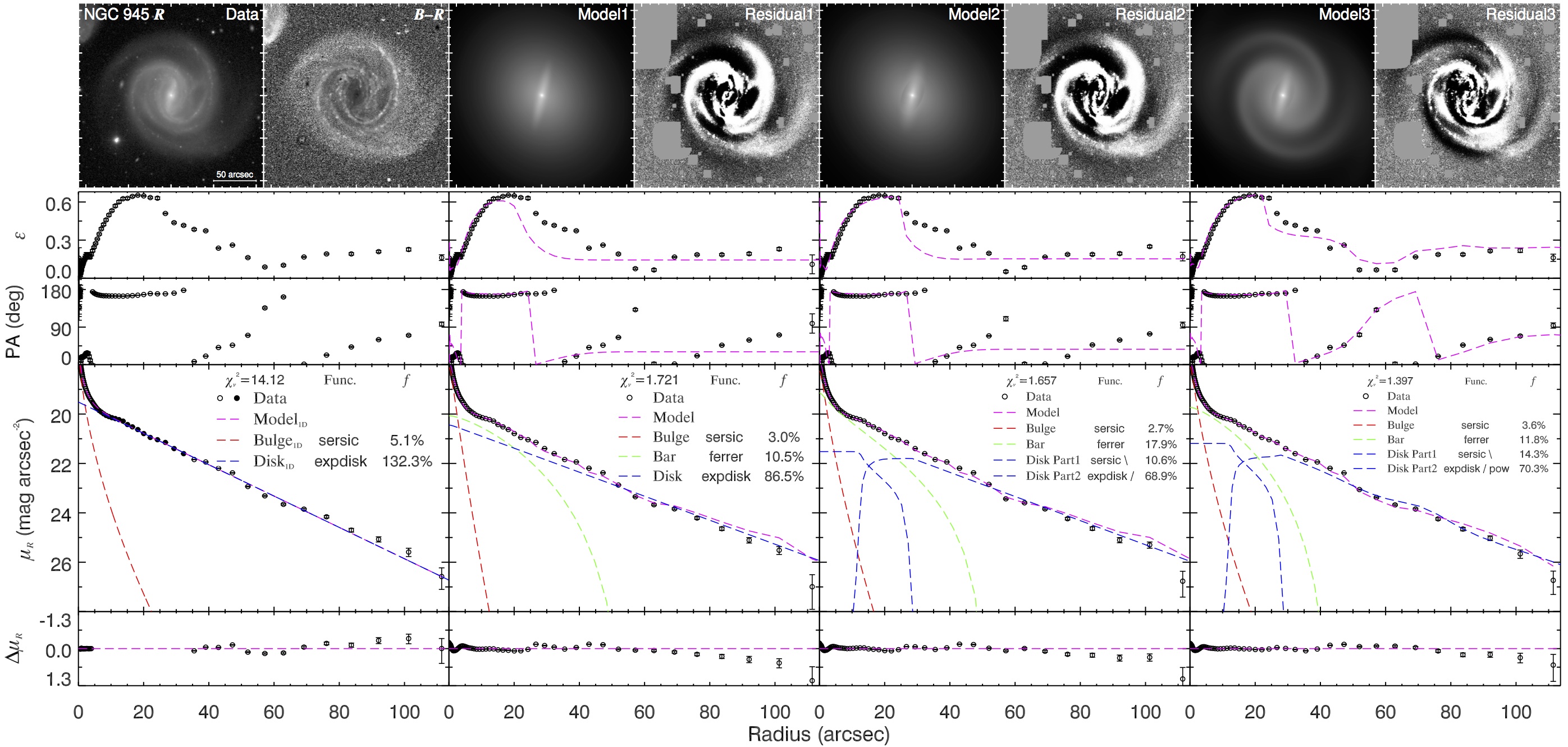}
  \caption{The best-fit 1D/2D models and isophotal analysis of NGC~945. Same convention as in Figure~\ref{fig:NGC1411}. \label{fig:NGC945}}
\end{figure*}

The structure of NGC~945 is consistent with that of NGC~7329, so we refer readers to Section~\ref{sec:ngc-7329} for a detailed description of their
common structural components. Note that, despite the presence of fragments (see Figure~\ref{fig:NGC945}), the grand-design spiral pattern of NGC~945 is
more prominent than that of NGC~7329. There is a companion galaxy, NGC~948, to the northeast of NGC~945. Since the two galaxies are clearly separated
on the image, we simply mask NGC~948 instead of modeling it simultaneously during the decompositions.

The galaxy was decomposed (Figure~\ref{fig:NGC945}; Table~\ref{tab:NGC945}) similarly to NGC~7329. We decompose the 1D profile of the galaxy by
excluding data from 4\arcsec{} to 34\arcsec{}, which is dominated by the bar and the inner ring. The uncertainties are obtained by expanding and
contracting the excluded range by shifting the start point by 1\arcsec{} and the end point by 5\arcsec{}.  We find that the empty region around the
bar is much less prominent compared with previous cases (see the Residual1 panel).  Nevertheless, we consistently break the disk into two parts to correct
for the systematic negative residuals around the bulge in Model2.  Spiral arms are accounted for in Model3, as usual. Similar to NGC~7329, part of the
surface brightness profile of NGC~945 that is dominated by the bar smoothly merges into the rest of the profile and would be unnoticed if it were not
for the ellipticity and PA profiles. Therefore, excluding the radii containing the bar has little to no impact on the 1D fit. The relative errors are
large because the bulge is quite weak. The bulge derived from 1D fitting is systematically stronger than that obtained from all three 2D fits; it has
larger size, larger \sersic{} index, smaller ellipticity, and therefore brighter apparent magnitude. Moreover, the orientation derived in 1D is quite
far from those obtained from 2D decomposition.

The unrealistically large bar sizes in Model1 and Model2 have the same causes as explained in Sections~\ref{sec:ic-5240} and \ref{sec:ngc-7329}.
We achieve a realistic-looking bar in Model3 by properly modeling the inner and outer disk. We note that the tendency for \(B/T\) to increase after
correcting for the empty region around the bar, a trend found in the previous four barred galaxies, is not borne out here. Besides having an
unrealistic bar size, the bar component of Model2 also has a considerably brighter central surface brightness compared with the other two models,
thereby suppressing the bulge to be fainter than it should be. Moreover, the ellipticity of the bulge component in Model2 is exceptionally higher than
those of the other two fits. This behavior was not observed in the other cases studied here.

\begin{deluxetable*}{lccccccc}
  \tablecaption{Best-fit Parameters for the Bulge of NGC~945 \label{tab:NGC945}}
  \tablecolumns{8}
  \tablehead{\multicolumn{1}{c}{\multirow{2}{*}{Model}} & \colhead{\(m_{R}\)} & \colhead{\(B/T\)} & \colhead{\(\mu_{{e},R}\)} &
    \colhead{\(n\)} & \colhead{\(r_{{e}}\)} & \colhead{\(\epsilon\)} & \colhead{PA} \\ \colhead{} & \colhead{(mag)} & \colhead{} &
    \colhead{(mag~arcsec\(^{-2}\))} & \colhead{} & \colhead{(\arcsec{})} & \colhead{} & \colhead{(\arcdeg{})}\\ \colhead{(1)} & \colhead{(2)} &
    \colhead{(3)} & \colhead{(4)} & \colhead{(5)} & \colhead{(6)} & \colhead{(7)} & \colhead{(8)}}
  \startdata
  Model0\tablenotemark{a} & \(15.09\pm0.43\) & \(0.051\pm0.015\) & \(20.27\pm0.39\) & \(1.73\pm0.35\) & \(2.89\pm0.80\) & \(0.073\pm0.015\) &
  \(11.85\pm3.25\) \\[0.05cm]
  \hline
  \rule{0pt}{3ex}Model1\tablenotemark{b} & \(15.50\substack{+0.00\\-0.00}\) & \(0.030\substack{+0.000\\-0.001}\) & \(19.48\substack{+0.01\\-0.00}\) &
  \(1.20\substack{+0.01\\-0.00}\) & \(1.87\substack{+0.01\\-0.00}\) & \(0.141\substack{+0.001\\-0.000}\) & \(49.75\substack{+0.48\\-0.20}\) \\[0.1cm]
  Model2\tablenotemark{c} & \(15.62\substack{+0.00\\-0.03}\) & \(0.027\substack{+0.001\\-0.000}\) & \(19.86\substack{+0.00\\-0.06}\) &
  \(1.46\substack{+0.01\\-0.04}\) & \(2.28\substack{+0.01\\-0.07}\) & \(0.324\substack{+0.000\\-0.032}\) & \(69.14\substack{+0.00\\-1.21}\) \\[0.1cm]
  Model3\tablenotemark{d} & \(15.30\substack{+0.00\\-0.04}\) & \(0.036\substack{+0.001\\-0.000}\) & \(19.88\substack{+0.00\\-0.01}\) &
  \(1.57\substack{+0.00\\-0.00}\) & \(2.43\substack{+0.00\\-0.01}\) & \(0.214\substack{+0.000\\-0.029}\) & \(61.14\substack{+0.00\\-1.88}\)
  \enddata
  \tablecomments{See Table~\ref{tab:NGC1411} for details.}
  \tablenotetext{a}{Model configuration: Bulge+Disk.}
  \tablenotetext{b}{Model configuration: Bulge+Bar+Disk.}
  \tablenotetext{c}{Model configuration: Bulge+Bar+Broken Disk(Part1+Part2).}
  \tablenotetext{d}{Model configuration: Bulge+Bar+Broken Disk(Part1+Spiral Part2).}
\end{deluxetable*}

\section{Discussion}
\label{sec:discussion}

\subsection{1D vs. 2D}
\label{sec:1d-vs.-2d}

Despite the inherent limitations of 1D bulge-to-disk decomposition discussed in
Section~\ref{sec:introduction}, we still perform traditional 1D fitting of the surface brightness
profiles for every galaxy in the sample. By comparing the bulge parameters derived from 1D analysis
with the corresponding ones from 2D fitting, we find that the results of 1D decomposition often
deviate significantly from those of 2D fitting, or show exceptionally large uncertainties. The
latter are caused by further loss of information when we exclude part of the surface brightness
profile, which may leave insufficient data to constrain the models. It is worth noting that such
cases are not exclusively barred galaxies. This shortcoming can be overcome by fitting more complex
multi-component models to the surface brightness profile, as advocated by
\citet{2016ApJS+Savorgnan+SBH_spheroids}. In addition to a bulge and a disk, they fit a bar, rings,
even nuclear components to the disk galaxies in their sample. We do not consider their 1D approach
superior to that adopted here or in \citet{2008AJ+Fisher+Sersic_Index}. Multi-component
decomposition in 1D, without constraints from geometric information, can be highly degenerate,
resulting in severely underestimated, unquantifiable uncertainties.

A central result of our analysis is that the structural parameters of bulges in general are very
sensitive to a variety of internal substructure, especially those located in the inner portions of
the galaxy. These substructures must be modeled properly, and the most direct and effective way of
doing is through detailed multi-component 2D fitting, as illustrated for the 10 prototype disk
galaxies studied here. In Section~\ref{sec:bulge-disk-decomp}, we show that most of the
morphological features (beyond axisymmetric models of bulge and disk), including bars, lenses,
rings, disk breaks, and spiral arms, can be modeled consistently by combinations of just a limited
set of analytic functions.  In order to successfully reproduce the complicated appearances of CGS
disk galaxies, we took a series of intermediate steps of model refinement, which was extremely
time-consuming.  A large number of free parameters were involved in this process, some of them
having to be fixed to reasonable values; analytic functions were chopped into blocks by
truncations. These methods are not traditionally used for bulge-to-disk decomposition.  However, we
demonstrate that they are necessary in order to obtain reliable bulge parameters with robust
estimates of their uncertainties.  Fortunately, not all of the features affect the bulge
measurements equally.  Some are critical but others less so.  In the rest of this section, we
discuss the relative importance and priorities of these features, aiming to answer two questions:
which ones can be neglected and, if so, how much difference will they make?

\subsection{Bars}
\label{sec:bars}

There is no doubt that bars should be modeled, or else bulge measurements will be significantly biased \citep{2004MNRAS+Laurikainen+BDBAR,
  2005MNRAS+Laurikainen+multi_comp_dec_S0,2008MNRAS+Gadotti+BUDDA2}. We always include a bar in the model whenever one is seen. However, there is no
consensus on how to model bars. The most commonly adopted assumptions for bars in the literature are the Ferrer function and the \sersic{} function
(e.g., Ferrer bars: \citealp{2004MNRAS+Laurikainen+BDBAR,2005MNRAS+Laurikainen+multi_comp_dec_S0,2006AJ+Laurikainen+BD_dec_S0,
  2010MNRAS+Laurikainen+phot_scal_rel_s0_sp, 2015ApJS+Salo+S4G_multi_comp_dec}; \sersic{} bars: \citealp{2008MNRAS+Gadotti+BUDDA2,
  2009MNRAS+Gadotti+SDSS_FP,2009ApJ+Weinzirl+BD_dec_OSUBGS_H,2014ApJ+Kim+Disk_Break,2015MNRAS+Head+multi_decom_coma}). We choose the modified Ferrer
function over the \sersic{} function as the default assumption for bars, in consideration of the extra freedom that its analytic form allows for the
central slope (\(\alpha\)) to decouple from the outer slope (\(\beta\)). Nevertheless, \sersic{} bars can offer equivalently good fits. To
quantify the impact of these two different assumptions on bulge parameters, we substitute the modified Ferrer bar in the final best fit with a
\sersic{} bar and refit the model. We find that these two different functional forms of bars yield consistent bulge parameters. Specifically,
  we notice that \(m_{R}\) and \(B/T\) show systematic trends when the bar model changes from the modified Ferrer function to the \sersic{} function,
  namely that \(m_{R}\) decreases and \(B/T\) increases. However, we find no clear trends for other bulge parameters. In general, the impact of the
  different functional forms of the bar model is small. As measured from the five barred cases, substitution of the best-fit modified Ferrer bar with a
  \sersic{} bar affects $m_{R}$ by \(\sim 0.05\,\mathrm{mag}\), $\mu_{e,R}$ by \(\sim0.05\,\mathrm{mag~arcsec^{-2}}\), and $B/T$, $n$, $r_e$, and
  $\epsilon$ by $\sim$5\%, 2\%, 3\%, and 2\%, respectively\footnote{We express variations of bulge parameters in terms of absolute variations for
    \(m_{R}\) and \(\mu_{e,R}\), but in terms of fractional variations for \(B/T\), \(n\), \(r_{e}\), and \(\epsilon\), in
    consideration that absolute variations of magnitude parameters are actually proxies for fractional variations of flux parameters.}.

We demonstrate that our choice of analytic function for the bar component is not crucial for bulge measurements. Therefore, we will consistently
model bars with the modified Ferrer function in future studies of CGS galaxies.

\subsection{Lenses and Rings}
\label{sec:lenses-rings}

We group lenses and rings into one category, for the reasons given in Section~\ref{sec:ngc-1533}. Both lenses and rings appear as shelves or ends of
shelves on surface brightness profiles, although lenses have a radial extent toward the galaxy center while rings do not. In fact, lenses and rings
have same implications on our model construction (see Model3 and Model4 of NGC~1411, and Model2 and Model3 of NGC~1533). As lenses and rings perturb
disk surface brightness in a similar way, it is not a surprise that they affect bulge parameters in a similar fashion.

NGC~1411 and NGC~2784 are two prototypical lensed galaxies in our sample. In addition, we would like to take NGC~1357 into consideration, because its
inner disk shares a similar surface brightness profile as the inner lenses of NGC~1411 and NGC~2784, even though this ``lens'' bears spiral arms. In
the case of NGC~1411, we show that the nuclear lens and the inner lens need to be modeled simultaneously, if we desire to get all bulge parameters
correctly. If the nuclear lens is not properly accounted for, it at least has a significant effect on the bulge \sersic{} index. In NGC~2784, we
show that if its inner lens is not properly modeled, it becomes incorporated into the bulge component, thereby leading to an overestimate of
\(B/T\). The same thing happens in the case of NGC~1357, if we consider the inner disk of this galaxy as an inner lens. By contrast, the inner lens of
NGC~1411 does not affect the bulge in the same way. Taking all three examples into account, we conclude that both nuclear lenses and inner lenses need
to be modeled because their influence on bulge structural parameters cannot be well predicted. Each case must be treated individually. As for the
outer lens, we show that in NGC~2784 this component only has a secondary effect on the bulge; neglecting the outer lens induces an error of only
0.06\,mag, 6.0\%, 0.08\,mag~arcsec\(^{-2}\), 3.6\%, 6.3\%, and 5.5\%, for \(m_{R}\), \(B/T\), \(\mu_{{e},R}\), \(n\), \(r_{{e}}\), and \(\epsilon\),
respectively.

Four out of the five barred galaxies contain inner rings that separate an inner redder disk from an outer bluer disk.  The only exception is NGC~1533,
which shows similar colors for both parts of its disk. We find that these rings demarcate a transition in surface brightness profile slope, from an
inner disk with a shallower profile to an outer disk with steeper slope.  We attribute these features to the dynamical influence of the bar.  The inner
rings of barred galaxies must be modeled, not for reproducing the rings themselves, but for correctly describing variations of surface brightness
across the rings to prevent the bulges from being systematically underestimated. Their disk profiles should fall in the category of Type \Rmnum{2}
profiles, but they can be easily missed, since azimuthally averaged profiles smooth out the deficiency of inner disk light by compensating it with the
excess emission from the bar.  We will revisit this issue along with other galaxies that show Type \Rmnum{2} profiles in
Section~\ref{sec:disk-breaks}.

We are not the first ones to notice or try to correct for such a bias. \citet{2008MNRAS+Gadotti+BUDDA2} found well-defined regions of conspicuous
negative residuals when fitting models that adopt exponential disks to NGC~4608 and NGC~5701. These two galaxies are morphologically similar to our
barred galaxies. Gadotti tried to improve their fits by breaking the disk components into two parts: an inner part with constant surface brightness
and an outer part that remains exponential (see their Figure~8). Gadotti's strategy is similar to ours, although we implement it differently.
\citet{2016MNRAS+Kim+bar_light_deficit} also noticed the depressed regions around bars, which they refer to as a \(\theta\)-shaped
morphology. They modeled the broken disks as two exponential subsections with different scale lengths and observationally confirmed the link between
bar strength and light deficit of the inner disks.

There is only one galaxy in our sample---NGC~1326---that has both a nuclear ring and an outer ring.  Consistent with the situation regarding the outer
lenses (e.g., as in NGC~2784), we find that modeling the outer ring makes little difference for the bulge parameters. Fractional errors caused by
omitting the outer ring from the 2D model are 0.01\,mag, 4.4\%, 0.02\,mag~arcsec\(^{-2}\), 2.3\%, 0.2\%, and 2.9\%, for \(m_{R}\), \(B/T\),
\(\mu_{{e},R}\), \(n\), \(r_{{e}}\), and \(\epsilon\), respectively. The nuclear ring is the only key morphological feature that we do not model in our
sample, although we do mask it to quantify its impact on the bulge parameters. Putting aside the issue of whether the nuclear ring is part of the
photometric bulge or not, we find that it affects \(n\) more than \(B/T\).

To verify the robustness of the above results, we further analyze another four galaxies that show visible outer lenses/rings: NGC~254, NGC~1302,
NGC~4984, and NGC~6893. In total we have a representative sample of six galaxies that exhibit outer lens/ring features: two of them barred that show
outer rings (NGC~1302 and NGC~1326), two unbarred but that show outer rings (NGC~254 and NGC~4984), and two unbarred with outer lenses (NGC~2784 and
NGC~6893). Our conclusion that outer lenses and rings have a minor impact on bulge measurements still holds with this enlarged sample. Their impact on
$B/T$ ranges from 2.7\% to 13.5\%, with a mean of 7.1\%, which is in fair agreement with the two cases discussed more extensively in this study.
Finally, we estimate that bypassing outer lenses/rings will cause a small error of 0.05\,mag, 7.1\%, 0.09\,mag~arcsec\(^{-2}\), 5.8\%, 5.3\%, and
4.8\%, for \(m_{R}\), \(B/T\), \(\mu_{{e},R}\), \(n\), \(r_{{e}}\), and \(\epsilon\), respectively.

To summarize: the inner lens/ring and the nuclear lens/ring deserve proper treatment. They need to be either included explicitly in the 2D model or
masked. The outer lens/ring only has relatively minor influence on bulge parameters compared with their counterparts on smaller sizes, and thus will
no longer be treated in the rest of the CGS disk galaxy sample.  Although bypassing the outer lens/ring will introduce some degree of uncertainty into
the bulge parameters, their typical values can be inferred from the case studies highlighted in this paper.

\subsection{Disk Breaks}
\label{sec:disk-breaks}

Disk breaks are readily recognized in the outer surface brightness profiles of NGC~7083 and NGC~6118.  Among barred galaxies, we identify another type
of ``disk break''; these occur at inner rings. The two are not the same: the former occur at the faint outskirt of their disks
(\(\sim23\)\,mag~arcsec\(^{-2}\)), whereas the latter are seen in the inner, bright regions (\(\sim21\text{--}22\)\,mag~arcsec\(^{-2}\)). Disk breaks
in barred galaxies are plausibly associated with bar-induced secular evolution, while disk breaks in unbarred galaxies occur at the edges of
spiral arms, suggesting that the underluminous and red outer disks may be caused by suppression of star formation therein and stellar migration
\citep[and references therein]{2016A&A+Marino+outer_disk_redden}. From the point of view of image decomposition, their effect on bulge structural
parameters is similar: \(B/T\) will be underestimated if disk breaks are not properly modeled.  The degree to which the bulge luminosity is
underestimated depends on where the break occurs. \citet{2014ApJ+Kim+Disk_Break} stressed the importance of accounting for disk break, otherwise the flux
of both the bulge and the bar will be underestimated. In our study, disk breaks in barred galaxies cause \(B/T\) to vary by \(\sim10\text{--}50\%\),
with a mean value of 32.5\%. For unbarred galaxies in our sample, the corresponding values are 31.7\% for NGC~7083 and 176\% for NGC~6118.  Note that
even though our sample does not include any Type \Rmnum{3} disk galaxies, we expect such disk breaks to be equally important if they were to occur at
similar positions in the disk as the disk breaks of Type \Rmnum{2} profiles.

\subsection{Spiral Arms}
\label{sec:spiral-arms}

Spiral arms are common features of galaxies in our sample. In total we have six galaxies that show recognizable spiral patterns in their disks:
NGC~1357, NGC~6118, and NGC~7083 are unbarred; IC~5240, NGC~945, and NGC~7329 are barred. We successfully reproduce their spiral patterns up to
three-arm features by applying coordinate rotation and, optionally, Fourier modes to their disk components. Note that spiral arms in our models are
not add-on components to disks.  They are actually azimuthally distorted disks.

We find that disk scale lengths become larger when spiral arms are invoked. The cause of this was discussed in detail in Section~\ref{sec:ngc-1357},
and will not be repeated here.  Our primary concern, however, is with the bulge, not the disk.  In general, spiral arms make only minor perturbations
to the bulge parameters.  This is especially true for barred galaxies, whose spiral arms stop at the inner ring, whereas they extend to the center in
unbarred systems.  For unbarred galaxies, ignoring spiral arms introduces to the bulge component an uncertainty of 0.14\,mag, 11.7\%,
0.24\,mag~arcsec\(^{-2}\), 10.1\%, 13.6\%, and 0.4\% for \(m_{R}\), \(B/T\), \(\mu_{e,R}\), \(n\), \(r_{{e}}\), and \(\epsilon\), respectively; for
barred galaxies, the corresponding values are 0.03\,mag, 2.2\%, 0.03\,mag~arcsec\(^{-2}\), 1.1\%, 1.4\%, and 0.4\% for \(m_{R}\), \(B/T\),
\(\mu_{{e},R}\), \(n\), \(r_{{e}}\), and \(\epsilon\), respectively.

It is worth remarking that our sample only includes galaxies that show clear two-arm or multiple-arm patterns.  There are galaxies with spiral
patterns that are so flocculent that they lie beyond the capabilities of available image-fitting tools. Thus, their impact on bulge parameters cannot
be quantified straightforwardly. Nevertheless, flocculent spiral patterns are weaker non-axisymmetric perturbation to disk surface brightness than
continuous spiral arms. So we hypothesize that flocculent spiral patterns should affect bulge parameters even less than the stronger
spiral patterns investigated here.

\subsection{Prescription to Estimate Uncertainties of 2D Fits of Bulges}
\label{sec:prescr-estim-uncert}

In the previous sections, we have systematically examined the relative importance of including all major morphological features in 2D image
decomposition of galaxies.  We are most concerned with knowing which features need to be treated, and which can be ignored, for the purposes of
obtaining robust photometric parameters of bulges. The aim is to arrive at a set of useful guidelines that can be applied to future studies of larger
samples, for which the extremely detailed and time-consuming approach adopted in this pilot study would be impractical.

First, we show that outer lenses and rings have the lowest priority. Ignoring such features in the outskirts of disks typically will cost only precisions
of 0.05\,mag, 7.1\%, 0.09\,mag~arcsec\(^{-2}\), 5.8\%, 5.3\%, and 4.8\% for \(m_{R}\), \(B/T\), \(\mu_{{e},R}\), \(n\), \(r_{{e}}\), and \(\epsilon\),
respectively.  Whenever galaxies show outer lenses or rings, we will simply ignore them in future decomposition of CGS galaxies aimed at bulge
studies. Second, spiral arms are also found not to be crucial in constructing surface brightness models. The consequences of ignoring spiral arms
differ between unbarred and barred galaxies. For unbarred galaxies, spiral arms affect bulge parameters at the level of 0.14\,mag, 11.7\%,
0.24\,mag~arcsec\(^{-2}\), 10.1\%, 13.6\%, and 0.4\% for \(m_{R}\), \(B/T\), \(\mu_{{e},R}\), \(n\), \(r_{{e}}\), and \(\epsilon\), respectively; for
barred galaxies, the corresponding effects are even milder, namely 0.03\,mag, 2.2\%, 0.03\,mag~arcsec\(^{-2}\), 1.1\%, 1.4\%, and 0.4\% for \(m_{R}\),
\(B/T\), \(\mu_{{e},R}\), \(n\), \(r_{{e}}\), and \(\epsilon\), respectively. Ignoring these complicated features will greatly speed up the 2D fitting. Apart from
the outer lenses/rings and spiral arms, we find that all other major secondary morphological components need to be properly considered.

We also reveal some hidden uncertainties of bulge parameters when modeling lenses and rings. As shown in the cases of NGC~1411 and NGC~1533,
disk breaks, lenses, and rings, along with the underlying disk, can be modeled mathematically interchangeably. And different
mathematical representations of disk surface brightness certainly introduce variations in bulge parameters, as estimated from these two cases:
0.09\,mag, 6.7\%, 0.15\,mag~arcsec\(^{-2}\), 8.0\%, 6.9\%, and 7.7\% for \(m_{R}\), \(B/T\), \(\mu_{e,R}\), \(n\), \(r_{{e}}\), and \(\epsilon\),
respectively. These numbers will be applied to future studies whenever lenses and disk breaks (with or without rings) are modeled. The typical
  variations as measured in these two cases should be included in the parameter error budget as extra uncertainties added in quadrature.

Suppose that we are dealing with a barred galaxy that possesses an inner ring and spiral arms. Based
on the lessons learned in this study, we would prepare a 2D model that includes a \sersic{} bulge, a
modified Ferrer bar, and a broken axisymmetric disk. After obtaining the best-fit parameters of this
model, we estimate two sources of model-dependent uncertainties: (1) omission of spiral
arms, and (2) assumption of the mathematical representation of the inner ring along
  with the underlying disk. H. Gao et al. (2017, in preparation) will adopt a similar strategy for
investigating the bulge properties of the entire sample of S0 and spiral galaxies in CGS. We will
not explore multiple 2D models for every galaxy. Instead, for each galaxy we will construct a single
model with the minimum number of necessary components based on the particular morphological
attributes of the galaxy, guided by the experience gained from this study.  The final error budget
of the bulge parameters will take into account the model-induced systematics described in this paper,
as well as uncertainties due to sky subtraction.

\section{Summary}
\label{sec:summary}

The main goal of this study is to investigate the degree to which the photometric parameters of the
bulge are influenced by the manner in which we model the various complex morphological features
typically seen in high-quality, well-resolved optical/near-infrared images of nearby
galaxies. Without carefully and systematically treating each feature in turn, it is impossible to
predict which will matter and which will not.  Of course, it is not our intention to dissect every
morphological detail; that would have little significance.  Nor can we realistically explore the
full array of morphological diversity displayed throughout the Hubble sequence.  Yet, the vast
majority of normal (i.e.\ isolated, non-interacting) disk galaxies do contain a sufficiently well-defined 
set of basic ``building block'' components---the very ones that justify detailed morphological
classification (e.g., \citealp{1991Springer+de_Vaucouleurs+RC3,2015ApJS+Buta+Morphology_S4G})---such
that we can gain some useful insights from a detailed investigation of a limited set of
prototypes. Based on case studies of these prototypes, we determine which morphological components
are essential and which are peripheral to the robust measurement of bulge parameters, with the
intention of developing a set of guidelines that would enable us to more efficiently perform
bulge-to-disk decomposition of a large sample of galaxies, without loss of accuracy.

Toward this end, we present 1D and 2D bulge-to-disk decompositions of $R$-band images of 10
representative disk galaxies selected from the CGS, spanning Hubble types S0 to Scd and SB0 to
SBc. We find that the 1D approach is not appropriate for most cases in our sample. We perform 2D
multi-model and multi-component decomposition using the latest version of \galfit{}. Thanks to the
great flexibility provided by \galfit{}, we are able to reproduce in a consistent manner not only
the bulge, bar, and disk components of the galaxies, but also all extra principal morphological
features, including lenses, rings, and disk breaks, on both small and large scales, as well as
spiral arms. By exploring different input surface brightness models for \galfit{}, we identify
morphological features that are considered to be essential constituents of adequate surface
brightness models and also identify those that can be ignored. The typical variations of bulge
structural parameters measured across different surface brightness models serve as estimates of
typical model-induced uncertainties.

Our main conclusions are as follows:

\begin{itemize}
\item Under most circumstances, outer lenses, outer rings, and spiral arms can be excluded from the model.  These components affect bulge magnitudes
  only at the level of \(\lesssim 0.1\,\mathrm{mag}\) and \(B/T\), \(r_{e}\), and \sersic{} \(n\) at the level of \(\lesssim 10\%\).

\item Components that intimately overlap with the bulge, such as nuclear lenses/rings and inner lenses/rings, must be treated properly.  Specifically,
  inner lenses/rings have a considerable impact on \(B/T\) and \sersic{} \(n\), while nuclear lenses/rings, when they are not regarded as part of the
  photometric bulge, at least affect \sersic{} \(n\).

\item We confirm that bars and disk breaks, including inner disk breaks induced by bars, need to be modeled.
\end{itemize}

\acknowledgments

We thank the referee for a thorough and helpful review of this manuscript. We thank Aaron Barth,
Zhao-Yu Li, and Chien Peng for their contributions to the CGS project. We are grateful to Song
Huang, Chien Peng and Zhao-Yu Li for helpful discussions and suggestions throughout this study. We
appreciate the generosity of Taehyun Kim and Minjin Kim for sharing their experience in modeling
disk breaks and spiral arms, respectively. Financial support for the work was provided by the
National Key Program for Science and Technology Research and Development grant 2016YFA0400702. This
research has made use of the NASA/IPAC Extragalactic Database (NED) which is operated by the Jet
Propulsion Laboratory, California Institute of Technology, under contract with the National
Aeronautics and Space Administration.

\appendix

\section{Technical Biases of 1D Decomposition}
\label{sec:technical-biases-1d}

We derive radial surface brightness profiles of the galaxies using the IRAF task \ellipse{}. The task fits elliptical isophotes to sky-subtracted
images and then outputs the azimuthally averaged surface brightness intensity along each isophotal ellipse, along with the ellipticity and position
angle of each isophote, as a function of the semi-major axis length. The major axis length of the isophotes grows logarithmically (geometrically) or
linearly, as specified by the user. Intuitively one might expect that a logarithmically sampled profile puts more weight on the brighter, central
region of the galaxy. But rigorously speaking, how the surface brightness profile is sampled (logarithmically or linearly), in conjunction with other
factors, affects the fitting results of the surface brightness profile in an intricate way. In this section, starting with considerations on how the
surface brightness profile should be sampled, we explore several aspects of 1D bulge-to-disk decomposition and biases that are introduced during this
process.

To produce a sparsely sampled profile is equivalent to merging adjacent data of a more intensively
sampled profile. However, how these adjacent data are merged actually depends on how the task
assigns pixel values along the elliptical isophotal path. The task provides three methods to sample
the image: bi-linear interpolation and either the mean or median over elliptical annulus
sectors. The bi-linear interpolation method is the default option; even if either of the other two
methods is selected, it is still enforced by the task in the central region of the galaxy. The
bi-linear method only extracts a one pixel wide sample of the image for each isophote, which misses
many pixels for isophotes with increment of semi-major axis length larger than 2 pixels. The mean or
median method makes full use of the pixels in successive annuli but is computationally more
expensive than the bi-linear interpolation method.

Returning to the issue of merging data in the intensively sampled profile to produce a more sparsely sampled one, assume that we adopt the mean method
for the task. If error propagation is properly done, supposing that \(\sigma^{*}\) is the new error of a merged data point, then
\(\sigma^{*}=\sigma/\sqrt{N}\), where \(N\) is the number of data points to be merged in the intensively sampled profile, and \(\sigma\) is their
error. By weighting the merged data in the sparsely sampled profile with its new error and the original data that have not been merged in the
intensively sampled profile with their errors, one finds that they contribute equally to \(\chi_{\nu}^{2}\), assuming a good fit is achieved (i.e.,
data deviations from the model are comparable to their error bars). So, ideally speaking, how the surface brightness profile is sampled
(logarithmically or linearly) does not matter. However, this reasoning only applies to a local flat part of the profile.  In reality the profile may
be highly curved. Curvature in the profile will introduce non-Poisson fluctuations into the error budget of a sparsely sampled profile, which makes
this issue more complicated. Thus, to be rigorous, the fitting results of profiles that are sampled in different ways should differ from one
another, and the differences depend on various aspects of the fitting procedure, such as which image-sampling technique is adopted, how error
estimation is conducted, the detailed shape of the profiles (highly curved or flat), and how data are weighted in the least \(\chi^{2}\) fitting
(S/N-based weighting or non-S/N-based weighting). From a statistical point of view, there is no rigorously perfect choice.

Here, we simply make a choice that we deem to be realistic. To save computation time, the surface brightness profiles in this paper are
logarithmically spaced in order to sample the bright (bulge) region intensively and the faint outskirts sparsely.  We adopt the default bi-linear
interpolation sampling method, and all data are weighted by their measurement errors. \citet{2016ApJS+Savorgnan+SBH_spheroids} argued that a linearly
sampled profile and a logarithmically sampled profile implicitly imply different weighting schemes.  We disagree, for the reasons given above.  They
then adopted linearly sampled surface brightness profiles and assigned equivalent weights to their data, according to their definition of what
constitutes a good fit.

Having established the method to extract surface brightness profiles from galaxy images, the next step is to perform the least \(\chi^{2}\) fitting.
Should \(\chi_{\nu}^{2}\) be based on surface brightness flux or on surface brightness magnitude? Both can be expressed in a similar way:
\begin{equation}
  \label{eq:chi_flux}
  \chi_{\nu}^{2}=\frac{1}{N_{\mathrm{dof}}}\sum_{i=1}^{N_{\mathrm{data}}}\frac{\left(f_{\mathrm{data},i}-f_{\mathrm{model},i}\right)^{2}}{\sigma_{f,\mathrm{data},i}^{2}},
\end{equation}
\begin{equation}
  \label{eq:chi_mag}
  \chi_{\nu}^{2}=\frac{1}{N_{\mathrm{dof}}}\sum_{i=1}^{N_{\mathrm{data}}}\frac{\left(\mu_{\mathrm{data},i}-\mu_{\mathrm{model},i}\right)^{2}}{\sigma_{\mu,\mathrm{data},i}^{2}},
\end{equation}
where \(\chi_{\nu}^{2}\) is the reduced \(\chi^{2}\), \(N_{\mathrm{dof}}\) is the number of degrees of freedom, \(N_{\mathrm{data}}\) is the number of
data points in the surface brightness profile, \(f\) and \(\mu\) represent surface brightness flux and surface brightness magnitude at a certain
radius, for both data and model, respectively, and \(\sigma_{f}\) and \(\sigma_{\mu}\) are their corresponding errors. These two definitions are
formally different from a mathematical point of view, but both seem to be reasonable for the purposes of bulge-to-disk decomposition. As discussed
above, there is no way to guarantee absolute rigorousness in 1D fitting.  Here we add that even the choice of \(\chi_{\nu}^{2}\) is unclear.
Nevertheless, they are consistent with each other in high-S/N regions. Our tests indicate that, in the high-S/N regime of our data, in practice these
two definitions of \(\chi_{\nu}^{2}\) produce consistent fitting results.  We adopt Equation~(\ref{eq:chi_flux}) throughout this paper, in view of the
convenience of estimating errors in terms of flux rather than magnitude.

\section{Sky Level Measurements}
\label{sec:sky-level-meas}

Bulge-to-disk decomposition of these bright galaxies is not sensitive to sky measurements errors, because the bulge is a high-surface-brightness
  component that lies significantly above the sky. Our approach to sky measurement is qualitatively similar to that of \citet{2011ApJS+Li+CGS2} and
\citet{2013ApJ+Huang+CGS3}. We measure the sky level in two ways: (1) a direct/model-independent approach for 1D bulge-to-disk decomposition (similar
to \citealp{2011ApJS+Li+CGS2}), and (2) an indirect/model-dependent approach for 2D bulge-to-disk decomposition (similar to
\citealp{2013ApJ+Huang+CGS3}). The indirect approach is necessary for galaxies whose sizes are sufficiently large relative to the field-of-view of the
image such that not enough blank sky area is available for direct sky measurement.  Below we prove that these two approaches are internally
consistent.  For convenience and simplicity, our future 2D decomposition analysis will adopt the indirect approach.

\subsection{Direct Approach to Measure the Sky Level}
\label{sec:direct-appr-meas}

This approach begins by masking all pixels containing signal from the science target and other field objects.  We generally mask the central galaxy
with an elliptical region of semi-major axis length \(3-5R_{80}\) of the galaxy, where $R_{80}$ is the radius enclosing 80\% of the total flux. After
further masking other field objects (for a description of the procedure, see \citealp{2011ApJS+Li+CGS2}), any remaining unmasked pixels are considered
to be sky pixels. We perform \(3\sigma\) clipping for the sky pixel values, and then compute the median of the cleaned sample as the sky level.

To characterize large-scale spatial fluctuations in the sky, we drop random boxes onto the sky region (i.e., beyond the elliptical mask of the central
galaxy) of the sky-subtracted image (see Appendix~B of \citealp{2013ApJ+Huang+CGS3}). If the sky level has been adequately determined and subtracted,
the average pixel values in each box should oscillate around 0. The root mean square of the
box-averaged residuals serves as a measurement of
uncertainty of the sky level \(\sigma=\sqrt{\sum_{i=1}^{N}{\rm res}_{i}^{2}/N}\), where \(N\) is the number of boxes and res$_i$ is the the average
pixel value in the \(i\)th box. We measure the sky level uncertainty from the sky-subtracted residuals in order to be consistent with our indirect
approach of sky measurement, wherein the sky level uncertainty is also determined from the sky-subtracted image. The number of random boxes is limited
by the box size (the larger the box size, the fewer useful, non-overlapping ones left). We explore a range of varying box sizes, starting from
\(20\times20\) pixels and increasing until \(\sigma\) ceases to decrease, but at no time do we allow there to be fewer than 20 useful boxes. The final
largest box size is adopted, and we repeat the process of generating random boxes at least 30 times, with this given box size. The mean of all
\(\sigma\) is finally adopted as the uncertainty of the measured sky level. In this paper, the sky level measured from this direct approach is used to
prepare sky-subtracted images for 1D decomposition, and it also serves as the initial guess of the sky level parameter in 2D decomposition.

\subsection{Indirect Approach to Measure the Sky Level}
\label{sec:indir-appr-meas}

In this approach, the sky level is solved as an additional component---modeled as a tilted plane---in the 2D fitting process.  As such, the best-fit
sky level depends on how the galaxy surface brightness model is constructed.  In general, if the surface brightness model of the galaxy is reasonably
accurate, the precision of the sky level is limited by the intrinsic large-scale variations of the sky, which can be inferred from the original (not
sky-subtracted) image \citep{2013ApJ+Huang+CGS3}. Here, we consider a more general approach to quantify how well the best-fit sky level approximates
the ``true'' sky level, irrespective of the accuracy of the surface brightness model of the galaxy.  Whether or not the best-fit sky level describes
well the ``true'' sky is immediately reflected in the sky-subtracted residuals in the sky region, as only the best-fit sky component should be
subtracted from the original image since galaxy components may help to compensate for the ``true'' sky level (e.g., a high-\(n\) \sersic{} component
with an extended wing). We measure the sky level uncertainty using the method described in Section~\ref{sec:direct-appr-meas}, by dropping random
boxes onto the same sky region of the sky-subtracted image, only that in this case the subtracted sky is determined from a best-fit global 2D
model. The choice of box size is kept the same as that determined from the direct approach. The sky level uncertainty estimated in this way not
only contains the large-scale fluctuations of the sky, but also incorporates possible biases that may be introduced by an improper surface brightness
model.

\subsection{Internal Consistency of Direct and Indirect Sky Level Measurements}
\label{sec:intern-cons-direct}

\begin{figure}
  \epsscale{1.16}
  \plotone{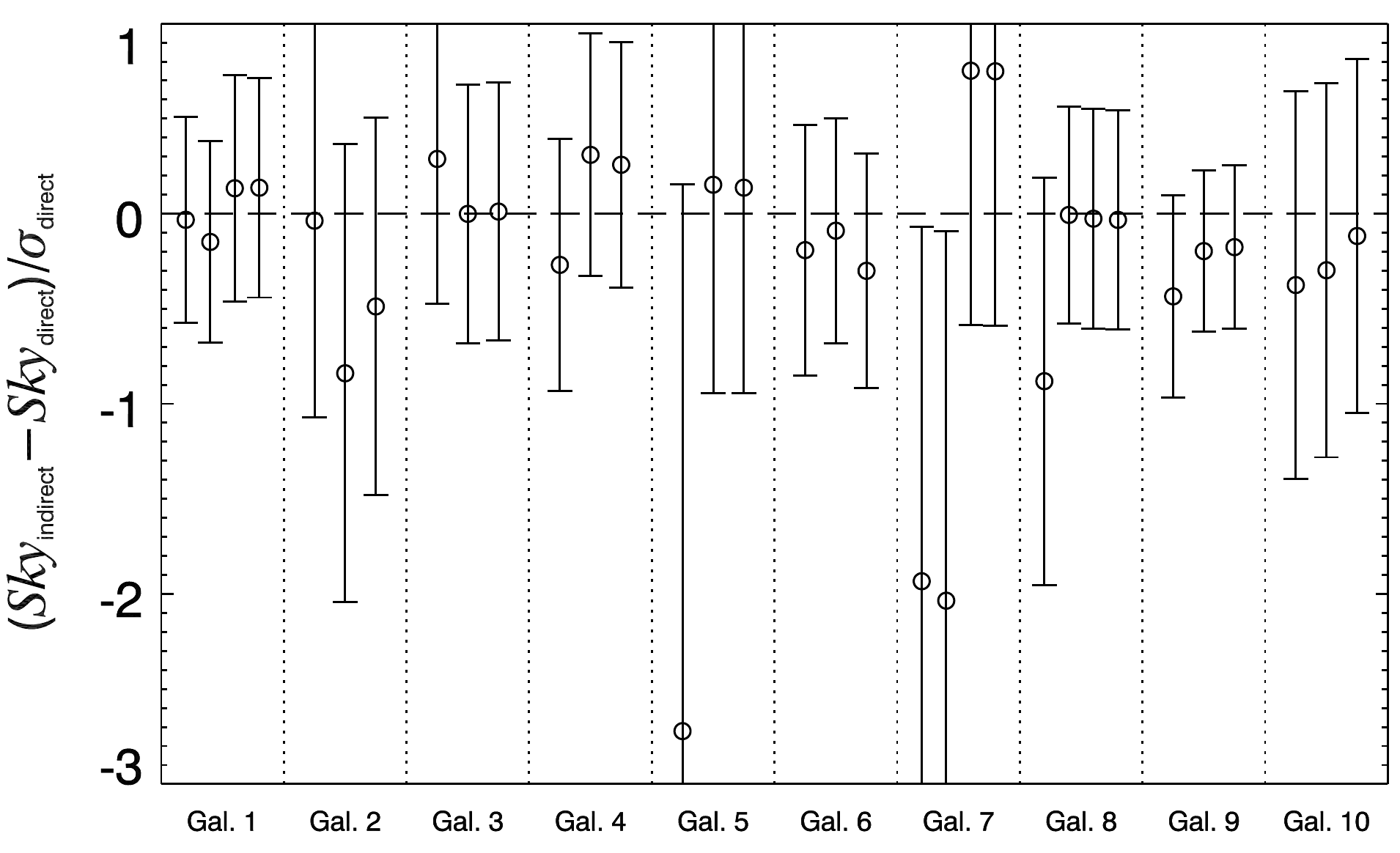}
  \caption{Comparison of sky levels measured by the indirect and direct approaches. The scatter of the indirect sky level measurements around their
    corresponding direct measurements are divided by the uncertainties of the direct measurements.  The uncertainties of the indirect measurements are
    also scaled accordingly.  The comparison is conducted in groups, one per galaxy. Groups of indirect measurements are separated by dotted
    lines. Within any given group, the measurements are slightly offset horizontally for clarity, ordered by the increasing complexity of their
    corresponding 2D models (see Figure~\ref{fig:NGC1411}--\ref{fig:NGC945} in this paper). The numbered sequence of galaxies along the horizontal
    axis is consistent with the top-to-bottom sequence of galaxies in Table~\ref{tab:sample_property}. \label{fig:sky_comp}}
\end{figure}

As we use different approaches to measure the sky level in our 1D and 2D analysis, it is important
to consider the extent to which this might affect our comparison between the 1D and 2D models, as
well as among the different 2D models. Figure~\ref{fig:sky_comp} compares the sky levels derived by
these two approaches. The sky levels obtained by the direct approach are adopted as the ``ground
truth,'' and the scatter of the indirect measurements around their corresponding ``ground truth''
are normalized by the uncertainty of the direct measurement. We find that most of the indirect
measurements agree well with their corresponding direct measurements (within
\(1\sigma_{\mathrm{direct}}\)), especially when considering their error bars. Only three indirect
measurements deviate from their direct measurements more than \(1\sigma_{\mathrm{direct}}\), one for
NGC~6118 and two for NGC~1326 (aliased as Gal. 5 and Gal. 7 in Figure~\ref{fig:sky_comp}). We trace
these discrepancies back to the 2D models whose sky levels have been underestimated due to
inadequate treatment of the truncation of the outer disks. On the other hand, their error bars are
also correspondingly large.  Figure~\ref{fig:sky_comp} shows that all indirect measurements, even
the three extreme ones, overlap, within their uncertainties, with their corresponding direct
measurements.  This ensures that whatever differences between the 1D and 2D bulge parameters that
are caused by biases in sky level measurements have been incorporated in the uncertainties of bulge
parameters. Thus, we conclude that the differences between the 1D and 2D bulge parameters are
genuine, as they cannot be fully accounted for by the uncertainties. The same holds for comparison
of bulge parameters among the various 2D models for the same galaxy.

\end{CJK*}
\bibliographystyle{aasjournal}
\bibliography{myref1}
\end{document}